\documentstyle[multicol,aps,prl,epsf]{revtex}
\begin{document}
\title{\bf Series Expansion Calculation of Persistence Exponents}

\author{George C. M. A Ehrhardt and Alan J. Bray}

\address{Department   of   Physics   and  Astronomy,   University   of
Manchester, Manchester, M13 9PL, UK}

\date{\today}

\maketitle

\begin{abstract}
\noindent We consider an  arbitrary Gaussian Stationary Process $X(T)$
with  known  correlator  $C(T)$,   sampled  at  discrete  times  $T_n=
n\Delta\!T$.  The  probability that $(n+1)$ consecutive  values of $X$
have  the same  sign decays  as  $P_n \sim  \exp(-\theta_D T_n)$.   We
calculate  the discrete  persistence exponent  $\theta_D$ as  a series
expansion in the  correlator $C(\Delta T)$ up to  $14^{th}$ order, and
extrapolate to $\Delta T=0$ using  constrained Pad\'e  approximants to
obtain the continuum persistence  exponent $\theta$. For the diffusion
equation  our results  are in exceptionally good agreement with recent  
numerical estimates.

\medskip\noindent {PACS numbers: 05.70.Ln, 05.40.-a, 02.50.-r, 81.10.Aj}
\end{abstract}

\begin{multicols}{2}

Persistence of a continuous stochastic process has been the subject of
considerable   recent   interest    among   both   theoreticians   and
experimentalists   in   the   field   of   nonequilibrium   processes.
Persistence  is  the probability  $P(t)$  that  a stochastic  variable
$x(t)$ of zero  mean does not change sign up to  time $t$.  In systems
which  do not  possess a definite  length  or time  scale, $P(t)  \sim
t^{-\theta}$ at late times, where the persistence exponent $\theta$ is
non-trivial.   Systems studied  include  reaction-diffusion processes,
phase-ordering kinetics,  fluctuating interfaces and  simple diffusion
from   random    initial   conditions   \cite{review}.    Experimental
measurements have  been carried  out on breath  figures \cite{marcos},
liquid crystals \cite{yurke}, soap froths \cite{tam}, and diffusion of
Xe gas in one dimension \cite{wong}.

Even     for    Gaussian     processes,  few   exact results for 
$\theta$ are  known, the random walk and  random acceleration problems
being two  exceptions \cite{review}.  For a  general Gaussian process,
the  main analytic  method is  the Independent  Interval Approximation
(IIA) \cite{IIA1,IIA2}, which is based on the assumption that the time
intervals between zeros of $x(t)$, measured on a logarithmic time scale, 
are independently distributed.  This is the case for Markov processes  
such as the random walk, but not for most processes of physical interest 
such as simple diffusion from a random initial condition.  Despite
this,  the  IIA gives  surprisingly  good  results  for the  diffusion
equation when compared with numerical data \cite{IIA1,IIA2,NL}. However,  
the IIA method is not systematically improvable, nor is there a way of 
estimating the errors involved in the approximation.

In this Letter we introduce a technique for calculating $\theta$ using
a  systematically  improvable  series  expansion.  This  expansion  is
motivated by work on ``discrete'' persistence \cite{dpers1,dpers2}, so
a brief explanation of this  is necessary. First, however, let us note
that the  non-stationary variable $x(t)$  may be mapped to  a Gaussian
stationary  process  (GSP)  for the  variable  $X(T)=x(t)/\sqrt{\left<
x^2(t)\right>}$ by use of the log-time transformation, $T=\ln t$. Then
the persistence  has the  form $P(T) \sim  \exp(-\theta T)$  for large
$T$, and the correlator of  the GSP, $C(T) = \left< X(T) X(0)\right>$,
is normalized  to unity at $T=0$.  Discrete  persistence addresses the
following problem: Although the underlying process governing $X(T)$ is
continuous, in many experimental  measurements of persistence one only
samples the process at discrete  times.  It is therefore possible that
$X(T)$ may  cross and then re-cross zero  between samplings, resulting
in a false positive classification of the persistence of $X(T)$.  Such
undetected  crossings   will  result  in   the  persistence  exponent,
$\theta_D$, of  the discretely sampled process being  smaller than the
continuum exponent $\theta$.  Let us consider a system that is sampled
logarithmically  in time  $t$, and  hence uniformly  in $T$,  with the
spacing  between  samplings  being  $\Delta\!T$.   Let  $P_n$  be  the
persistence probability after $n$ samplings.  Then for large $n$, $P_n
\sim  \rho^n$  where  $\rho=\exp(-\theta_D \Delta\!T)$  \cite{dpers1}.
Notice that for  $\Delta\!T \to 0$ we have $\rho  \to 1$ and $\theta_D
\to \theta$, while  for $\Delta\!T \to \infty$ we  have $\rho \to 1/2$
(since the measured values,  $X(n\Delta\!T)$, are uncorrelated in this
limit \cite{dpers1}) and hence $\theta_D \to 0$.

In previous work the study of discrete persistence has been limited to
Markovian  \cite{dpers1}, and  ``weakly  non-Markovian'' \cite{dpers2}
processes, where the latter are defined here as processes which can be
written as Markovian in a finite number of variables. For example, the
random acceleration process, $\ddot{x}  = \eta(t)$, where $\eta(t)$ is
Gaussian white noise, is non-Markovian  but can be re-expressed as the
two  Markov processes  $\dot{x}  =  v$ and  $\dot{v}  = \eta(t)$.   In
practice, the methods developed in \cite{dpers1,dpers2} become 
impractical for more than two Markov processes.

By  contrast,  the approach  developed  here  can  be applied  to  any
Gaussian  stationary  process, including  those  which are  ``strongly
non-Markovian'', meaning that re-expressing them in terms of Markovian
processes  requires an  infinite  number of  variables.   This is  the
generic case. Our results show, for example, that the discrete 
measurement regime employed in the diffusion experiment of 
ref.\ \cite{wong} has a negligible effect on the measured exponent. 

Our approach is to develop a series expansion for the quantity $\rho =
\exp(-\theta_D\,\Delta  T)$  in   powers  of  the  correlator  between
neighboring  discrete  times, $C(\Delta  T)$,  to  give very  accurate
estimates of $\theta_D$ for almost  all non-zero values of $\Delta T$.
Extrapolating the series  to infinity using Pad\'e approximants,
and imposing the constraint $\rho(\Delta T=0) = 1$, gives further
improvement for  the discrete  case, and an  estimate of  the exponent
$\theta$ for the continuum limit. Furthermore, for sufficiently smooth
processes (which includes those  described by the diffusion equation),
we  derive  a   second  constraint,  $[d\theta_D/d(\Delta  T)]_{\Delta
T=0}=0$,  which,  imposed  on  the Pad\'e  approximants,  yields  very
accurate  estimates   of  $\theta$  for   diffusive  persistence  (and
presumably  for  other sufficiently  smooth  processes).  The  results
obtained compare favorably with the IIA, and have the added advantage  
that by  calculating more terms in the series one may systematically 
improve the result,  while also providing an indication of the errors 
associated with the calculated value of $\theta$.

The expansion starts from the obvious identity
\begin{equation} 
P_n = \left< \prod_{i=1}^n \Theta [X(i \Delta \! T)] \right>
\label{10} 
\end{equation}
where $\Theta(x)$  is the Heaviside step function  and the expectation
value is taken in the  stationary state.  One may write $\Theta(X_i) =
(1+\sigma_i)/2$,  where $\sigma_i  \equiv  {\rm sign}[X(i\Delta\!T)]$,
and expand the product to give,
\begin{equation} 
\!\!\!\!    P_n   ={1\over    2^n}   \left(\!1   +\!    \sum_{1=i<j}^n
\left<\sigma_i     \sigma_j     \right>     +     \!\!\!\!\!\!\!\!\!\!
\sum_{1=i<j<k<l}^n  \!\!\!\!\!\!    \left<\sigma_i  \sigma_j  \sigma_k
\sigma_l \right> +\ldots \right)
\label{20} 
\end{equation}
where the terms with odd numbers of $\sigma$s vanish since the process
is  symmetric under  $X \to  -X$.  To  evaluate the  terms we  use the
representation
\begin{equation} 
\sigma_l= {1\over  i \pi} \lim_{\epsilon  \to 0} \int_{-\infty}^\infty
{dz_l\,z_l\, e^{i z_l X_l}  \over (z_l-i \epsilon) (z_l+i \epsilon)} \
.  \label{30}
\end{equation} 
Evaluating  the required averages  of the  Gaussian process  gives the
correlation functions appearing in (\ref{20}):
\begin{equation}
\left     <\sigma_{l_1}\ldots\sigma_{l_m}\right     >    =
\int\prod_{i=1}^m\left({dz_i  \over z_i}\right)\exp\left(-{1\over 2}
z_{\alpha}\,C_{\alpha\beta}\, z_{\beta}\right),
\label{40}
\end{equation}
where  $C_{\alpha\beta}= \langle X[\alpha\Delta\!T]\,X[\beta\Delta\!T]
\rangle  =  C(|\alpha-\beta|\Delta  T)$,   and  there  is  an  implied
summation over $\alpha$ and $\beta$ from $1$ to $m$. Note also that we
have already taken the limit  $\epsilon \to 0$ in (\ref{40}), with the
understanding that all integrals are now principal part integrals.

We  now expand  the exponential  in equation  (\ref{40}) in  powers of
$C_{\alpha,\beta}$  ($\alpha  \ne \beta$)  and  leave  the terms  with
$\alpha=\beta$  unexpanded  (noting  that  $C_{\alpha\alpha}=1$).   By
symmetry, only  terms which generate odd powers  of every $z_{\alpha}$
in  the expansion  (to  give  even powers  overall  in the  integrand,
through the factors $1/z_i$)  give a non-zero integral.  This suggests
a simple  diagrammatic representation for the terms  in (\ref{20}), as
given  by  (\ref{40}). On  a  one-dimensional  lattice containing  $n$
sites,  with lattice  spacing $\Delta\!T$,  draw $m$  vertices  at the
locations $l_1,l_2,\ldots,l_m$.  Connect the  vertices by lines in all
possible ways (summing over  these different possibilities) subject to
the  constraint that  each vertex  is connected  to an  odd  number of
lines.   Associate  a  factor  $\sqrt{2  \pi}  (p-2)!!$  (coming  from
evaluating the  Gaussian integrals) with  each vertex of order  $p$, a
factor $(-C_{l_il_j})^r/r!$  with the $r$ lines  connecting site $l_i$
to site $l_j$, and an  overall factor $(\pi i)^{-m}$ with the diagram.
This suffices to evaluate the integrals in (\ref{40}).  Evaluating the
summations  in (\ref{20}) involves  enumerating all  configurations of
the vertices  on the lattice for  a given ordering of  the points, and
noting that  the factor $C_{l_il_j}$  associated with a given  line is
equal  to $C(q\Delta\!T)$, where  $q=|l_i-l_j|$ is  the length  of the
line in units of $\Delta\!T$.   The configurations we need to consider
at  a  given  order are  determined  by  noting  that for  almost  all
processes of physical interest  (including all those considered here),
the  correlator  $C(q\Delta\!T)$  decreases  exponentially  for  large
argument.  Since  our expansion is  a large $\Delta\!T$  expansion, we
define $C(\Delta\!T)$  as $1^{st}$ order small  and $C(q\Delta\!T)$ as
$q^{th}$ order  small. The  order of  a diagram is  then equal  to the
total length  of its lines (measured  in units of  the lattice spacing
$\Delta\!T$).

To  illustrate this  approach we  show in  Figure 1  all  the diagrams
contributing to $\langle
\sigma_i\sigma_{i+1}\sigma_{i+2}\sigma_{i+3}\rangle$  up  to  $4^{th}$
order. The first  diagram is $2^{nd}$ order, while  the remaining five
are  $4^{th}$ order.

\begin{figure} 
\narrowtext\centerline{\epsfxsize\columnwidth
\epsfbox{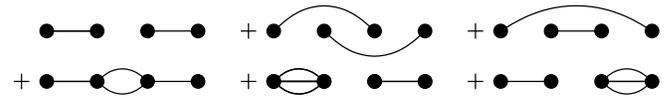}}
\caption{Contributions                   to                   $\langle
\sigma_i\sigma_{i+1}\sigma_{i+2}\sigma_{i+3}\rangle$   up   to  fourth
order.}
\label{f10}
\end{figure}

In Figure 2, we show all the contributions to the term $\sum_{1=i<j}^n
\langle \sigma_i \sigma_j \rangle$  in (\ref{20}), together with their
embedding  factors (the  number  of ways  they  can be  placed on  the
lattice), up to third order.

\begin{figure} 
\narrowtext\centerline{\epsfxsize\columnwidth
\epsfbox{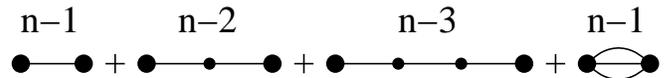}}
\caption{Contributions to  $\sum_ {1=i<j}^n \langle  \sigma_i \sigma_j
\rangle$, with  embedding factors, up  to third order. Large  dots are
vertices, small dots intermediate sites.}
\label{f20}
\end{figure}

\begin{figure} 
\narrowtext\centerline{\epsfxsize\columnwidth
\epsfbox{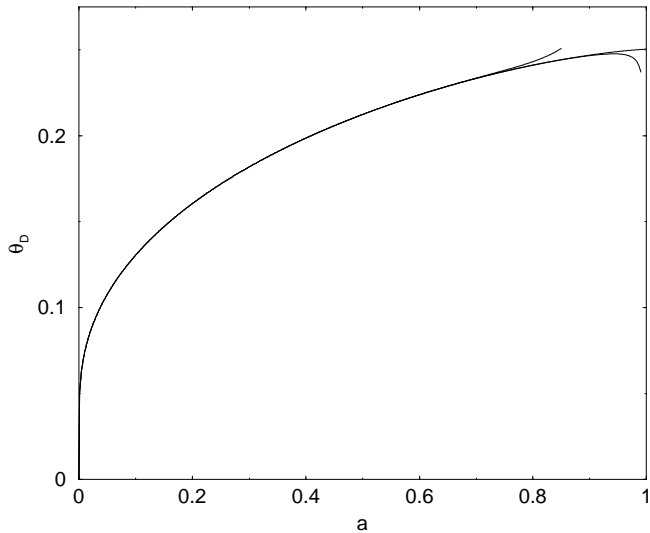}}
\caption{Plot of $\theta_D$  against $a \equiv \exp(-\Delta\!T/2)$ for
the random acceleration  process to 14th order in  $a$. The raw series
in $a$ can be seen to separate at $a \approx 0.75$,  the unconstrained
(6,8) Pad\'e at $a \approx 0.9$,  while  the  (8,7) Pad\'e  with  one
constraint is well-behaved up to the continuum limit.}
\label{f30}
\end{figure}

Using a computer we have calculated all terms in $P_n$ up to $14^{th}$
order.   Noting  that  $\rho  =  P_{n+1}/P_n$ (where  we  recall  that
$\theta_D = -\ln\rho/\Delta\!T$), and expanding in powers of $C$ up to
$C(14 \Delta \!  T)$ and  $C(\Delta \!  T)^{14}$ (and all intermediate
combinations) gives  a series expansion for $\rho$.   To second order,
for example, we obtain \cite{expansion}
\begin{equation} 
\rho = {1\over 2}+ {C[\Delta \!T]  \over \pi} + {C[2 \Delta \!T] \over
\pi} - {2 C[\Delta \!T]^2 \over \pi^2} +\ldots
\label{60} 
\end{equation} 
The correlation functions for  the random walk and random acceleration
processes   in  one   dimension  are   \cite{review} $\exp(-T/2)$  and
$[3\exp(-T/2)-\exp(-3T/2)]/2$ respectively.   Substituting these forms
into   the  expansion   for   $\rho$  gives   series   in  $a   \equiv
\exp(-\Delta\!T/2)$  to  $14^{th}$ order,  the  coefficients of  which
agree  with  those  obtained  by  a  different  method  \cite{dpers2},
providing  a powerful check  of our  results.  Figure  \ref{f30} shows
$\theta_D$ against $a$ for the random acceleration problem.  Note that
the series  in $a$ gives diverging  results for $a  \to 1$ ($\Delta\!T
\to  0$).   The  reason  is  clear.  The  relation  $\theta_D  =  -\ln
\rho/\Delta\!T$ will give $\theta_D \to \pm \infty$ for $\Delta\!T \to
0$ unless $\rho \to 1$ in this limit. The latter condition must indeed
be satisfied  in an exact calculation,  but our finite  series gives a
result close to, but not exactly  equal to, unity.  To remedy this, we
employ a technique commonly used  on series expansions in the field of
critical  phenomena, the  Pad\'e  approximant \cite{DombAndGreenVol3}.
This extends the accuracy of the  series to larger $a$ (Figure 3), but
does not yet  eliminate the divergence at $a=1$. To do  this we add an
extra term to the Pad\'e  polynomial in the numerator (or denominator)
to force the condition $\rho=1$ at $a=1$.

In  Figure  3  the  unforced  (6,8) Pad\'e  (i.e.\  a  $6^{th}$  order
polynomial in  the numerator, $8^{th}$  order in the  denominator) and
the forced  $(8,7)$ Pad\'e are  shown. The two Pade's  are essentially
indistinguishable from each other and from the raw series for $a<0.7$.
The  extrapolation $a  \to  1$, however,  required  for the  continuum
limit, is more sensitive to the choice of Pad\'e.  We typically choose
Pad\'es in  the range ($n-2,n$) to  $(n+2,n)$, i.e.\ not  too far from
``diagonal'',  and discard  any Pade's  which give  poles on  the real
$a$-axis  in   the  interval  $(-0.5,1.5)$.   Poles   in  $(0,1)$  are
unphysical,  and  poles  close   to  this  interval  can  distort  the
results.  The remaining  Pad\'es are  used to  estimate  the continuum
values -- see  Table 1. The quoted errors  reflect the scatter between
different Pad\'es  and the  observed (weak) trends  as more  terms are
added to the series. They remain, however, somewhat subjective.

This completes  our general  method for finding  $\theta_D$ up  to and
including the continuum limit. Our  final estimate (Figure 3 and Table
1),  $\theta =  0.2506(5)$,  for  the continuum  limit  of the  random
acceleration  process  is in  good  agreement  with  the exact  result
$\theta = 1/4$ \cite{RA}.

\begin{center}
\begin{tabular}{||l||l|l|l|l||} \hline
&Pad\'e1CR &Pad\'e2CR &numerical\cite{NL} &IIA\cite{IIA1,IIA2}\\ \hline 
$\ddot x  = \eta(t)$ & 0.2506(5) & -----------& 1/4 (exact) & 0.2647 \\ 
1-d diff & 0.119(1) & 0.1201(3)  & 0.12050(5) & 0.1203 \\ 
2-d  diff & 0.187(1) & 0.1875(1) & 0.1875(1) & 0.1862 \\  
3-d diff & 0.24(3) & 0.237(1) & 0.2382(1) & 0.2358 \\ \hline
\end{tabular} 
\end{center}
\begin{small}
\noindent  Table1. Continuum persistence exponent $\theta$ for the  
random  acceleration  process,  and  the  diffusion equation for 
$d=1,2,3$. The first two columns are the mean and standard deviations 
of the Pad\'e estimates (with 1 and 2 constraints) from the 4 highest 
orders of the expansion.  
\end{small}
\smallskip

We now consider  the simple diffusion equation, $\partial\phi/\partial
t = {\nabla^2}  \phi$, where $\phi({\bf x},t)$ is  the diffusion field
and the  initial condition $\phi({\bf  x},0)$ is taken to  be Gaussian
white  noise   with  zero  mean.   We  consider   the  persistence  of
$\phi(0,t)$, the  field at a single site  which we can take  to be the
origin.   On  changing  to  log  time, the  normalized  correlator  is
$C(T)={\rm  sech}(T/2)^{d/2}$   in  $d$  dimensions  \cite{IIA1,IIA2}.
Substituting this  form into our expression for  $\rho$, and expanding
as before  in powers  of $a \equiv  \exp(-\Delta\!T/2)$ ($d=2$)  or $a
\equiv \exp(-\Delta\!T/4)$ ($ d=1$ or  3), we obtain an expression for
$\theta_D$  as a  power  series in  $a$.  Note that  since $C(T)  \sim
\exp(-dT/4)$  for  $T  \to  \infty$,  these  definitions  of  $a$  for
$d=1,2,3$ imply that the leading term in the expansion is of order $a$
for  $d=1,2$, but  of order  $a^3$ for  $d=3$. Hence  an  expansion to
``$14^{th}$'' order  means to order $a^{14}$ for  $d=1,2$ and $a^{42}$
for $d=3$. The reasoning behind  these definitions of $a$ is to ensure
that, in each case, every  term in the expansion contains only integer
powers of $a$, since this greatly simplifies the Pad\'e analysis.

\begin{figure} 
\narrowtext\centerline{\epsfxsize\columnwidth
\epsfbox{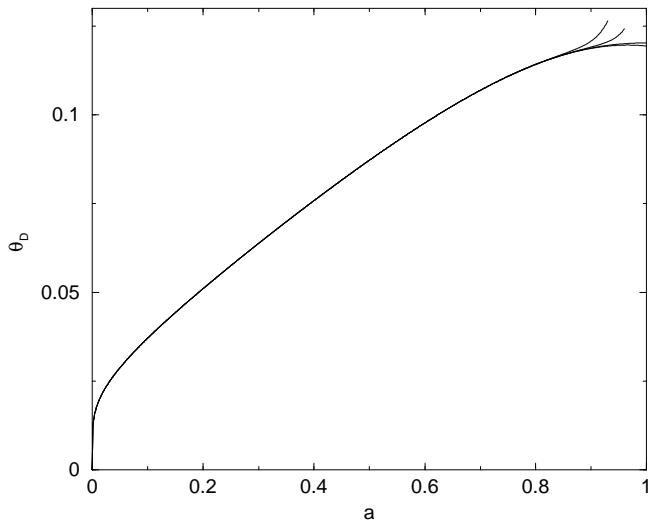}}
\caption{Plot of $\theta_D$  against $a \equiv \exp(-\Delta\!T/4)$ for
the one-dimensional diffusion equation to  14th order in $a$.  The raw
series in $a$  begins to separate at $a \approx 0.85$, the unconstrained
Pad\'e at $a \approx 0.9$. Both constrained Pad\'es  are well behaved up
to the continuum  limit, the more accurate (upper) curve being obtained 
with two constraints.}
\label{f40}
\end{figure}

The results for $d=1$ are shown in Figure \ref{f40}. Similar plots are
obtained for  $d=2$ and  3. As before,  we use Pad\'e  approximants to
improve  the  series  and  constrained  Pad\'e  approximants  (forcing
$\rho=1$ at $a=1$) to enable us to approach the continuum limit $a=1$.
Estimates for the continuum persistent exponents are given in Table 1,
where ``Pad\'e1CR''  indicates that we have imposed  one constraint on
the series, as discussed above.

While  this approach  already gives  rather accurate  results  for the
continuum limit (see Table 1),  for the case of the diffusion equation
(and  other sufficiently  smooth  processes) a  further refinement  is
possible, leading to even more  accurate results.  To see what we mean
by ``sufficiently  smooth'', consider the short-time  expansion of the
correlator $C(T)$. For the random acceleration process it has the form
$C_{ra}(T) = 1 - b_2 T^2 +  b_3 T^3 - \cdots$, while for the diffusion
equation one  obtains $C_{diff}(T)  = 1 -  b_2T^2 + b_4T^4  - \cdots$.
The presence of an odd (or noninteger) power implies nonanalyticity at
$T=0$ since  $C(T)$ must  be even in  $T$ (i.e.\ $T^3$  actually means
$|T|^3$).  For  processes having a  $C(T)$ with a  small-$T$ expansion
like the  diffusion equation (i.e.\ sufficiently smooth),  there is an
additional constraint on the series.
 
Let $P_1(T)$ be the  probability distribution of the intervals between
zero crossings of $X(T)$. With the assumption that these intervals are
independent (the IIA),  it is easy to show that  $P_1(T) \to \alpha T$
for small  $T$ for processes described by  $C_{diff}(T)$. Indeed, this
can  be proved  exactly, and  even the  coefficient $\alpha$  is given
correctly  by  the  IIA  \cite{Zeitak}.   Now  consider  the  discrete
persistence, $P_D(T)$,  close to  the continuum limit  ($\Delta\!T \ll
1$). The first correction to  the continuum persistence $P(T)$ is from
paths for which $X(T)$ is positive  at all times apart from one double
crossing  between samplings.   In this  limit we  can use  the  IIA to
estimate $P_D(T)$. For $\Delta\!T \to 0$ we have $P_D(T) = P(T) + {\rm
Prob(1\  double\ crossing)}$.   Using the  IIA and  making use  of the
result $P_1(T)  \propto T$ for small  $T$, it  is   straightforward to
show   that $P_D(T)  =   Ae^{-\theta   T}\,(1  +
Bn\Delta\!T^3)  +O(\Delta\!T^4)$  where   $A$  and  $B$  are  positive
constants.   Thus,  using   $T=n\Delta\!T$,  we   find  $P_D(T)   =  A
\exp-T[\theta   -    B\Delta\!T^2   +   O(\Delta\!T^3)]$,    to   give
$\theta_D=\theta   -  B  \Delta\!T^2   +O(\Delta\!T^3)$  and   so  $(d
\theta_D/d \Delta\!T) |_{\Delta\!T=0} =0$.  By adding one more term to
the Pad\'e to enforce this constraint we achieve improved results (see
Table 1, Pad\'e2CR).  The same calculation for the random acceleration
process,  for which  $P_1(0) \ne  0$, gives  $\theta_D =  \theta  - B'
\Delta\!T       +        O(\Delta\!T^2)$       and       $(d\theta_D/d
\Delta\!T)|_{\Delta\!T=0} = - B' \ne 0$.

From Table 1  (Pad\'e2CR) we see that the  series expansion gives very
accurate results. Although the IIA is slightly better for diffusion in
$d=1$, the present  approach is more accurate for  other values of $d$
(and can be improved by adding more terms).  For discrete persistence,
the measurement regime employed  in the experiments on one-dimensional
gas diffusion  \cite{wong} is equivalent to  $\Delta\!T \approx 0.24$,
i.e.\ $a  \approx 0.94$.   For this value  we find $\theta  - \theta_D
\approx 10^{-4}$,  which is  negligible compared to  the error  quoted in
\cite{wong}.

An obvious question concerns the  convergence of the estimates for the
continuum persistence exponent $\theta$ with increasing order, $N$, of
the  expansion.  In  figure  5  we plot  $\theta$  against $1/N$,  for
diffusion in $d=1,2,3$ (for $d=3$,  we define $N$ as the highest power
of  $a^3$  retained,  as  discussed  above). No  systematic  trend  is
observable at large $N$.

\begin{figure} 
\narrowtext\centerline{\epsfxsize\columnwidth
\epsfbox{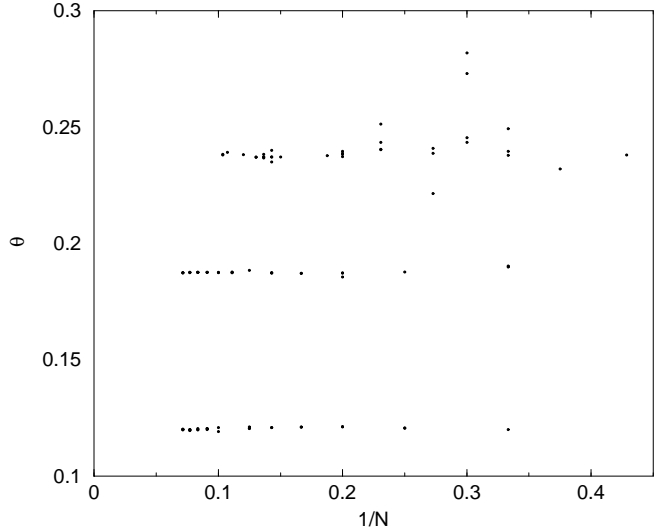}}
\caption{Plot of $\theta$ against $1/N$, where $N$ is the order of the
expansion, for the diffusion equation in $d=1,2,3$ (bottom to top).  }
\label{f50}
\end{figure} 

All the Pad\'es (with 2 constraints) satisfying the criteria discussed
above are plotted,  so there is often more than one  point for a given
$N$ (though these are sometimes so close as to be indistinguishable on
the plot). For reasons we do  not fully understand, the $d=3$ data are
more  erarratic than for  $d=1,2$, which  accounts for  the relatively
large uncertainties quoted in Table 1.  Note that we cannot go to such
high  order  in  $d=3$,  due   to  the  difficulty  of  computing  the
coefficients   for  the   constrained   Pad\'es  for   series  up   to
$a^{42}$. Instead  we stopped at  $a^{29}$. For $d=1,2$,  by contrast,
the estimates  are quite steady as  a function of  $N$, though further
work is needed  to clarify the precise nature  of the convergence with
increasing $N$.

In summary,  we have developed  a powerful series  expansion technique
for  calculating   persistence  exponents  of   non-Markov  processes,
providing accurate estimates of discrete persistence exponents for any
Gaussian stationary process.  Using constrained Pad\'e approximants to
extrapolate  the  series to  the  continuum  limit,  we have  achieved
accurate  estimates  of  the  persistence exponent  $\theta$  for  the
diffusion equation.  The results compare favorably with those computed
using the  IIA, and are systematically improvable  by calculating more
terms in the expansion.

This work was supported by EPSRC (UK).

\end{multicols}

\widetext 

\newpage

\noindent{\bf Appendix}

\medskip

We display below the complete $14{th}$ order expansion for ${\rho}$, 
with the compact notion $c_n \equiv C(n\Delta\!T)$.

\vspace{0.3cm}

$
\rho=\frac{1}{2}+\frac{ {c_1}}{\pi } -\frac{2 c_{1}^{2}}{{{\pi }^2}}+\frac{{c_2}}{\pi }+
\frac{8 c_{1}^{3}}{{{\pi }^3}}+\frac{c_{1}^{3}}{6 \pi }-\frac{6 {c_1} {c_2}}{{{\pi }^2}}+\frac{{c_3}}{\pi }
-\frac{40 c_{1}^{4}}{{{\pi }^4}}+\frac{c_{1}^{4}}{3 {{\pi }^2}}+\frac{40 c_{1}^{2}
{c_2}}{{{\pi }^3}}-\frac{4 c_{1}^{2} {c_2}}{{{\pi }^2}}-\frac{2 c_{2}^{2}}{{{\pi }^2}}-\frac{6 {c_1} {c_3}}{{{\pi
}^2}}+\frac{{c_4}}{\pi }+  {}
{}
\frac{224 c_{1}^{5}}{{{\pi }^5}}-\frac{4 c_{1}^{5}}{{{\pi }^3}}+\frac{3 c_{1}^{5}}{40
\pi }-\frac{280 c_{1}^{3} {c_2}}{{{\pi }^4}}+\frac{32 c_{1}^{3} {c_2}}{{{\pi }^3}}+\frac{c_{1}^{3} {c_2}}{{{\pi
}^2}}+  {}
{}
 \frac{44 {c_1} c_{2}^{2}}{{{\pi }^3}}-\frac{4 {c_1} c_{2}^{2}}{{{\pi }^2}}+\frac{44 c_{1}^{2}
{c_3}}{{{\pi }^3}}-\frac{4 c_{1}^{2} {c_3}}{{{\pi }^2}}-\frac{6 {c_2} {c_3}}{{{\pi }^2}}-\frac{6 {c_1}
{c_4}}{{{\pi }^2}}+\frac{{c_5}}{\pi }
-\frac{1344 c_{1}^{6}}{{{\pi }^6}}+\frac{100 c_{1}^{6}}{3 {{\pi }^4}}+\frac{251
c_{1}^{6}}{180 {{\pi }^2}}+\frac{2016 c_{1}^{4} {c_2}}{{{\pi }^5}}-\frac{240 c_{1}^{4} {c_2}}{{{\pi }^4}}-  {}
{}
 \frac{50 c_{1}^{4} {c_2}}{3 {{\pi }^3}}-\frac{6 c_{1}^{4} {c_2}}{{{\pi }^2}}-\frac{552 c_{1}^{2}
c_{2}^{2}}{{{\pi }^4}}+\frac{80 c_{1}^{2} c_{2}^{2}}{{{\pi }^3}}+\frac{7 c_{1}^{2} c_{2}^{2}}{{{\pi }^2}}+\frac{8
c_{2}^{3}}{{{\pi }^3}}+\frac{c_{2}^{3}}{6 \pi }-  {}
{}
 \frac{328 c_{1}^{3} {c_3}}{{{\pi }^4}}+\frac{40 c_{1}^{3} {c_3}}{{{\pi }^3}}+\frac{3 c_{1}^{3}
{c_3}}{{{\pi }^2}}+\frac{108 {c_1} {c_2} {c_3}}{{{\pi }^3}}-\frac{16 {c_1} {c_2} {c_3}}{{{\pi }^2}}-  {}
{}
 \frac{2 c_{3}^{2}}{{{\pi }^2}}+\frac{44 c_{1}^{2} {c_4}}{{{\pi }^3}}-\frac{4 c_{1}^{2} {c_4}}{{{\pi
}^2}}-\frac{6 {c_2} {c_4}}{{{\pi }^2}}-\frac{6 {c_1} {c_5}}{{{\pi }^2}}+\frac{{c_6}}{\pi }+  {}
{}
\frac{8448 c_{1}^{7}}{{{\pi }^7}}-\frac{784 c_{1}^{7}}{3 {{\pi }^5}}-\frac{178
c_{1}^{7}}{15 {{\pi }^3}}+\frac{5 c_{1}^{7}}{112 \pi }-\frac{14784 c_{1}^{5} {c_2}}{{{\pi }^6}}+\frac{1792
c_{1}^{5} {c_2}}{{{\pi }^5}}+  {}
{}
 \frac{196 c_{1}^{5} {c_2}}{{{\pi }^4}}+\frac{136 c_{1}^{5} {c_2}}{3 {{\pi }^3}}+\frac{41
c_{1}^{5} {c_2}}{20 {{\pi }^2}}+\frac{5824 c_{1}^{3} c_{2}^{2}}{{{\pi }^5}}-\frac{1008 c_{1}^{3} c_{2}^{2}}{{{\pi
}^4}}-  {}
{}
 \frac{188 c_{1}^{3} c_{2}^{2}}{3 {{\pi }^3}}-\frac{4 c_{1}^{3} c_{2}^{2}}{{{\pi }^2}}-\frac{336
{c_1} c_{2}^{3}}{{{\pi }^4}}+\frac{40 {c_1} c_{2}^{3}}{{{\pi }^3}}+\frac{3 {c_1} c_{2}^{3}}{{{\pi }^2}}+\frac{2464
c_{1}^{4} {c_3}}{{{\pi }^5}}-  {}
{}
 \frac{336 c_{1}^{4} {c_3}}{{{\pi }^4}}-\frac{88 c_{1}^{4} {c_3}}{3 {{\pi }^3}}-\frac{4 c_{1}^{4}
{c_3}}{{{\pi }^2}}-\frac{1344 c_{1}^{2} {c_2} {c_3}}{{{\pi }^4}}+\frac{248 c_{1}^{2} {c_2} {c_3}}{{{\pi
}^3}}+  {}
{}
 \frac{6 c_{1}^{2} {c_2} {c_3}}{{{\pi }^2}}+\frac{44 c_{2}^{2} {c_3}}{{{\pi }^3}}-\frac{4
c_{2}^{2} {c_3}}{{{\pi }^2}}+\frac{44 {c_1} c_{3}^{2}}{{{\pi }^3}}-\frac{4 {c_1} c_{3}^{2}}{{{\pi }^2}}-  {}
{}
 \frac{336 c_{1}^{3} {c_4}}{{{\pi }^4}}+\frac{40 c_{1}^{3} {c_4}}{{{\pi }^3}}+\frac{3 c_{1}^{3}
{c_4}}{{{\pi }^2}}+\frac{116 {c_1} {c_2} {c_4}}{{{\pi }^3}}-\frac{16 {c_1} {c_2} {c_4}}{{{\pi }^2}}-  {}
{}
 \frac{6 {c_3} {c_4}}{{{\pi }^2}}+\frac{44 c_{1}^{2} {c_5}}{{{\pi }^3}}-\frac{4 c_{1}^{2}
{c_5}}{{{\pi }^2}}-\frac{6 {c_2} {c_5}}{{{\pi }^2}}-\frac{6 {c_1} {c_6}}{{{\pi }^2}}+\frac{{c_7}}{\pi }
-\frac{54912 c_{1}^{8}}{{{\pi }^8}}+\frac{2016 c_{1}^{8}}{{{\pi }^6}}+\frac{265
c_{1}^{8}}{3 {{\pi }^4}}+\frac{741 c_{1}^{8}}{280 {{\pi }^2}}+\frac{109824 c_{1}^{6} {c_2}}{{{\pi }^7}}-  {}
{}
 \frac{13440 c_{1}^{6} {c_2}}{{{\pi }^6}}-\frac{2016 c_{1}^{6} {c_2}}{{{\pi }^5}}-\frac{296 c_{1}^{6}
{c_2}}{{{\pi }^4}}-\frac{643 c_{1}^{6} {c_2}}{18 {{\pi }^3}}-\frac{61 c_{1}^{6} {c_2}}{6 {{\pi }^2}}- 
{}
{}
 \frac{56640 c_{1}^{4} c_{2}^{2}}{{{\pi }^6}}+\frac{10752 c_{1}^{4} c_{2}^{2}}{{{\pi }^5}}+\frac{560
c_{1}^{4} c_{2}^{2}}{{{\pi }^4}}+\frac{272 c_{1}^{4} c_{2}^{2}}{3 {{\pi }^3}}+\frac{25 c_{1}^{4} c_{2}^{2}}{2
{{\pi }^2}}+  {}
{}
 \frac{6528 c_{1}^{2} c_{2}^{3}}{{{\pi }^5}}-\frac{1152 c_{1}^{2} c_{2}^{3}}{{{\pi }^4}}-\frac{292
c_{1}^{2} c_{2}^{3}}{3 {{\pi }^3}}-\frac{4 c_{1}^{2} c_{2}^{3}}{{{\pi }^2}}-\frac{40 c_{2}^{4}}{{{\pi }^4}}+\frac{c_{2}^{4}}{3
{{\pi }^2}}-  {}
{}
 \frac{18624 c_{1}^{5} {c_3}}{{{\pi }^6}}+\frac{2688 c_{1}^{5} {c_3}}{{{\pi }^5}}+\frac{284 c_{1}^{5}
{c_3}}{{{\pi }^4}}+\frac{116 c_{1}^{5} {c_3}}{3 {{\pi }^3}}+\frac{101 c_{1}^{5} {c_3}}{20 {{\pi }^2}}+
 {}
{}
 \frac{14400 c_{1}^{3} {c_2} {c_3}}{{{\pi }^5}}-\frac{2976 c_{1}^{3} {c_2} {c_3}}{{{\pi }^4}}-\frac{82
c_{1}^{3} {c_2} {c_3}}{{{\pi }^3}}-\frac{16 c_{1}^{3} {c_2} {c_3}}{{{\pi }^2}}-  {}
{}
 \frac{1408 {c_1} c_{2}^{2} {c_3}}{{{\pi }^4}}+\frac{272 {c_1} c_{2}^{2} {c_3}}{{{\pi }^3}}+\frac{6
{c_1} c_{2}^{2} {c_3}}{{{\pi }^2}}-\frac{624 c_{1}^{2} c_{3}^{2}}{{{\pi }^4}}+\frac{96 c_{1}^{2} c_{3}^{2}}{{{\pi
}^3}}+  {}
{}
 \frac{8 c_{1}^{2} c_{3}^{2}}{{{\pi }^2}}+\frac{44 {c_2} c_{3}^{2}}{{{\pi }^3}}-\frac{4 {c_2}
c_{3}^{2}}{{{\pi }^2}}+\frac{2592 c_{1}^{4} {c_4}}{{{\pi }^5}}-\frac{352 c_{1}^{4} {c_4}}{{{\pi }^4}}-  {}
{}
 \frac{106 c_{1}^{4} {c_4}}{3 {{\pi }^3}}-\frac{4 c_{1}^{4} {c_4}}{{{\pi }^2}}-\frac{1464
c_{1}^{2} {c_2} {c_4}}{{{\pi }^4}}+\frac{280 c_{1}^{2} {c_2} {c_4}}{{{\pi }^3}}+\frac{8 c_{1}^{2}
{c_2} {c_4}}{{{\pi }^2}}+  {}
{}
 \frac{40 c_{2}^{2} {c_4}}{{{\pi }^3}}-\frac{4 c_{2}^{2} {c_4}}{{{\pi }^2}}+\frac{108 {c_1}
{c_3} {c_4}}{{{\pi }^3}}-\frac{16 {c_1} {c_3} {c_4}}{{{\pi }^2}}-\frac{2 c_{4}^{2}}{{{\pi }^2}}-  {}
{}
 \frac{336 c_{1}^{3} {c_5}}{{{\pi }^4}}+\frac{40 c_{1}^{3} {c_5}}{{{\pi }^3}}+\frac{3 c_{1}^{3}
{c_5}}{{{\pi }^2}}+\frac{116 {c_1} {c_2} {c_5}}{{{\pi }^3}}-\frac{16 {c_1} {c_2} {c_5}}{{{\pi }^2}}-  {}
{}
 \frac{6 {c_3} {c_5}}{{{\pi }^2}}+\frac{44 c_{1}^{2} {c_6}}{{{\pi }^3}}-\frac{4 c_{1}^{2}
{c_6}}{{{\pi }^2}}-\frac{6 {c_2} {c_6}}{{{\pi }^2}}-\frac{6 {c_1} {c_7}}{{{\pi }^2}}+\frac{{c_8}}{\pi }+  {}
{}
\frac{366080 c_{1}^{9}}{{{\pi }^9}}-\frac{15488 c_{1}^{9}}{{{\pi }^7}}-\frac{5740
c_{1}^{9}}{9 {{\pi }^5}}-\frac{7645 c_{1}^{9}}{378 {{\pi }^3}}+\frac{35 c_{1}^{9}}{1152 \pi }-\frac{823680
c_{1}^{7} {c_2}}{{{\pi }^8}}+  {}
{}
 \frac{101376 c_{1}^{7} {c_2}}{{{\pi }^7}}+\frac{19360 c_{1}^{7} {c_2}}{{{\pi }^6}}+\frac{1792 c_{1}^{7}
{c_2}}{{{\pi }^5}}+\frac{1253 c_{1}^{7} {c_2}}{3 {{\pi }^4}}+\frac{926 c_{1}^{7} {c_2}}{15 {{\pi }^3}}+
 {}
{}
 \frac{167 c_{1}^{7} {c_2}}{56 {{\pi }^2}}+\frac{525888 c_{1}^{5} c_{2}^{2}}{{{\pi }^7}}-\frac{105600
c_{1}^{5} c_{2}^{2}}{{{\pi }^6}}-\frac{5568 c_{1}^{5} c_{2}^{2}}{{{\pi }^5}}-\frac{1064 c_{1}^{5} c_{2}^{2}}{{{\pi
}^4}}-  {}
{}
 \frac{2041 c_{1}^{5} c_{2}^{2}}{30 {{\pi }^3}}-\frac{6 c_{1}^{5} c_{2}^{2}}{{{\pi }^2}}-\frac{95040
c_{1}^{3} c_{2}^{3}}{{{\pi }^6}}+\frac{20352 c_{1}^{3} c_{2}^{3}}{{{\pi }^5}}+\frac{3604 c_{1}^{3} c_{2}^{3}}{3
{{\pi }^4}}+  {}
{}
 \frac{4 c_{1}^{3} c_{2}^{3}}{{{\pi }^3}}+\frac{35 c_{1}^{3} c_{2}^{3}}{6 {{\pi }^2}}+\frac{2608
{c_1} c_{2}^{4}}{{{\pi }^5}}-\frac{352 {c_1} c_{2}^{4}}{{{\pi }^4}}-\frac{106 {c_1} c_{2}^{4}}{3 {{\pi
}^3}}-\frac{4 {c_1} c_{2}^{4}}{{{\pi }^2}}+  {}
{}
 \frac{141504 c_{1}^{6} {c_3}}{{{\pi }^7}}-\frac{21120 c_{1}^{6} {c_3}}{{{\pi }^6}}-\frac{8128 c_{1}^{6}
{c_3}}{3 {{\pi }^5}}-\frac{280 c_{1}^{6} {c_3}}{{{\pi }^4}}-\frac{2123 c_{1}^{6} {c_3}}{45 {{\pi }^3}}-
 {}
{}
 \frac{14 c_{1}^{6} {c_3}}{3 {{\pi }^2}}-\frac{142560 c_{1}^{4} {c_2} {c_3}}{{{\pi }^6}}+\frac{31616
c_{1}^{4} {c_2} {c_3}}{{{\pi }^5}}+\frac{888 c_{1}^{4} {c_2} {c_3}}{{{\pi }^4}}+  {}
{}
 \frac{572 c_{1}^{4} {c_2} {c_3}}{3 {{\pi }^3}}+\frac{15 c_{1}^{4} {c_2} {c_3}}{2
{{\pi }^2}}+\frac{25824 c_{1}^{2} c_{2}^{2} {c_3}}{{{\pi }^5}}-\frac{6080 c_{1}^{2} c_{2}^{2} {c_3}}{{{\pi
}^4}}-  {}
{}
 \frac{80 c_{1}^{2} c_{2}^{2} {c_3}}{{{\pi }^3}}-\frac{6 c_{1}^{2} c_{2}^{2} {c_3}}{{{\pi
}^2}}-\frac{336 c_{2}^{3} {c_3}}{{{\pi }^4}}+\frac{40 c_{2}^{3} {c_3}}{{{\pi }^3}}+\frac{3 c_{2}^{3} {c_3}}{{{\pi
}^2}}+\frac{7264 c_{1}^{3} c_{3}^{2}}{{{\pi }^5}}-  {}
{}
 \frac{1392 c_{1}^{3} c_{3}^{2}}{{{\pi }^4}}-\frac{272 c_{1}^{3} c_{3}^{2}}{3 {{\pi }^3}}-\frac{2
c_{1}^{3} c_{3}^{2}}{{{\pi }^2}}-\frac{1432 {c_1} {c_2} c_{3}^{2}}{{{\pi }^4}}+\frac{272 {c_1} {c_2}
c_{3}^{2}}{{{\pi }^3}}+  {}
{}
 \frac{6 {c_1} {c_2} c_{3}^{2}}{{{\pi }^2}}+\frac{8 c_{3}^{3}}{{{\pi }^3}}+\frac{c_{3}^{3}}{6 \pi
}-\frac{20064 c_{1}^{5} {c_4}}{{{\pi }^6}}+\frac{2944 c_{1}^{5} {c_4}}{{{\pi }^5}}+\frac{352 c_{1}^{5}
{c_4}}{{{\pi }^4}}+  {}
{}
 \frac{92 c_{1}^{5} {c_4}}{3 {{\pi }^3}}+\frac{91 c_{1}^{5} {c_4}}{20 {{\pi }^2}}+\frac{15872
c_{1}^{3} {c_2} {c_4}}{{{\pi }^5}}-\frac{3456 c_{1}^{3} {c_2} {c_4}}{{{\pi }^4}}-\frac{206 c_{1}^{3}
{c_2} {c_4}}{3 {{\pi }^3}}-  {}
{}
 \frac{12 c_{1}^{3} {c_2} {c_4}}{{{\pi }^2}}-\frac{1488 {c_1} c_{2}^{2} {c_4}}{{{\pi }^4}}+\frac{280
{c_1} c_{2}^{2} {c_4}}{{{\pi }^3}}+\frac{8 {c_1} c_{2}^{2} {c_4}}{{{\pi }^2}}-\frac{1432 c_{1}^{2}
{c_3} {c_4}}{{{\pi }^4}}+  {}
{}
 \frac{272 c_{1}^{2} {c_3} {c_4}}{{{\pi }^3}}+\frac{6 c_{1}^{2} {c_3} {c_4}}{{{\pi }^2}}+\frac{116
{c_2} {c_3} {c_4}}{{{\pi }^3}}-\frac{16 {c_2} {c_3} {c_4}}{{{\pi }^2}}+\frac{44 {c_1} c_{4}^{2}}{{{\pi
}^3}}-  {}
{}
 \frac{4 {c_1} c_{4}^{2}}{{{\pi }^2}}+\frac{2608 c_{1}^{4} {c_5}}{{{\pi }^5}}-\frac{352 c_{1}^{4}
{c_5}}{{{\pi }^4}}-\frac{106 c_{1}^{4} {c_5}}{3 {{\pi }^3}}-\frac{4 c_{1}^{4} {c_5}}{{{\pi }^2}}-\frac{1488
c_{1}^{2} {c_2} {c_5}}{{{\pi }^4}}+  {}
{}
 \frac{280 c_{1}^{2} {c_2} {c_5}}{{{\pi }^3}}+\frac{8 c_{1}^{2} {c_2} {c_5}}{{{\pi }^2}}+\frac{44
c_{2}^{2} {c_5}}{{{\pi }^3}}-\frac{4 c_{2}^{2} {c_5}}{{{\pi }^2}}+\frac{116 {c_1} {c_3} {c_5}}{{{\pi }^3}}-
 {}
{}
 \frac{16 {c_1} {c_3} {c_5}}{{{\pi }^2}}-\frac{6 {c_4} {c_5}}{{{\pi }^2}}-\frac{336 c_{1}^{3}
{c_6}}{{{\pi }^4}}+\frac{40 c_{1}^{3} {c_6}}{{{\pi }^3}}+\frac{3 c_{1}^{3} {c_6}}{{{\pi }^2}}+\frac{116 {c_1}
{c_2} {c_6}}{{{\pi }^3}}-  {}
{}
 \frac{16 {c_1} {c_2} {c_6}}{{{\pi }^2}}-\frac{6 {c_3} {c_6}}{{{\pi }^2}}+\frac{44 c_{1}^{2}
{c_7}}{{{\pi }^3}}-\frac{4 c_{1}^{2} {c_7}}{{{\pi }^2}}-\frac{6 {c_2} {c_7}}{{{\pi }^2}}-\frac{6 {c_1}
{c_8}}{{{\pi }^2}}+\frac{{c_9}}{\pi }  {}
{}
-\frac{2489344 c_{1}^{10}}{{{\pi }^{10}}}+\frac{118976 c_{1}^{10}}{{{\pi }^8}}+\frac{22776
c_{1}^{10}}{5 {{\pi }^6}}+\frac{50759 c_{1}^{10}}{378 {{\pi }^4}}+\frac{471341 c_{1}^{10}}{100800 {{\pi }^2}}+
 {}
{}
 \frac{6223360 c_{1}^{8} {c_2}}{{{\pi }^9}}-\frac{768768 c_{1}^{8} {c_2}}{{{\pi }^8}}-\frac{178464
c_{1}^{8} {c_2}}{{{\pi }^7}}-\frac{9920 c_{1}^{8} {c_2}}{{{\pi }^6}}-  {}
{}
 \frac{20856 c_{1}^{8} {c_2}}{5 {{\pi }^5}}-\frac{758 c_{1}^{8} {c_2}}{3 {{\pi }^4}}-\frac{4439
c_{1}^{8} {c_2}}{84 {{\pi }^3}}-\frac{75 c_{1}^{8} {c_2}}{4 {{\pi }^2}}-\frac{4740736 c_{1}^{6}
c_{2}^{2}}{{{\pi }^8}}+  {}
{}
 \frac{988416 c_{1}^{6} c_{2}^{2}}{{{\pi }^7}}+\frac{58960 c_{1}^{6} c_{2}^{2}}{{{\pi }^6}}+\frac{9600
c_{1}^{6} c_{2}^{2}}{{{\pi }^5}}+\frac{4813 c_{1}^{6} c_{2}^{2}}{15 {{\pi }^4}}+\frac{370 c_{1}^{6} c_{2}^{2}}{3
{{\pi }^3}}+  {}
{}
 \frac{107 c_{1}^{6} c_{2}^{2}}{4 {{\pi }^2}}+\frac{1189760 c_{1}^{4} c_{2}^{3}}{{{\pi }^7}}-\frac{286080
c_{1}^{4} c_{2}^{3}}{{{\pi }^6}}-\frac{11360 c_{1}^{4} c_{2}^{3}}{{{\pi }^5}}-\frac{472 c_{1}^{4} c_{2}^{3}}{3
{{\pi }^4}}-  {}
{}
 \frac{1105 c_{1}^{4} c_{2}^{3}}{9 {{\pi }^3}}-\frac{46 c_{1}^{4} c_{2}^{3}}{3 {{\pi }^2}}-\frac{70880
c_{1}^{2} c_{2}^{4}}{{{\pi }^6}}+\frac{14208 c_{1}^{2} c_{2}^{4}}{{{\pi }^5}}+\frac{1132 c_{1}^{2} c_{2}^{4}}{{{\pi
}^4}}+  {}
{}
 \frac{104 c_{1}^{2} c_{2}^{4}}{3 {{\pi }^3}}+\frac{27 c_{1}^{2} c_{2}^{4}}{4 {{\pi }^2}}+\frac{224
c_{2}^{5}}{{{\pi }^5}}-\frac{4 c_{2}^{5}}{{{\pi }^3}}+\frac{3 c_{2}^{5}}{40 \pi }-\frac{1079936 c_{1}^{7} {c_3}}{{{\pi
}^8}}+  {}
{}
 \frac{164736 c_{1}^{7} {c_3}}{{{\pi }^7}}+\frac{25280 c_{1}^{7} {c_3}}{{{\pi }^6}}+\frac{1728 c_{1}^{7}
{c_3}}{{{\pi }^5}}+\frac{6763 c_{1}^{7} {c_3}}{15 {{\pi }^4}}+\frac{103 c_{1}^{7} {c_3}}{3 {{\pi }^3}}+
 {}
{}
 \frac{433 c_{1}^{7} {c_3}}{56 {{\pi }^2}}+\frac{1345344 c_{1}^{5} {c_2} {c_3}}{{{\pi }^7}}-\frac{312960
c_{1}^{5} {c_2} {c_3}}{{{\pi }^6}}-\frac{9792 c_{1}^{5} {c_2} {c_3}}{{{\pi }^5}}-  {}
{}
 \frac{5680 c_{1}^{5} {c_2} {c_3}}{3 {{\pi }^4}}-\frac{1937 c_{1}^{5} {c_2} {c_3}}{30
{{\pi }^3}}-\frac{23 c_{1}^{5} {c_2} {c_3}}{{{\pi }^2}}-\frac{370880 c_{1}^{3} c_{2}^{2} {c_3}}{{{\pi }^6}}+
 {}
{}
 \frac{97728 c_{1}^{3} c_{2}^{2} {c_3}}{{{\pi }^5}}+\frac{1396 c_{1}^{3} c_{2}^{2} {c_3}}{3
{{\pi }^4}}+\frac{124 c_{1}^{3} c_{2}^{2} {c_3}}{{{\pi }^3}}+\frac{14 c_{1}^{3} c_{2}^{2} {c_3}}{{{\pi
}^2}}+  {}
{}
 \frac{16048 {c_1} c_{2}^{3} {c_3}}{{{\pi }^5}}-\frac{3520 {c_1} c_{2}^{3} {c_3}}{{{\pi }^4}}-\frac{90
{c_1} c_{2}^{3} {c_3}}{{{\pi }^3}}-\frac{12 {c_1} c_{2}^{3} {c_3}}{{{\pi }^2}}-  {}
{}
 \frac{76160 c_{1}^{4} c_{3}^{2}}{{{\pi }^6}}+\frac{16512 c_{1}^{4} c_{3}^{2}}{{{\pi }^5}}+\frac{868
c_{1}^{4} c_{3}^{2}}{{{\pi }^4}}+\frac{40 c_{1}^{4} c_{3}^{2}}{{{\pi }^3}}+\frac{8 c_{1}^{4} c_{3}^{2}}{{{\pi
}^2}}+  {}
{}
 \frac{27312 c_{1}^{2} {c_2} c_{3}^{2}}{{{\pi }^5}}-\frac{6512 c_{1}^{2} {c_2} c_{3}^{2}}{{{\pi
}^4}}-\frac{108 c_{1}^{2} {c_2} c_{3}^{2}}{{{\pi }^3}}-\frac{4 c_{1}^{2} {c_2} c_{3}^{2}}{{{\pi }^2}}-\frac{624
c_{2}^{2} c_{3}^{2}}{{{\pi }^4}}+  {}
{}
 \frac{96 c_{2}^{2} c_{3}^{2}}{{{\pi }^3}}+\frac{8 c_{2}^{2} c_{3}^{2}}{{{\pi }^2}}-\frac{336 {c_1}
c_{3}^{3}}{{{\pi }^4}}+\frac{40 {c_1} c_{3}^{3}}{{{\pi }^3}}+\frac{3 {c_1} c_{3}^{3}}{{{\pi }^2}}+\frac{155584
c_{1}^{6} {c_4}}{{{\pi }^7}}-  {}
{}
 \frac{24000 c_{1}^{6} {c_4}}{{{\pi }^6}}-\frac{3336 c_{1}^{6} {c_4}}{{{\pi }^5}}-\frac{592 c_{1}^{6}
{c_4}}{3 {{\pi }^4}}-\frac{4891 c_{1}^{6} {c_4}}{90 {{\pi }^3}}-\frac{14 c_{1}^{6} {c_4}}{3 {{\pi
}^2}}-  {}
{}
 \frac{159040 c_{1}^{4} {c_2} {c_4}}{{{\pi }^6}}+\frac{37248 c_{1}^{4} {c_2} {c_4}}{{{\pi
}^5}}+\frac{716 c_{1}^{4} {c_2} {c_4}}{{{\pi }^4}}+\frac{580 c_{1}^{4} {c_2} {c_4}}{3 {{\pi }^3}}+
 {}
{}
 \frac{19 c_{1}^{4} {c_2} {c_4}}{2 {{\pi }^2}}+\frac{28464 c_{1}^{2} c_{2}^{2} {c_4}}{{{\pi
}^5}}-\frac{6864 c_{1}^{2} c_{2}^{2} {c_4}}{{{\pi }^4}}-\frac{124 c_{1}^{2} c_{2}^{2} {c_4}}{{{\pi }^3}}-\frac{2
c_{1}^{2} c_{2}^{2} {c_4}}{{{\pi }^2}}-  {}
{}
 \frac{280 c_{2}^{3} {c_4}}{{{\pi }^4}}+\frac{32 c_{2}^{3} {c_4}}{{{\pi }^3}}+\frac{c_{2}^{3} {c_4}}{{{\pi
}^2}}+\frac{16336 c_{1}^{3} {c_3} {c_4}}{{{\pi }^5}}-\frac{3552 c_{1}^{3} {c_3} {c_4}}{{{\pi }^4}}-  {}
{}
 \frac{102 c_{1}^{3} {c_3} {c_4}}{{{\pi }^3}}-\frac{12 c_{1}^{3} {c_3} {c_4}}{{{\pi }^2}}-\frac{3496
{c_1} {c_2} {c_3} {c_4}}{{{\pi }^4}}+\frac{800 {c_1} {c_2} {c_3} {c_4}}{{{\pi }^3}}+  {}
{}
 \frac{44 c_{3}^{2} {c_4}}{{{\pi }^3}}-\frac{4 c_{3}^{2} {c_4}}{{{\pi }^2}}-\frac{624 c_{1}^{2}
c_{4}^{2}}{{{\pi }^4}}+\frac{96 c_{1}^{2} c_{4}^{2}}{{{\pi }^3}}+\frac{8 c_{1}^{2} c_{4}^{2}}{{{\pi }^2}}+\frac{44
{c_2} c_{4}^{2}}{{{\pi }^3}}-  {}
{}
 \frac{4 {c_2} c_{4}^{2}}{{{\pi }^2}}-\frac{20384 c_{1}^{5} {c_5}}{{{\pi }^6}}+\frac{2976 c_{1}^{5}
{c_5}}{{{\pi }^5}}+\frac{368 c_{1}^{5} {c_5}}{{{\pi }^4}}+\frac{92 c_{1}^{5} {c_5}}{3 {{\pi }^3}}+\frac{91
c_{1}^{5} {c_5}}{20 {{\pi }^2}}+  {}
{}
 \frac{16336 c_{1}^{3} {c_2} {c_5}}{{{\pi }^5}}-\frac{3552 c_{1}^{3} {c_2} {c_5}}{{{\pi }^4}}-\frac{242
c_{1}^{3} {c_2} {c_5}}{3 {{\pi }^3}}-\frac{12 c_{1}^{3} {c_2} {c_5}}{{{\pi }^2}}-  {}
{}
 \frac{1552 {c_1} c_{2}^{2} {c_5}}{{{\pi }^4}}+\frac{304 {c_1} c_{2}^{2} {c_5}}{{{\pi }^3}}+\frac{8
{c_1} c_{2}^{2} {c_5}}{{{\pi }^2}}-\frac{1552 c_{1}^{2} {c_3} {c_5}}{{{\pi }^4}}+\frac{304 c_{1}^{2}
{c_3} {c_5}}{{{\pi }^3}}+  {}
{}
 \frac{8 c_{1}^{2} {c_3} {c_5}}{{{\pi }^2}}+\frac{108 {c_2} {c_3} {c_5}}{{{\pi }^3}}-\frac{16
{c_2} {c_3} {c_5}}{{{\pi }^2}}+\frac{108 {c_1} {c_4} {c_5}}{{{\pi }^3}}-\frac{16 {c_1} {c_4}
{c_5}}{{{\pi }^2}}-  {}
{}
 \frac{2 c_{5}^{2}}{{{\pi }^2}}+\frac{2608 c_{1}^{4} {c_6}}{{{\pi }^5}}-\frac{352 c_{1}^{4} {c_6}}{{{\pi
}^4}}-\frac{106 c_{1}^{4} {c_6}}{3 {{\pi }^3}}-\frac{4 c_{1}^{4} {c_6}}{{{\pi }^2}}-\frac{1488 c_{1}^{2}
{c_2} {c_6}}{{{\pi }^4}}+  {}
{}
 \frac{280 c_{1}^{2} {c_2} {c_6}}{{{\pi }^3}}+\frac{8 c_{1}^{2} {c_2} {c_6}}{{{\pi }^2}}+\frac{44
c_{2}^{2} {c_6}}{{{\pi }^3}}-\frac{4 c_{2}^{2} {c_6}}{{{\pi }^2}}+\frac{116 {c_1} {c_3} {c_6}}{{{\pi }^3}}-
 {}
{}
 \frac{16 {c_1} {c_3} {c_6}}{{{\pi }^2}}-\frac{6 {c_4} {c_6}}{{{\pi }^2}}-\frac{336 c_{1}^{3}
{c_7}}{{{\pi }^4}}+\frac{40 c_{1}^{3} {c_7}}{{{\pi }^3}}+\frac{3 c_{1}^{3} {c_7}}{{{\pi }^2}}+\frac{116 {c_1}
{c_2} {c_7}}{{{\pi }^3}}-  {}
{}
 \frac{16 {c_1} {c_2} {c_7}}{{{\pi }^2}}-\frac{6 {c_3} {c_7}}{{{\pi }^2}}+\frac{44 c_{1}^{2}
{c_8}}{{{\pi }^3}}-\frac{4 c_{1}^{2} {c_8}}{{{\pi }^2}}-\frac{6 {c_2} {c_8}}{{{\pi }^2}}-\frac{6 {c_1}
{c_9}}{{{\pi }^2}}+\frac{{c_{10}}}{\pi }+  {}
{}
\frac{17199104 c_{1}^{11}}{{{\pi }^{11}}}-\frac{915200 c_{1}^{11}}{{{\pi }^9}}-\frac{161744
c_{1}^{11}}{5 {{\pi }^7}}-\frac{113098 c_{1}^{11}}{135 {{\pi }^5}}-  {}
{}
 \frac{1537 c_{1}^{11}}{50 {{\pi }^3}}+\frac{63 c_{1}^{11}}{2816 \pi }-\frac{47297536 c_{1}^{9}
{c_2}}{{{\pi }^{10}}}+\frac{5857280 c_{1}^{9} {c_2}}{{{\pi }^9}}+\frac{1601600 c_{1}^{9} {c_2}}{{{\pi }^8}}+  {}
{}
 \frac{46464 c_{1}^{9} {c_2}}{{{\pi }^7}}+\frac{114928 c_{1}^{9} {c_2}}{3 {{\pi }^6}}+\frac{576
c_{1}^{9} {c_2}}{5 {{\pi }^5}}+\frac{157051 c_{1}^{9} {c_2}}{270 {{\pi }^4}}+\frac{24859 c_{1}^{9}
{c_2}}{315 {{\pi }^3}}+  {}
{}
 \frac{827 c_{1}^{9} {c_2}}{192 {{\pi }^2}}+\frac{41879552 c_{1}^{7} c_{2}^{2}}{{{\pi }^9}}-\frac{8968960
c_{1}^{7} c_{2}^{2}}{{{\pi }^8}}-\frac{631488 c_{1}^{7} c_{2}^{2}}{{{\pi }^7}}-  {}
{}
 \frac{72160 c_{1}^{7} c_{2}^{2}}{{{\pi }^6}}-\frac{11848 c_{1}^{7} c_{2}^{2}}{5 {{\pi }^5}}-\frac{5234
c_{1}^{7} c_{2}^{2}}{5 {{\pi }^4}}-\frac{7109 c_{1}^{7} c_{2}^{2}}{210 {{\pi }^3}}-\frac{10 c_{1}^{7}
c_{2}^{2}}{{{\pi }^2}}-  {}
{}
 \frac{13581568 c_{1}^{5} c_{2}^{3}}{{{\pi }^8}}+\frac{3527040 c_{1}^{5} c_{2}^{3}}{{{\pi }^7}}+\frac{99616
c_{1}^{5} c_{2}^{3}}{{{\pi }^6}}+\frac{14144 c_{1}^{5} c_{2}^{3}}{3 {{\pi }^5}}+  {}
{}
 \frac{7472 c_{1}^{5} c_{2}^{3}}{15 {{\pi }^4}}-\frac{97 c_{1}^{5} c_{2}^{3}}{{{\pi }^3}}+\frac{457
c_{1}^{5} c_{2}^{3}}{40 {{\pi }^2}}+\frac{1344640 c_{1}^{3} c_{2}^{4}}{{{\pi }^7}}-\frac{330880 c_{1}^{3}
c_{2}^{4}}{{{\pi }^6}}-  {}
{}
 \frac{17080 c_{1}^{3} c_{2}^{4}}{{{\pi }^5}}+\frac{880 c_{1}^{3} c_{2}^{4}}{{{\pi }^4}}+\frac{257
c_{1}^{3} c_{2}^{4}}{18 {{\pi }^3}}-\frac{22 c_{1}^{3} c_{2}^{4}}{3 {{\pi }^2}}-\frac{20416 {c_1}
c_{2}^{5}}{{{\pi }^6}}+  {}
{}
 \frac{2976 {c_1} c_{2}^{5}}{{{\pi }^5}}+\frac{368 {c_1} c_{2}^{5}}{{{\pi }^4}}+\frac{92 {c_1}
c_{2}^{5}}{3 {{\pi }^3}}+\frac{91 {c_1} c_{2}^{5}}{20 {{\pi }^2}}+\frac{8273408 c_{1}^{8} {c_3}}{{{\pi
}^9}}-  {}
{}
 \frac{1281280 c_{1}^{8} {c_3}}{{{\pi }^8}}-\frac{230912 c_{1}^{8} {c_3}}{{{\pi }^7}}-\frac{8800
c_{1}^{8} {c_3}}{{{\pi }^6}}-\frac{63832 c_{1}^{8} {c_3}}{15 {{\pi }^5}}-\frac{1426 c_{1}^{8} {c_3}}{15
{{\pi }^4}}-  {}
{}
 \frac{5198 c_{1}^{8} {c_3}}{105 {{\pi }^3}}-\frac{6 c_{1}^{8} {c_3}}{{{\pi }^2}}-\frac{12300288
c_{1}^{6} {c_2} {c_3}}{{{\pi }^8}}+\frac{2961024 c_{1}^{6} {c_2} {c_3}}{{{\pi }^7}}+  {}
{}
 \frac{110880 c_{1}^{6} {c_2} {c_3}}{{{\pi }^6}}+\frac{15424 c_{1}^{6} {c_2} {c_3}}{{{\pi
}^5}}+\frac{1572 c_{1}^{6} {c_2} {c_3}}{5 {{\pi }^4}}+\frac{89 c_{1}^{6} {c_2} {c_3}}{{{\pi }^3}}+
 {}
{}
 \frac{45 c_{1}^{6} {c_2} {c_3}}{4 {{\pi }^2}}+\frac{4642880 c_{1}^{4} c_{2}^{2} {c_3}}{{{\pi
}^7}}-\frac{1320000 c_{1}^{4} c_{2}^{2} {c_3}}{{{\pi }^6}}+\frac{2824 c_{1}^{4} c_{2}^{2} {c_3}}{{{\pi
}^5}}-  {}
{}
 \frac{5120 c_{1}^{4} c_{2}^{2} {c_3}}{3 {{\pi }^4}}+\frac{341 c_{1}^{4} c_{2}^{2}
{c_3}}{2 {{\pi }^3}}-\frac{23 c_{1}^{4} c_{2}^{2} {c_3}}{2 {{\pi }^2}}-\frac{399520 c_{1}^{2} c_{2}^{3}
{c_3}}{{{\pi }^6}}+  {}
{}
 \frac{108032 c_{1}^{2} c_{2}^{3} {c_3}}{{{\pi }^5}}+\frac{628 c_{1}^{2} c_{2}^{3} {c_3}}{{{\pi
}^4}}-\frac{196 c_{1}^{2} c_{2}^{3} {c_3}}{3 {{\pi }^3}}+\frac{7 c_{1}^{2} c_{2}^{3} {c_3}}{{{\pi
}^2}}+  {}
{}
 \frac{2608 c_{2}^{4} {c_3}}{{{\pi }^5}}-\frac{352 c_{2}^{4} {c_3}}{{{\pi }^4}}-\frac{106 c_{2}^{4}
{c_3}}{3 {{\pi }^3}}-\frac{4 c_{2}^{4} {c_3}}{{{\pi }^2}}+\frac{749760 c_{1}^{5} c_{3}^{2}}{{{\pi }^7}}-  {}
{}
 \frac{176000 c_{1}^{5} c_{3}^{2}}{{{\pi }^6}}-\frac{8192 c_{1}^{5} c_{3}^{2}}{{{\pi }^5}}-\frac{1360
c_{1}^{5} c_{3}^{2}}{3 {{\pi }^4}}-\frac{1001 c_{1}^{5} c_{3}^{2}}{30 {{\pi }^3}}-\frac{3 c_{1}^{5}
c_{3}^{2}}{2 {{\pi }^2}}-  {}
{}
 \frac{406560 c_{1}^{3} {c_2} c_{3}^{2}}{{{\pi }^6}}+\frac{110112 c_{1}^{3} {c_2} c_{3}^{2}}{{{\pi
}^5}}+\frac{2080 c_{1}^{3} {c_2} c_{3}^{2}}{3 {{\pi }^4}}-\frac{68 c_{1}^{3} {c_2} c_{3}^{2}}{{{\pi
}^3}}+  {}
{}
 \frac{6 c_{1}^{3} {c_2} c_{3}^{2}}{{{\pi }^2}}+\frac{28112 {c_1} c_{2}^{2} c_{3}^{2}}{{{\pi
}^5}}-\frac{6896 {c_1} c_{2}^{2} c_{3}^{2}}{{{\pi }^4}}-\frac{54 {c_1} c_{2}^{2} c_{3}^{2}}{{{\pi }^3}}-\frac{4
{c_1} c_{2}^{2} c_{3}^{2}}{{{\pi }^2}}+  {}
{}
 \frac{7472 c_{1}^{2} c_{3}^{3}}{{{\pi }^5}}-\frac{1440 c_{1}^{2} c_{3}^{3}}{{{\pi }^4}}-\frac{278
c_{1}^{2} c_{3}^{3}}{3 {{\pi }^3}}-\frac{2 c_{1}^{2} c_{3}^{3}}{{{\pi }^2}}-\frac{336 {c_2} c_{3}^{3}}{{{\pi
}^4}}+\frac{40 {c_2} c_{3}^{3}}{{{\pi }^3}}+  {}
{}
 \frac{3 {c_2} c_{3}^{3}}{{{\pi }^2}}-\frac{1208064 c_{1}^{7} {c_4}}{{{\pi }^8}}+\frac{192896 c_{1}^{7}
{c_4}}{{{\pi }^7}}+\frac{30800 c_{1}^{7} {c_4}}{{{\pi }^6}}+\frac{1024 c_{1}^{7} {c_4}}{{{\pi }^5}}+  {}
{}
 \frac{2682 c_{1}^{7} {c_4}}{5 {{\pi }^4}}+\frac{31 c_{1}^{7} {c_4}}{3 {{\pi }^3}}+\frac{349
c_{1}^{7} {c_4}}{56 {{\pi }^2}}+\frac{1518528 c_{1}^{5} {c_2} {c_4}}{{{\pi }^7}}-\frac{373120 c_{1}^{5}
{c_2} {c_4}}{{{\pi }^6}}-  {}
{}
 \frac{8608 c_{1}^{5} {c_2} {c_4}}{{{\pi }^5}}-\frac{5696 c_{1}^{5} {c_2} {c_4}}{3
{{\pi }^4}}+\frac{277 c_{1}^{5} {c_2} {c_4}}{10 {{\pi }^3}}-\frac{15 c_{1}^{5} {c_2} {c_4}}{{{\pi
}^2}}-  {}
{}
 \frac{417120 c_{1}^{3} c_{2}^{2} {c_4}}{{{\pi }^6}}+\frac{114720 c_{1}^{3} c_{2}^{2} {c_4}}{{{\pi
}^5}}+\frac{208 c_{1}^{3} c_{2}^{2} {c_4}}{{{\pi }^4}}-\frac{236 c_{1}^{3} c_{2}^{2} {c_4}}{3 {{\pi
}^3}}+  {}
{}
 \frac{11 c_{1}^{3} c_{2}^{2} {c_4}}{{{\pi }^2}}+\frac{16400 {c_1} c_{2}^{3} {c_4}}{{{\pi
}^5}}-\frac{3552 {c_1} c_{2}^{3} {c_4}}{{{\pi }^4}}-\frac{242 {c_1} c_{2}^{3} {c_4}}{3 {{\pi }^3}}-
 {}
{}
 \frac{12 {c_1} c_{2}^{3} {c_4}}{{{\pi }^2}}-\frac{170720 c_{1}^{4} {c_3} {c_4}}{{{\pi }^6}}+\frac{40512
c_{1}^{4} {c_3} {c_4}}{{{\pi }^5}}+\frac{3608 c_{1}^{4} {c_3} {c_4}}{3 {{\pi }^4}}+  {}
{}
 \frac{280 c_{1}^{4} {c_3} {c_4}}{3 {{\pi }^3}}+\frac{9 c_{1}^{4} {c_3} {c_4}}{2
{{\pi }^2}}+\frac{64560 c_{1}^{2} {c_2} {c_3} {c_4}}{{{\pi }^5}}-\frac{17456 c_{1}^{2} {c_2} {c_3}
{c_4}}{{{\pi }^4}}+  {}
{}
 \frac{212 c_{1}^{2} {c_2} {c_3} {c_4}}{{{\pi }^3}}-\frac{8 c_{1}^{2} {c_2} {c_3}
{c_4}}{{{\pi }^2}}-\frac{1488 c_{2}^{2} {c_3} {c_4}}{{{\pi }^4}}+\frac{280 c_{2}^{2} {c_3} {c_4}}{{{\pi
}^3}}+  {}
{}
 \frac{8 c_{2}^{2} {c_3} {c_4}}{{{\pi }^2}}-\frac{1432 {c_1} c_{3}^{2} {c_4}}{{{\pi }^4}}+\frac{272
{c_1} c_{3}^{2} {c_4}}{{{\pi }^3}}+\frac{6 {c_1} c_{3}^{2} {c_4}}{{{\pi }^2}}+\frac{7472 c_{1}^{3}
c_{4}^{2}}{{{\pi }^5}}-  {}
{}
 \frac{1440 c_{1}^{3} c_{4}^{2}}{{{\pi }^4}}-\frac{278 c_{1}^{3} c_{4}^{2}}{3 {{\pi }^3}}-\frac{2
c_{1}^{3} c_{4}^{2}}{{{\pi }^2}}-\frac{1576 {c_1} {c_2} c_{4}^{2}}{{{\pi }^4}}+\frac{304 {c_1} {c_2}
c_{4}^{2}}{{{\pi }^3}}+  {}
{}
 \frac{8 {c_1} {c_2} c_{4}^{2}}{{{\pi }^2}}+\frac{44 {c_3} c_{4}^{2}}{{{\pi }^3}}-\frac{4
{c_3} c_{4}^{2}}{{{\pi }^2}}+\frac{159808 c_{1}^{6} {c_5}}{{{\pi }^7}}-\frac{24640 c_{1}^{6} {c_5}}{{{\pi }^6}}-
 {}
{}
 \frac{10768 c_{1}^{6} {c_5}}{3 {{\pi }^5}}-\frac{520 c_{1}^{6} {c_5}}{3 {{\pi }^4}}-\frac{4621
c_{1}^{6} {c_5}}{90 {{\pi }^3}}-\frac{14 c_{1}^{6} {c_5}}{3 {{\pi }^2}}-\frac{165440 c_{1}^{4} {c_2}
{c_5}}{{{\pi }^6}}+  {}  $

$
{}
 \frac{39072 c_{1}^{4} {c_2} {c_5}}{{{\pi }^5}}+\frac{796 c_{1}^{4} {c_2} {c_5}}{{{\pi }^4}}+\frac{484
c_{1}^{4} {c_2} {c_5}}{3 {{\pi }^3}}+\frac{9 c_{1}^{4} {c_2} {c_5}}{{{\pi }^2}}+  {}
{}
 \frac{29760 c_{1}^{2} c_{2}^{2} {c_5}}{{{\pi }^5}}-\frac{7328 c_{1}^{2} c_{2}^{2} {c_5}}{{{\pi
}^4}}-\frac{84 c_{1}^{2} c_{2}^{2} {c_5}}{{{\pi }^3}}-\frac{2 c_{1}^{2} c_{2}^{2} {c_5}}{{{\pi }^2}}-\frac{336
c_{2}^{3} {c_5}}{{{\pi }^4}}+  {}
{}
 \frac{40 c_{2}^{3} {c_5}}{{{\pi }^3}}+\frac{3 c_{2}^{3} {c_5}}{{{\pi }^2}}+\frac{17696 c_{1}^{3}
{c_3} {c_5}}{{{\pi }^5}}-\frac{4032 c_{1}^{3} {c_3} {c_5}}{{{\pi }^4}}-\frac{242 c_{1}^{3} {c_3}
{c_5}}{3 {{\pi }^3}}-  {}
{}
 \frac{8 c_{1}^{3} {c_3} {c_5}}{{{\pi }^2}}-\frac{3544 {c_1} {c_2} {c_3} {c_5}}{{{\pi
}^4}}+\frac{800 {c_1} {c_2} {c_3} {c_5}}{{{\pi }^3}}+\frac{44 c_{3}^{2} {c_5}}{{{\pi }^3}}-\frac{4
c_{3}^{2} {c_5}}{{{\pi }^2}}-  {}
{}
 \frac{1432 c_{1}^{2} {c_4} {c_5}}{{{\pi }^4}}+\frac{272 c_{1}^{2} {c_4} {c_5}}{{{\pi }^3}}+\frac{6
c_{1}^{2} {c_4} {c_5}}{{{\pi }^2}}+\frac{116 {c_2} {c_4} {c_5}}{{{\pi }^3}}-\frac{16 {c_2} {c_4}
{c_5}}{{{\pi }^2}}+  {}
{}
 \frac{44 {c_1} c_{5}^{2}}{{{\pi }^3}}-\frac{4 {c_1} c_{5}^{2}}{{{\pi }^2}}-\frac{20416 c_{1}^{5}
{c_6}}{{{\pi }^6}}+\frac{2976 c_{1}^{5} {c_6}}{{{\pi }^5}}+\frac{368 c_{1}^{5} {c_6}}{{{\pi }^4}}+\frac{92 c_{1}^{5}
{c_6}}{3 {{\pi }^3}}+  {}
{}
 \frac{91 c_{1}^{5} {c_6}}{20 {{\pi }^2}}+\frac{16400 c_{1}^{3} {c_2} {c_6}}{{{\pi }^5}}-\frac{3552
c_{1}^{3} {c_2} {c_6}}{{{\pi }^4}}-\frac{242 c_{1}^{3} {c_2} {c_6}}{3 {{\pi }^3}}-\frac{12 c_{1}^{3}
{c_2} {c_6}}{{{\pi }^2}}-  {}
{}
 \frac{1576 {c_1} c_{2}^{2} {c_6}}{{{\pi }^4}}+\frac{304 {c_1} c_{2}^{2} {c_6}}{{{\pi }^3}}+\frac{8
{c_1} c_{2}^{2} {c_6}}{{{\pi }^2}}-\frac{1576 c_{1}^{2} {c_3} {c_6}}{{{\pi }^4}}+\frac{304 c_{1}^{2}
{c_3} {c_6}}{{{\pi }^3}}+  {}
{}
 \frac{8 c_{1}^{2} {c_3} {c_6}}{{{\pi }^2}}+\frac{116 {c_2} {c_3} {c_6}}{{{\pi }^3}}-\frac{16
{c_2} {c_3} {c_6}}{{{\pi }^2}}+\frac{116 {c_1} {c_4} {c_6}}{{{\pi }^3}}-\frac{16 {c_1} {c_4}
{c_6}}{{{\pi }^2}}-  {}
{}
 \frac{6 {c_5} {c_6}}{{{\pi }^2}}+\frac{2608 c_{1}^{4} {c_7}}{{{\pi }^5}}-\frac{352 c_{1}^{4}
{c_7}}{{{\pi }^4}}-\frac{106 c_{1}^{4} {c_7}}{3 {{\pi }^3}}-\frac{4 c_{1}^{4} {c_7}}{{{\pi }^2}}-\frac{1488
c_{1}^{2} {c_2} {c_7}}{{{\pi }^4}}+  {}
{}
 \frac{280 c_{1}^{2} {c_2} {c_7}}{{{\pi }^3}}+\frac{8 c_{1}^{2} {c_2} {c_7}}{{{\pi }^2}}+\frac{44
c_{2}^{2} {c_7}}{{{\pi }^3}}-\frac{4 c_{2}^{2} {c_7}}{{{\pi }^2}}+\frac{116 {c_1} {c_3} {c_7}}{{{\pi }^3}}-
 {}
{}
 \frac{16 {c_1} {c_3} {c_7}}{{{\pi }^2}}-\frac{6 {c_4} {c_7}}{{{\pi }^2}}-\frac{336 c_{1}^{3}
{c_8}}{{{\pi }^4}}+\frac{40 c_{1}^{3} {c_8}}{{{\pi }^3}}+\frac{3 c_{1}^{3} {c_8}}{{{\pi }^2}}+\frac{116 {c_1}
{c_2} {c_8}}{{{\pi }^3}}-  {}
{}
 \frac{16 {c_1} {c_2} {c_8}}{{{\pi }^2}}-\frac{6 {c_3} {c_8}}{{{\pi }^2}}+\frac{44 c_{1}^{2}
{c_9}}{{{\pi }^3}}-\frac{4 c_{1}^{2} {c_9}}{{{\pi }^2}}-\frac{6 {c_2} {c_9}}{{{\pi }^2}}-\frac{6 {c_1}
{c_{10}}}{{{\pi }^2}}+\frac{{c_{11}}}{\pi }
-\frac{120393728 c_{1}^{12}}{{{\pi }^{12}}}+\frac{21159424 c_{1}^{12}}{3 {{\pi }^{10}}}+\frac{3429712
c_{1}^{12}}{15 {{\pi }^8}}+\frac{44276 c_{1}^{12}}{9 {{\pi }^6}}+  {}
{}
 \frac{1485521 c_{1}^{12}}{9072 {{\pi }^4}}+\frac{2216479 c_{1}^{12}}{266112 {{\pi }^2}}+\frac{361181184
c_{1}^{10} {c_2}}{{{\pi }^{11}}}-\frac{44808192 c_{1}^{10} {c_2}}{{{\pi }^{10}}}-  {}
{}
 \frac{42318848 c_{1}^{10} {c_2}}{3 {{\pi }^9}}-\frac{128128 c_{1}^{10} {c_2}}{{{\pi }^8}}-\frac{1669096
c_{1}^{10} {c_2}}{5 {{\pi }^7}}+\frac{12848 c_{1}^{10} {c_2}}{{{\pi }^6}}-  {}
{}
 \frac{78136 c_{1}^{10} {c_2}}{15 {{\pi }^5}}+\frac{3931 c_{1}^{10} {c_2}}{105 {{\pi }^4}}-\frac{748717
c_{1}^{10} {c_2}}{10080 {{\pi }^3}}-\frac{5863 c_{1}^{10} {c_2}}{160 {{\pi }^2}}-  {}
{}
 \frac{364575744 c_{1}^{8} c_{2}^{2}}{{{\pi }^{10}}}+\frac{79659008 c_{1}^{8} c_{2}^{2}}{{{\pi }^9}}+\frac{6662656
c_{1}^{8} c_{2}^{2}}{{{\pi }^8}}+\frac{439296 c_{1}^{8} c_{2}^{2}}{{{\pi }^7}}+  {}
{}
 \frac{30588 c_{1}^{8} c_{2}^{2}}{{{\pi }^6}}+\frac{16832 c_{1}^{8} c_{2}^{2}}{5 {{\pi }^5}}-\frac{320683
c_{1}^{8} c_{2}^{2}}{315 {{\pi }^4}}+\frac{3688 c_{1}^{8} c_{2}^{2}}{21 {{\pi }^3}}+\frac{1979 c_{1}^{8}
c_{2}^{2}}{32 {{\pi }^2}}+  {}
{}
 \frac{145739776 c_{1}^{6} c_{2}^{3}}{{{\pi }^9}}-\frac{39975936 c_{1}^{6} c_{2}^{3}}{{{\pi }^8}}-\frac{2727296
c_{1}^{6} c_{2}^{3}}{3 {{\pi }^7}}-\frac{62240 c_{1}^{6} c_{2}^{3}}{{{\pi }^6}}+  {}
{}
 \frac{139508 c_{1}^{6} c_{2}^{3}}{15 {{\pi }^5}}+\frac{2096 c_{1}^{6} c_{2}^{3}}{{{\pi }^4}}-\frac{85787
c_{1}^{6} c_{2}^{3}}{540 {{\pi }^3}}-\frac{899 c_{1}^{6} c_{2}^{3}}{18 {{\pi }^2}}-\frac{20908160 c_{1}^{4}
c_{2}^{4}}{{{\pi }^8}}+  {}
{}
 \frac{5857280 c_{1}^{4} c_{2}^{4}}{{{\pi }^7}}+\frac{496720 c_{1}^{4} c_{2}^{4}}{3 {{\pi }^6}}-\frac{24448
c_{1}^{4} c_{2}^{4}}{{{\pi }^5}}-\frac{341 c_{1}^{4} c_{2}^{4}}{{{\pi }^4}}+  {}
{}
 \frac{58 c_{1}^{4} c_{2}^{4}}{3 {{\pi }^3}}+\frac{99 c_{1}^{4} c_{2}^{4}}{4 {{\pi }^2}}+\frac{728832
c_{1}^{2} c_{2}^{5}}{{{\pi }^7}}-\frac{160000 c_{1}^{2} c_{2}^{5}}{{{\pi }^6}}-\frac{12496 c_{1}^{2} c_{2}^{5}}{{{\pi
}^5}}-  {}
{}
 \frac{248 c_{1}^{2} c_{2}^{5}}{3 {{\pi }^4}}-\frac{39 c_{1}^{2} c_{2}^{5}}{{{\pi }^3}}-\frac{4
c_{1}^{2} c_{2}^{5}}{{{\pi }^2}}-\frac{1344 c_{2}^{6}}{{{\pi }^6}}+\frac{100 c_{2}^{6}}{3 {{\pi }^4}}+\frac{251
c_{2}^{6}}{180 {{\pi }^2}}-  {}
{}
 \frac{63591424 c_{1}^{9} {c_3}}{{{\pi }^{10}}}+\frac{9957376 c_{1}^{9} {c_3}}{{{\pi }^9}}+\frac{6214208
c_{1}^{9} {c_3}}{3 {{\pi }^8}}+\frac{27456 c_{1}^{9} {c_3}}{{{\pi }^7}}+  {}
{}
 \frac{116528 c_{1}^{9} {c_3}}{3 {{\pi }^6}}-\frac{5856 c_{1}^{9} {c_3}}{5 {{\pi }^5}}+\frac{664603
c_{1}^{9} {c_3}}{1890 {{\pi }^4}}+\frac{2885 c_{1}^{9} {c_3}}{126 {{\pi }^3}}+\frac{801 c_{1}^{9}
{c_3}}{64 {{\pi }^2}}+  {}
{}
 \frac{109983744 c_{1}^{7} {c_2} {c_3}}{{{\pi }^9}}-\frac{27163136 c_{1}^{7} {c_2} {c_3}}{{{\pi
}^8}}-\frac{1249248 c_{1}^{7} {c_2} {c_3}}{{{\pi }^7}}-  {}
{}
 \frac{96960 c_{1}^{7} {c_2} {c_3}}{{{\pi }^6}}-\frac{2072 c_{1}^{7} {c_2} {c_3}}{{{\pi }^5}}+\frac{25292
c_{1}^{7} {c_2} {c_3}}{45 {{\pi }^4}}+\frac{2173 c_{1}^{7} {c_2} {c_3}}{140 {{\pi }^3}}-  {}
{}
 \frac{40 c_{1}^{7} {c_2} {c_3}}{{{\pi }^2}}-\frac{53254656 c_{1}^{5} c_{2}^{2} {c_3}}{{{\pi
}^8}}+\frac{15997696 c_{1}^{5} c_{2}^{2} {c_3}}{{{\pi }^7}}-\frac{102480 c_{1}^{5} c_{2}^{2} {c_3}}{{{\pi
}^6}}+  {}
{}
 \frac{18720 c_{1}^{5} c_{2}^{2} {c_3}}{{{\pi }^5}}-\frac{28028 c_{1}^{5} c_{2}^{2} {c_3}}{5
{{\pi }^4}}-\frac{1436 c_{1}^{5} c_{2}^{2} {c_3}}{15 {{\pi }^3}}+\frac{77 c_{1}^{5} c_{2}^{2} {c_3}}{2
{{\pi }^2}}+  {}
{}
 \frac{7329920 c_{1}^{3} c_{2}^{3} {c_3}}{{{\pi }^7}}-\frac{2242560 c_{1}^{3} c_{2}^{3} {c_3}}{{{\pi
}^6}}+\frac{65240 c_{1}^{3} c_{2}^{3} {c_3}}{3 {{\pi }^5}}+\frac{12400 c_{1}^{3} c_{2}^{3} {c_3}}{3
{{\pi }^4}}+  {}
{}
 \frac{413 c_{1}^{3} c_{2}^{3} {c_3}}{3 {{\pi }^3}}-\frac{92 c_{1}^{3} c_{2}^{3} {c_3}}{3
{{\pi }^2}}-\frac{169728 {c_1} c_{2}^{4} {c_3}}{{{\pi }^6}}+\frac{40384 {c_1} c_{2}^{4} {c_3}}{{{\pi }^5}}+
 {}
{}
 \frac{3596 {c_1} c_{2}^{4} {c_3}}{3 {{\pi }^4}}+\frac{304 {c_1} c_{2}^{4} {c_3}}{3
{{\pi }^3}}+\frac{5 {c_1} c_{2}^{4} {c_3}}{{{\pi }^2}}-\frac{7081984 c_{1}^{6} c_{3}^{2}}{{{\pi }^8}}+  {}
{}
 \frac{1757184 c_{1}^{6} c_{3}^{2}}{{{\pi }^7}}+\frac{239840 c_{1}^{6} c_{3}^{2}}{3 {{\pi }^6}}+\frac{3392
c_{1}^{6} c_{3}^{2}}{{{\pi }^5}}-\frac{368 c_{1}^{6} c_{3}^{2}}{5 {{\pi }^4}}-\frac{92 c_{1}^{6} c_{3}^{2}}{5
{{\pi }^3}}+  {}
{}
 \frac{73 c_{1}^{6} c_{3}^{2}}{8 {{\pi }^2}}+\frac{5245760 c_{1}^{4} {c_2} c_{3}^{2}}{{{\pi
}^7}}-\frac{1547520 c_{1}^{4} {c_2} c_{3}^{2}}{{{\pi }^6}}+\frac{4136 c_{1}^{4} {c_2} c_{3}^{2}}{{{\pi
}^5}}+  {}
{}
 \frac{4184 c_{1}^{4} {c_2} c_{3}^{2}}{3 {{\pi }^4}}+\frac{425 c_{1}^{4} {c_2} c_{3}^{2}}{3
{{\pi }^3}}-\frac{17 c_{1}^{4} {c_2} c_{3}^{2}}{2 {{\pi }^2}}-\frac{695904 c_{1}^{2} c_{2}^{2} c_{3}^{2}}{{{\pi
}^6}}+  {}
{}
 \frac{205056 c_{1}^{2} c_{2}^{2} c_{3}^{2}}{{{\pi }^5}}-\frac{2548 c_{1}^{2} c_{2}^{2} c_{3}^{2}}{{{\pi
}^4}}-\frac{192 c_{1}^{2} c_{2}^{2} c_{3}^{2}}{{{\pi }^3}}+\frac{37 c_{1}^{2} c_{2}^{2} c_{3}^{2}}{2
{{\pi }^2}}+  {}
{}
 \frac{7440 c_{2}^{3} c_{3}^{2}}{{{\pi }^5}}-\frac{1440 c_{2}^{3} c_{3}^{2}}{{{\pi }^4}}-\frac{278
c_{2}^{3} c_{3}^{2}}{3 {{\pi }^3}}-\frac{2 c_{2}^{3} c_{3}^{2}}{{{\pi }^2}}-\frac{122560 c_{1}^{3} c_{3}^{3}}{{{\pi
}^6}}+  {}
{}
 \frac{29472 c_{1}^{3} c_{3}^{3}}{{{\pi }^5}}+\frac{3916 c_{1}^{3} c_{3}^{3}}{3 {{\pi }^4}}-\frac{64
c_{1}^{3} c_{3}^{3}}{{{\pi }^3}}+\frac{23 c_{1}^{3} c_{3}^{3}}{6 {{\pi }^2}}+\frac{16576 {c_1} {c_2}
c_{3}^{3}}{{{\pi }^5}}-  {}
{}
 \frac{3616 {c_1} {c_2} c_{3}^{3}}{{{\pi }^4}}-\frac{102 {c_1} {c_2} c_{3}^{3}}{{{\pi }^3}}-\frac{12
{c_1} {c_2} c_{3}^{3}}{{{\pi }^2}}-\frac{40 c_{3}^{4}}{{{\pi }^4}}+\frac{c_{3}^{4}}{3 {{\pi }^2}}+  {}
{}
 \frac{9391616 c_{1}^{8} {c_4}}{{{\pi }^9}}-\frac{1537536 c_{1}^{8} {c_4}}{{{\pi }^8}}-\frac{279136
c_{1}^{8} {c_4}}{{{\pi }^7}}-\frac{2880 c_{1}^{8} {c_4}}{{{\pi }^6}}-  {}
{}
 \frac{24862 c_{1}^{8} {c_4}}{5 {{\pi }^5}}+\frac{8104 c_{1}^{8} {c_4}}{45 {{\pi }^4}}-\frac{8781
c_{1}^{8} {c_4}}{140 {{\pi }^3}}-\frac{6 c_{1}^{8} {c_4}}{{{\pi }^2}}-\frac{14035840 c_{1}^{6} {c_2}
{c_4}}{{{\pi }^8}}+  {}
{}
 \frac{3569280 c_{1}^{6} {c_2} {c_4}}{{{\pi }^7}}+\frac{317680 c_{1}^{6} {c_2} {c_4}}{3
{{\pi }^6}}+\frac{13888 c_{1}^{6} {c_2} {c_4}}{{{\pi }^5}}-\frac{3793 c_{1}^{6} {c_2} {c_4}}{5 {{\pi
}^4}}+  {}
{}
 \frac{1903 c_{1}^{6} {c_2} {c_4}}{15 {{\pi }^3}}+\frac{53 c_{1}^{6} {c_2} {c_4}}{4
{{\pi }^2}}+\frac{5291520 c_{1}^{4} c_{2}^{2} {c_4}}{{{\pi }^7}}-\frac{1579840 c_{1}^{4} c_{2}^{2} {c_4}}{{{\pi
}^6}}+  {}
{}
 \frac{14272 c_{1}^{4} c_{2}^{2} {c_4}}{{{\pi }^5}}+\frac{272 c_{1}^{4} c_{2}^{2} {c_4}}{{{\pi
}^4}}+\frac{68 c_{1}^{4} c_{2}^{2} {c_4}}{{{\pi }^3}}-\frac{6 c_{1}^{4} c_{2}^{2} {c_4}}{{{\pi }^2}}- 
{}
{}
 \frac{435616 c_{1}^{2} c_{2}^{3} {c_4}}{{{\pi }^6}}+\frac{119904 c_{1}^{2} c_{2}^{3} {c_4}}{{{\pi
}^5}}+\frac{836 c_{1}^{2} c_{2}^{3} {c_4}}{{{\pi }^4}}-\frac{128 c_{1}^{2} c_{2}^{3} {c_4}}{{{\pi }^3}}+
 {}
{}
 \frac{4 c_{1}^{2} c_{2}^{3} {c_4}}{{{\pi }^2}}+\frac{2016 c_{2}^{4} {c_4}}{{{\pi }^5}}-\frac{240
c_{2}^{4} {c_4}}{{{\pi }^4}}-\frac{50 c_{2}^{4} {c_4}}{3 {{\pi }^3}}-\frac{6 c_{2}^{4} {c_4}}{{{\pi }^2}}+
 {}
{}
 \frac{1686464 c_{1}^{5} {c_3} {c_4}}{{{\pi }^7}}-\frac{425600 c_{1}^{5} {c_3} {c_4}}{{{\pi
}^6}}-\frac{13224 c_{1}^{5} {c_3} {c_4}}{{{\pi }^5}}-\frac{544 c_{1}^{5} {c_3} {c_4}}{{{\pi }^4}}-  {}
{}
 \frac{309 c_{1}^{5} {c_3} {c_4}}{10 {{\pi }^3}}-\frac{11 c_{1}^{5} {c_3} {c_4}}{{{\pi
}^2}}-\frac{948320 c_{1}^{3} {c_2} {c_3} {c_4}}{{{\pi }^6}}+\frac{282880 c_{1}^{3} {c_2} {c_3}
{c_4}}{{{\pi }^5}}-  {}
{}
 \frac{15716 c_{1}^{3} {c_2} {c_3} {c_4}}{3 {{\pi }^4}}+\frac{80 c_{1}^{3} {c_2}
{c_3} {c_4}}{{{\pi }^3}}+\frac{66672 {c_1} c_{2}^{2} {c_3} {c_4}}{{{\pi }^5}}-  {}
{}
 \frac{18224 {c_1} c_{2}^{2} {c_3} {c_4}}{{{\pi }^4}}+\frac{192 {c_1} c_{2}^{2} {c_3}
{c_4}}{{{\pi }^3}}-\frac{4 {c_1} c_{2}^{2} {c_3} {c_4}}{{{\pi }^2}}+\frac{28640 c_{1}^{2} c_{3}^{2}
{c_4}}{{{\pi }^5}}-  {}
{}
 \frac{7040 c_{1}^{2} c_{3}^{2} {c_4}}{{{\pi }^4}}-\frac{60 c_{1}^{2} c_{3}^{2} {c_4}}{{{\pi
}^3}}-\frac{4 c_{1}^{2} c_{3}^{2} {c_4}}{{{\pi }^2}}-\frac{1552 {c_2} c_{3}^{2} {c_4}}{{{\pi }^4}}+\frac{304
{c_2} c_{3}^{2} {c_4}}{{{\pi }^3}}+  {}
{}
 \frac{8 {c_2} c_{3}^{2} {c_4}}{{{\pi }^2}}-\frac{81152 c_{1}^{4} c_{4}^{2}}{{{\pi }^6}}+\frac{17856
c_{1}^{4} c_{4}^{2}}{{{\pi }^5}}+\frac{984 c_{1}^{4} c_{4}^{2}}{{{\pi }^4}}+\frac{24 c_{1}^{4} c_{4}^{2}}{{{\pi
}^3}}+  {}
{}
 \frac{15 c_{1}^{4} c_{4}^{2}}{2 {{\pi }^2}}+\frac{30880 c_{1}^{2} {c_2} c_{4}^{2}}{{{\pi
}^5}}-\frac{7744 c_{1}^{2} {c_2} c_{4}^{2}}{{{\pi }^4}}-\frac{90 c_{1}^{2} {c_2} c_{4}^{2}}{{{\pi }^3}}-\frac{552
c_{2}^{2} c_{4}^{2}}{{{\pi }^4}}+  {}
{}
 \frac{80 c_{2}^{2} c_{4}^{2}}{{{\pi }^3}}+\frac{7 c_{2}^{2} c_{4}^{2}}{{{\pi }^2}}-\frac{1432 {c_1}
{c_3} c_{4}^{2}}{{{\pi }^4}}+\frac{272 {c_1} {c_3} c_{4}^{2}}{{{\pi }^3}}+\frac{6 {c_1} {c_3} c_{4}^{2}}{{{\pi
}^2}}+\frac{8 c_{4}^{3}}{{{\pi }^3}}+  {}
{}
 \frac{c_{4}^{3}}{6 \pi }-\frac{1254656 c_{1}^{7} {c_5}}{{{\pi }^8}}+\frac{201344 c_{1}^{7} {c_5}}{{{\pi
}^7}}+\frac{101008 c_{1}^{7} {c_5}}{3 {{\pi }^6}}+\frac{656 c_{1}^{7} {c_5}}{{{\pi }^5}}+  {}
{}
 \frac{2652 c_{1}^{7} {c_5}}{5 {{\pi }^4}}+\frac{37 c_{1}^{7} {c_5}}{3 {{\pi }^3}}+\frac{349
c_{1}^{7} {c_5}}{56 {{\pi }^2}}+\frac{1594944 c_{1}^{5} {c_2} {c_5}}{{{\pi }^7}}-\frac{397440 c_{1}^{5}
{c_2} {c_5}}{{{\pi }^6}}-  {}
{}
 \frac{9160 c_{1}^{5} {c_2} {c_5}}{{{\pi }^5}}-\frac{1552 c_{1}^{5} {c_2} {c_5}}{{{\pi }^4}}+\frac{31
c_{1}^{5} {c_2} {c_5}}{30 {{\pi }^3}}-\frac{15 c_{1}^{5} {c_2} {c_5}}{{{\pi }^2}}-  {}
{}
 \frac{439840 c_{1}^{3} c_{2}^{2} {c_5}}{{{\pi }^6}}+\frac{123200 c_{1}^{3} c_{2}^{2} {c_5}}{{{\pi
}^5}}-\frac{836 c_{1}^{3} c_{2}^{2} {c_5}}{3 {{\pi }^4}}-\frac{152 c_{1}^{3} c_{2}^{2} {c_5}}{3
{{\pi }^3}}+  {}
{}
 \frac{10 c_{1}^{3} c_{2}^{2} {c_5}}{{{\pi }^2}}+\frac{17872 {c_1} c_{2}^{3} {c_5}}{{{\pi
}^5}}-\frac{4096 {c_1} c_{2}^{3} {c_5}}{{{\pi }^4}}-\frac{242 {c_1} c_{2}^{3} {c_5}}{3 {{\pi }^3}}-
 {}
{}
 \frac{8 {c_1} c_{2}^{3} {c_5}}{{{\pi }^2}}-\frac{184864 c_{1}^{4} {c_3} {c_5}}{{{\pi }^6}}+\frac{45984
c_{1}^{4} {c_3} {c_5}}{{{\pi }^5}}+\frac{2480 c_{1}^{4} {c_3} {c_5}}{3 {{\pi }^4}}+  {}
{}
 \frac{104 c_{1}^{4} {c_3} {c_5}}{{{\pi }^3}}+\frac{13 c_{1}^{4} {c_3} {c_5}}{2 {{\pi
}^2}}+\frac{67664 c_{1}^{2} {c_2} {c_3} {c_5}}{{{\pi }^5}}-\frac{18384 c_{1}^{2} {c_2} {c_3}
{c_5}}{{{\pi }^4}}+  {}
{}
 \frac{164 c_{1}^{2} {c_2} {c_3} {c_5}}{{{\pi }^3}}-\frac{4 c_{1}^{2} {c_2} {c_3}
{c_5}}{{{\pi }^2}}-\frac{1432 c_{2}^{2} {c_3} {c_5}}{{{\pi }^4}}+\frac{272 c_{2}^{2} {c_3} {c_5}}{{{\pi
}^3}}+  {}
{}
 \frac{6 c_{2}^{2} {c_3} {c_5}}{{{\pi }^2}}-\frac{1552 {c_1} c_{3}^{2} {c_5}}{{{\pi }^4}}+\frac{304
{c_1} c_{3}^{2} {c_5}}{{{\pi }^3}}+\frac{8 {c_1} c_{3}^{2} {c_5}}{{{\pi }^2}}+\frac{16576 c_{1}^{3}
{c_4} {c_5}}{{{\pi }^5}}-  {}
{}
 \frac{3616 c_{1}^{3} {c_4} {c_5}}{{{\pi }^4}}-\frac{102 c_{1}^{3} {c_4} {c_5}}{{{\pi }^3}}-\frac{12
c_{1}^{3} {c_4} {c_5}}{{{\pi }^2}}-\frac{3672 {c_1} {c_2} {c_4} {c_5}}{{{\pi }^4}}+  {}
{}
 \frac{848 {c_1} {c_2} {c_4} {c_5}}{{{\pi }^3}}+\frac{116 {c_3} {c_4} {c_5}}{{{\pi
}^3}}-\frac{16 {c_3} {c_4} {c_5}}{{{\pi }^2}}-\frac{624 c_{1}^{2} c_{5}^{2}}{{{\pi }^4}}+\frac{96 c_{1}^{2}
c_{5}^{2}}{{{\pi }^3}}+  {}
{}
 \frac{8 c_{1}^{2} c_{5}^{2}}{{{\pi }^2}}+\frac{44 {c_2} c_{5}^{2}}{{{\pi }^3}}-\frac{4 {c_2}
c_{5}^{2}}{{{\pi }^2}}+\frac{160576 c_{1}^{6} {c_6}}{{{\pi }^7}}-\frac{24704 c_{1}^{6} {c_6}}{{{\pi }^6}}-  {}
{}
 \frac{10888 c_{1}^{6} {c_6}}{3 {{\pi }^5}}-\frac{520 c_{1}^{6} {c_6}}{3 {{\pi }^4}}-\frac{4621
c_{1}^{6} {c_6}}{90 {{\pi }^3}}-\frac{14 c_{1}^{6} {c_6}}{3 {{\pi }^2}}-\frac{166944 c_{1}^{4} {c_2}
{c_6}}{{{\pi }^6}}+  {}
{}
 \frac{39328 c_{1}^{4} {c_2} {c_6}}{{{\pi }^5}}+\frac{844 c_{1}^{4} {c_2} {c_6}}{{{\pi }^4}}+\frac{484
c_{1}^{4} {c_2} {c_6}}{3 {{\pi }^3}}+\frac{9 c_{1}^{4} {c_2} {c_6}}{{{\pi }^2}}+  {}
{}
 \frac{30288 c_{1}^{2} c_{2}^{2} {c_6}}{{{\pi }^5}}-\frac{7472 c_{1}^{2} c_{2}^{2} {c_6}}{{{\pi
}^4}}-\frac{90 c_{1}^{2} c_{2}^{2} {c_6}}{{{\pi }^3}}-\frac{2 c_{1}^{2} c_{2}^{2} {c_6}}{{{\pi }^2}}-\frac{328
c_{2}^{3} {c_6}}{{{\pi }^4}}+  {}
{}
 \frac{40 c_{2}^{3} {c_6}}{{{\pi }^3}}+\frac{3 c_{2}^{3} {c_6}}{{{\pi }^2}}+\frac{18160 c_{1}^{3}
{c_3} {c_6}}{{{\pi }^5}}-\frac{4128 c_{1}^{3} {c_3} {c_6}}{{{\pi }^4}}-\frac{278 c_{1}^{3} {c_3}
{c_6}}{3 {{\pi }^3}}-  {}
{}
 \frac{8 c_{1}^{3} {c_3} {c_6}}{{{\pi }^2}}-\frac{3672 {c_1} {c_2} {c_3} {c_6}}{{{\pi
}^4}}+\frac{848 {c_1} {c_2} {c_3} {c_6}}{{{\pi }^3}}+\frac{40 c_{3}^{2} {c_6}}{{{\pi }^3}}-\frac{4
c_{3}^{2} {c_6}}{{{\pi }^2}}-  {}
{}
 \frac{1552 c_{1}^{2} {c_4} {c_6}}{{{\pi }^4}}+\frac{304 c_{1}^{2} {c_4} {c_6}}{{{\pi }^3}}+\frac{8
c_{1}^{2} {c_4} {c_6}}{{{\pi }^2}}+\frac{108 {c_2} {c_4} {c_6}}{{{\pi }^3}}-\frac{16 {c_2} {c_4}
{c_6}}{{{\pi }^2}}+  {}
{}
 \frac{108 {c_1} {c_5} {c_6}}{{{\pi }^3}}-\frac{16 {c_1} {c_5} {c_6}}{{{\pi }^2}}-\frac{2
c_{6}^{2}}{{{\pi }^2}}-\frac{20416 c_{1}^{5} {c_7}}{{{\pi }^6}}+\frac{2976 c_{1}^{5} {c_7}}{{{\pi }^5}}+  {}
{}
 \frac{368 c_{1}^{5} {c_7}}{{{\pi }^4}}+\frac{92 c_{1}^{5} {c_7}}{3 {{\pi }^3}}+\frac{91 c_{1}^{5}
{c_7}}{20 {{\pi }^2}}+\frac{16400 c_{1}^{3} {c_2} {c_7}}{{{\pi }^5}}-\frac{3552 c_{1}^{3} {c_2}
{c_7}}{{{\pi }^4}}-  {}
{}
 \frac{242 c_{1}^{3} {c_2} {c_7}}{3 {{\pi }^3}}-\frac{12 c_{1}^{3} {c_2} {c_7}}{{{\pi
}^2}}-\frac{1576 {c_1} c_{2}^{2} {c_7}}{{{\pi }^4}}+\frac{304 {c_1} c_{2}^{2} {c_7}}{{{\pi }^3}}+  {}
{}
 \frac{8 {c_1} c_{2}^{2} {c_7}}{{{\pi }^2}}-\frac{1576 c_{1}^{2} {c_3} {c_7}}{{{\pi }^4}}+\frac{304
c_{1}^{2} {c_3} {c_7}}{{{\pi }^3}}+\frac{8 c_{1}^{2} {c_3} {c_7}}{{{\pi }^2}}+\frac{116 {c_2} {c_3}
{c_7}}{{{\pi }^3}}-  {}
{}
 \frac{16 {c_2} {c_3} {c_7}}{{{\pi }^2}}+\frac{116 {c_1} {c_4} {c_7}}{{{\pi }^3}}-\frac{16
{c_1} {c_4} {c_7}}{{{\pi }^2}}-\frac{6 {c_5} {c_7}}{{{\pi }^2}}+\frac{2608 c_{1}^{4} {c_8}}{{{\pi }^5}}-
 {}
{}
 \frac{352 c_{1}^{4} {c_8}}{{{\pi }^4}}-\frac{106 c_{1}^{4} {c_8}}{3 {{\pi }^3}}-\frac{4 c_{1}^{4}
{c_8}}{{{\pi }^2}}-\frac{1488 c_{1}^{2} {c_2} {c_8}}{{{\pi }^4}}+\frac{280 c_{1}^{2} {c_2} {c_8}}{{{\pi
}^3}}+  {}
{}
 \frac{8 c_{1}^{2} {c_2} {c_8}}{{{\pi }^2}}+\frac{44 c_{2}^{2} {c_8}}{{{\pi }^3}}-\frac{4
c_{2}^{2} {c_8}}{{{\pi }^2}}+\frac{116 {c_1} {c_3} {c_8}}{{{\pi }^3}}-\frac{16 {c_1} {c_3} {c_8}}{{{\pi
}^2}}-  {}
{}
 \frac{6 {c_4} {c_8}}{{{\pi }^2}}-\frac{336 c_{1}^{3} {c_9}}{{{\pi }^4}}+\frac{40 c_{1}^{3}
{c_9}}{{{\pi }^3}}+\frac{3 c_{1}^{3} {c_9}}{{{\pi }^2}}+\frac{116 {c_1} {c_2} {c_9}}{{{\pi }^3}}-\frac{16
{c_1} {c_2} {c_9}}{{{\pi }^2}}-  {}
{}
 \frac{6 {c_3} {c_9}}{{{\pi }^2}}+\frac{44 c_{1}^{2} {c_{10}}}{{{\pi }^3}}-\frac{4 c_{1}^{2}
{c_{10}}}{{{\pi }^2}}-\frac{6 {c_2} {c_{10}}}{{{\pi }^2}}-\frac{6 {c_1} {c_{11}}}{{{\pi }^2}}+\frac{{c_{12}}}{\pi }+
 {}
{}
\frac{852017152 c_{1}^{13}}{{{\pi }^{13}}}-\frac{163391488 c_{1}^{13}}{3 {{\pi }^{11}}}-\frac{4823104
c_{1}^{13}}{3 {{\pi }^9}}-\frac{1188176 c_{1}^{13}}{45 {{\pi }^7}}-  {}
{}
 \frac{1103119 c_{1}^{13}}{1620 {{\pi }^5}}-\frac{23990941 c_{1}^{13}}{554400 {{\pi }^3}}+\frac{231
c_{1}^{13}}{13312 \pi }-\frac{2769055744 c_{1}^{11} {c_2}}{{{\pi }^{12}}}+  {}
{}
 \frac{343982080 c_{1}^{11} {c_2}}{{{\pi }^{11}}}+\frac{122543616 c_{1}^{11} {c_2}}{{{\pi }^{10}}}-\frac{2342912
c_{1}^{11} {c_2}}{3 {{\pi }^9}}+  {}
{}
 \frac{2802800 c_{1}^{11} {c_2}}{{{\pi }^8}}-\frac{198880 c_{1}^{11} {c_2}}{{{\pi }^7}}+\frac{366476
c_{1}^{11} {c_2}}{9 {{\pi }^6}}-\frac{670544 c_{1}^{11} {c_2}}{135 {{\pi }^5}}+  {}
{}
 \frac{34045 c_{1}^{11} {c_2}}{48 {{\pi }^4}}+\frac{37963 c_{1}^{11} {c_2}}{504 {{\pi }^3}}+\frac{9029
c_{1}^{11} {c_2}}{1408 {{\pi }^2}}+\frac{3138836480 c_{1}^{9} c_{2}^{2}}{{{\pi }^{11}}}-  {}
{}
 \frac{696563712 c_{1}^{9} c_{2}^{2}}{{{\pi }^{10}}}-\frac{205936640 c_{1}^{9} c_{2}^{2}}{3 {{\pi
}^9}}-\frac{1665664 c_{1}^{9} c_{2}^{2}}{{{\pi }^8}}-\frac{424424 c_{1}^{9} c_{2}^{2}}{{{\pi }^7}}+  {}
{}
 \frac{54296 c_{1}^{9} c_{2}^{2}}{{{\pi }^6}}+\frac{1923184 c_{1}^{9} c_{2}^{2}}{135 {{\pi }^5}}-\frac{9769
c_{1}^{9} c_{2}^{2}}{15 {{\pi }^4}}+\frac{306697 c_{1}^{9} c_{2}^{2}}{2016 {{\pi }^3}}-  {}
{}
 \frac{35 c_{1}^{9} c_{2}^{2}}{2 {{\pi }^2}}-\frac{1496322048 c_{1}^{7} c_{2}^{3}}{{{\pi }^{10}}}+\frac{427528192
c_{1}^{7} c_{2}^{3}}{{{\pi }^9}}+\frac{27589952 c_{1}^{7} c_{2}^{3}}{3 {{\pi }^8}}+  {}
{}
 \frac{1314368 c_{1}^{7} c_{2}^{3}}{3 {{\pi }^7}}-\frac{683672 c_{1}^{7} c_{2}^{3}}{3 {{\pi
}^6}}-\frac{1180936 c_{1}^{7} c_{2}^{3}}{45 {{\pi }^5}}-\frac{19031 c_{1}^{7} c_{2}^{3}}{10 {{\pi }^4}}-
 {}
{}
 \frac{15923 c_{1}^{7} c_{2}^{3}}{30 {{\pi }^3}}+\frac{2627 c_{1}^{7} c_{2}^{3}}{112 {{\pi
}^2}}+\frac{286540800 c_{1}^{5} c_{2}^{4}}{{{\pi }^9}}-\frac{87825920 c_{1}^{5} c_{2}^{4}}{{{\pi }^8}}-  {}
{}
 \frac{965600 c_{1}^{5} c_{2}^{4}}{{{\pi }^7}}+\frac{1117600 c_{1}^{5} c_{2}^{4}}{3 {{\pi }^6}}+\frac{231314
c_{1}^{5} c_{2}^{4}}{15 {{\pi }^5}}+\frac{13970 c_{1}^{5} c_{2}^{4}}{3 {{\pi }^4}}+  {}
{}
 \frac{6107 c_{1}^{5} c_{2}^{4}}{15 {{\pi }^3}}-\frac{18 c_{1}^{5} c_{2}^{4}}{{{\pi }^2}}-\frac{17308928
c_{1}^{3} c_{2}^{5}}{{{\pi }^8}}+\frac{4711680 c_{1}^{3} c_{2}^{5}}{{{\pi }^7}}+  {}
{}
 \frac{635296 c_{1}^{3} c_{2}^{5}}{3 {{\pi }^6}}-\frac{73040 c_{1}^{3} c_{2}^{5}}{3 {{\pi
}^5}}-\frac{4459 c_{1}^{3} c_{2}^{5}}{3 {{\pi }^4}}-\frac{148 c_{1}^{3} c_{2}^{5}}{{{\pi }^3}}+\frac{417
c_{1}^{3} c_{2}^{5}}{40 {{\pi }^2}}+  {}
{}
 \frac{160640 {c_1} c_{2}^{6}}{{{\pi }^7}}-\frac{24704 {c_1} c_{2}^{6}}{{{\pi }^6}}-\frac{10888 {c_1}
c_{2}^{6}}{3 {{\pi }^5}}-\frac{520 {c_1} c_{2}^{6}}{3 {{\pi }^4}}-\frac{4621 {c_1} c_{2}^{6}}{90
{{\pi }^3}}-  {}
{}
 \frac{14 {c_1} c_{2}^{6}}{3 {{\pi }^2}}+\frac{490174464 c_{1}^{10} {c_3}}{{{\pi }^{11}}}-\frac{77395968
c_{1}^{10} {c_3}}{{{\pi }^{10}}}-\frac{54925312 c_{1}^{10} {c_3}}{3 {{\pi }^9}}+  {}
{}
 \frac{128128 c_{1}^{10} {c_3}}{{{\pi }^8}}-\frac{1710896 c_{1}^{10} {c_3}}{5 {{\pi }^7}}+\frac{25784
c_{1}^{10} {c_3}}{{{\pi }^6}}-\frac{348248 c_{1}^{10} {c_3}}{135 {{\pi }^5}}+  {}
{}
 \frac{1443 c_{1}^{10} {c_3}}{5 {{\pi }^4}}-\frac{568523 c_{1}^{10} {c_3}}{25200 {{\pi }^3}}-\frac{83
c_{1}^{10} {c_3}}{10 {{\pi }^2}}-\frac{967449600 c_{1}^{8} {c_2} {c_3}}{{{\pi }^{10}}}+  {}
{}
 \frac{243716096 c_{1}^{8} {c_2} {c_3}}{{{\pi }^9}}+\frac{13732992 c_{1}^{8} {c_2} {c_3}}{{{\pi
}^8}}+\frac{934208 c_{1}^{8} {c_2} {c_3}}{3 {{\pi }^7}}+  {}
{}
 \frac{102256 c_{1}^{8} {c_2} {c_3}}{3 {{\pi }^6}}-\frac{339392 c_{1}^{8} {c_2} {c_3}}{15
{{\pi }^5}}-\frac{83239 c_{1}^{8} {c_2} {c_3}}{45 {{\pi }^4}}-\frac{48011 c_{1}^{8} {c_2} {c_3}}{210
{{\pi }^3}}+  {}
{}
 \frac{595 c_{1}^{8} {c_2} {c_3}}{32 {{\pi }^2}}+\frac{575131648 c_{1}^{6} c_{2}^{2}
{c_3}}{{{\pi }^9}}-\frac{180124672 c_{1}^{6} c_{2}^{2} {c_3}}{{{\pi }^8}}+  {}
{}
 \frac{4334656 c_{1}^{6} c_{2}^{2} {c_3}}{3 {{\pi }^7}}-\frac{123200 c_{1}^{6} c_{2}^{2}
{c_3}}{{{\pi }^6}}+\frac{1485484 c_{1}^{6} c_{2}^{2} {c_3}}{15 {{\pi }^5}}+  {}
{}
 \frac{246238 c_{1}^{6} c_{2}^{2} {c_3}}{45 {{\pi }^4}}+\frac{3941 c_{1}^{6} c_{2}^{2}
{c_3}}{4 {{\pi }^3}}-\frac{93 c_{1}^{6} c_{2}^{2} {c_3}}{4 {{\pi }^2}}-\frac{112461440 c_{1}^{4}
c_{2}^{3} {c_3}}{{{\pi }^8}}+  {}
{}
 \frac{37470080 c_{1}^{4} c_{2}^{3} {c_3}}{{{\pi }^7}}-\frac{2488576 c_{1}^{4} c_{2}^{3} {c_3}}{3
{{\pi }^6}}-\frac{82672 c_{1}^{4} c_{2}^{3} {c_3}}{{{\pi }^5}}-\frac{27527 c_{1}^{4} c_{2}^{3} {c_3}}{3
{{\pi }^4}}-  {}
{}
 \frac{2528 c_{1}^{4} c_{2}^{3} {c_3}}{3 {{\pi }^3}}+\frac{69 c_{1}^{4} c_{2}^{3} {c_3}}{4
{{\pi }^2}}+\frac{5436672 c_{1}^{2} c_{2}^{4} {c_3}}{{{\pi }^7}}-\frac{1615168 c_{1}^{2} c_{2}^{4} {c_3}}{{{\pi
}^6}}-  {}
{}
 \frac{1376 c_{1}^{2} c_{2}^{4} {c_3}}{{{\pi }^5}}+\frac{10768 c_{1}^{2} c_{2}^{4} {c_3}}{3
{{\pi }^4}}+\frac{1817 c_{1}^{2} c_{2}^{4} {c_3}}{6 {{\pi }^3}}-\frac{10 c_{1}^{2} c_{2}^{4} {c_3}}{{{\pi
}^2}}-  {}
{}
 \frac{20416 c_{2}^{5} {c_3}}{{{\pi }^6}}+\frac{2976 c_{2}^{5} {c_3}}{{{\pi }^5}}+\frac{368 c_{2}^{5}
{c_3}}{{{\pi }^4}}+\frac{92 c_{2}^{5} {c_3}}{3 {{\pi }^3}}+\frac{91 c_{2}^{5} {c_3}}{20 {{\pi }^2}}+  {}
{}
 \frac{64989184 c_{1}^{7} c_{3}^{2}}{{{\pi }^9}}-\frac{16796416 c_{1}^{7} c_{3}^{2}}{{{\pi }^8}}-\frac{807104
c_{1}^{7} c_{3}^{2}}{{{\pi }^7}}-\frac{28160 c_{1}^{7} c_{3}^{2}}{3 {{\pi }^6}}+  {}
{}
 \frac{45056 c_{1}^{7} c_{3}^{2}}{15 {{\pi }^5}}+\frac{36176 c_{1}^{7} c_{3}^{2}}{45 {{\pi
}^4}}+\frac{26281 c_{1}^{7} c_{3}^{2}}{210 {{\pi }^3}}-\frac{5 c_{1}^{7} c_{3}^{2}}{4 {{\pi }^2}}-\frac{61711104
c_{1}^{5} {c_2} c_{3}^{2}}{{{\pi }^8}}+  {}
{}
 \frac{19369984 c_{1}^{5} {c_2} c_{3}^{2}}{{{\pi }^7}}-\frac{174336 c_{1}^{5} {c_2} c_{3}^{2}}{{{\pi
}^6}}-\frac{64864 c_{1}^{5} {c_2} c_{3}^{2}}{3 {{\pi }^5}}-\frac{85661 c_{1}^{5} {c_2} c_{3}^{2}}{15
{{\pi }^4}}-  {}
{} 
 \frac{10301 c_{1}^{5} {c_2} c_{3}^{2}}{15 {{\pi }^3}}+\frac{17 c_{1}^{5} {c_2} c_{3}^{2}}{2
{{\pi }^2}}+\frac{12873984 c_{1}^{3} c_{2}^{2} c_{3}^{2}}{{{\pi }^7}}  $

$-\frac{4241216 c_{1}^{3} c_{2}^{2}
c_{3}^{2}}{{{\pi }^6}}+  {}
{}
 \frac{334672 c_{1}^{3} c_{2}^{2} c_{3}^{2}}{3 {{\pi }^5}}+\frac{24176 c_{1}^{3} c_{2}^{2}
c_{3}^{2}}{3 {{\pi }^4}}+\frac{1915 c_{1}^{3} c_{2}^{2} c_{3}^{2}}{3 {{\pi }^3}}-\frac{9 c_{1}^{3}
c_{2}^{2} c_{3}^{2}}{{{\pi }^2}}-  {}
{}
 \frac{438400 {c_1} c_{2}^{3} c_{3}^{2}}{{{\pi }^6}}+\frac{123840 {c_1} c_{2}^{3} c_{3}^{2}}{{{\pi
}^5}}-\frac{920 {c_1} c_{2}^{3} c_{3}^{2}}{3 {{\pi }^4}}-\frac{132 {c_1} c_{2}^{3} c_{3}^{2}}{{{\pi
}^3}}+  {}
{}
 \frac{7 {c_1} c_{2}^{3} c_{3}^{2}}{{{\pi }^2}}+\frac{1691584 c_{1}^{4} c_{3}^{3}}{{{\pi }^7}}-\frac{465280
c_{1}^{4} c_{3}^{3}}{{{\pi }^6}}-\frac{37112 c_{1}^{4} c_{3}^{3}}{3 {{\pi }^5}}+\frac{1632 c_{1}^{4} c_{3}^{3}}{{{\pi
}^4}}+  {}
{}
 \frac{571 c_{1}^{4} c_{3}^{3}}{9 {{\pi }^3}}-\frac{10 c_{1}^{4} c_{3}^{3}}{3 {{\pi }^2}}-\frac{439008
c_{1}^{2} {c_2} c_{3}^{3}}{{{\pi }^6}}+\frac{121728 c_{1}^{2} {c_2} c_{3}^{3}}{{{\pi }^5}}+  {}
{}
 \frac{728 c_{1}^{2} {c_2} c_{3}^{3}}{{{\pi }^4}}-\frac{216 c_{1}^{2} {c_2} c_{3}^{3}}{{{\pi
}^3}}+\frac{6 c_{1}^{2} {c_2} c_{3}^{3}}{{{\pi }^2}}+\frac{7472 c_{2}^{2} c_{3}^{3}}{{{\pi }^5}}-\frac{1440
c_{2}^{2} c_{3}^{3}}{{{\pi }^4}}-  {}
{}
 \frac{278 c_{2}^{2} c_{3}^{3}}{3 {{\pi }^3}}-\frac{2 c_{2}^{2} c_{3}^{3}}{{{\pi }^2}}+\frac{2608
{c_1} c_{3}^{4}}{{{\pi }^5}}-\frac{352 {c_1} c_{3}^{4}}{{{\pi }^4}}-\frac{106 {c_1} c_{3}^{4}}{3 {{\pi
}^3}}-\frac{4 {c_1} c_{3}^{4}}{{{\pi }^2}}-  {}
{}
 \frac{73096192 c_{1}^{9} {c_4}}{{{\pi }^{10}}}+\frac{12193792 c_{1}^{9} {c_4}}{{{\pi }^9}}+\frac{2492672
c_{1}^{9} {c_4}}{{{\pi }^8}}-\frac{71104 c_{1}^{9} {c_4}}{3 {{\pi }^7}}+  {}
{}
 \frac{44396 c_{1}^{9} {c_4}}{{{\pi }^6}}-\frac{18408 c_{1}^{9} {c_4}}{5 {{\pi }^5}}+\frac{2498
c_{1}^{9} {c_4}}{5 {{\pi }^4}}-\frac{4633 c_{1}^{9} {c_4}}{126 {{\pi }^3}}+\frac{563 c_{1}^{9} {c_4}}{64
{{\pi }^2}}+  {}
{}
 \frac{126756864 c_{1}^{7} {c_2} {c_4}}{{{\pi }^9}}-\frac{33080320 c_{1}^{7} {c_2} {c_4}}{{{\pi
}^8}}-\frac{1262560 c_{1}^{7} {c_2} {c_4}}{{{\pi }^7}}-  {}
{}
 \frac{193600 c_{1}^{7} {c_2} {c_4}}{3 {{\pi }^6}}+\frac{102128 c_{1}^{7} {c_2} {c_4}}{15
{{\pi }^5}}+\frac{22816 c_{1}^{7} {c_2} {c_4}}{45 {{\pi }^4}}+\frac{127699 c_{1}^{7} {c_2} {c_4}}{420
{{\pi }^3}}-  {}
{}
 \frac{45 c_{1}^{7} {c_2} {c_4}}{2 {{\pi }^2}}-\frac{61326720 c_{1}^{5} c_{2}^{2} {c_4}}{{{\pi
}^8}}+\frac{19377024 c_{1}^{5} c_{2}^{2} {c_4}}{{{\pi }^7}}-  {}
{}
 \frac{290320 c_{1}^{5} c_{2}^{2} {c_4}}{{{\pi }^6}}+\frac{2720 c_{1}^{5} c_{2}^{2} {c_4}}{{{\pi
}^5}}-\frac{100304 c_{1}^{5} c_{2}^{2} {c_4}}{15 {{\pi }^4}}-\frac{9562 c_{1}^{5} c_{2}^{2} {c_4}}{15
{{\pi }^3}}+  {}
{}
 \frac{87 c_{1}^{5} c_{2}^{2} {c_4}}{4 {{\pi }^2}}+\frac{8240512 c_{1}^{3} c_{2}^{3}
{c_4}}{{{\pi }^7}}-\frac{2605952 c_{1}^{3} c_{2}^{3} {c_4}}{{{\pi }^6}}+\frac{93848 c_{1}^{3} c_{2}^{3}
{c_4}}{3 {{\pi }^5}}+  {}
{}
 \frac{25312 c_{1}^{3} c_{2}^{3} {c_4}}{3 {{\pi }^4}}+\frac{3017 c_{1}^{3} c_{2}^{3}
{c_4}}{9 {{\pi }^3}}-\frac{53 c_{1}^{3} c_{2}^{3} {c_4}}{3 {{\pi }^2}}-\frac{167104 {c_1} c_{2}^{4}
{c_4}}{{{\pi }^6}}+  {}
{}
 \frac{39328 {c_1} c_{2}^{4} {c_4}}{{{\pi }^5}}+\frac{844 {c_1} c_{2}^{4} {c_4}}{{{\pi }^4}}+\frac{484
{c_1} c_{2}^{4} {c_4}}{3 {{\pi }^3}}+\frac{9 {c_1} c_{2}^{4} {c_4}}{{{\pi }^2}}-  {}
{}
 \frac{16027648 c_{1}^{6} {c_3} {c_4}}{{{\pi }^8}}+\frac{4233984 c_{1}^{6} {c_3} {c_4}}{{{\pi
}^7}}+\frac{432992 c_{1}^{6} {c_3} {c_4}}{3 {{\pi }^6}}-\frac{384 c_{1}^{6} {c_3} {c_4}}{{{\pi }^5}}-
 {}
{}
 \frac{1616 c_{1}^{6} {c_3} {c_4}}{45 {{\pi }^4}}-\frac{2662 c_{1}^{6} {c_3} {c_4}}{15
{{\pi }^3}}+\frac{15 c_{1}^{6} {c_3} {c_4}}{4 {{\pi }^2}}+\frac{12181952 c_{1}^{4} {c_2} {c_3}
{c_4}}{{{\pi }^7}}-  {}
{}
 \frac{3890624 c_{1}^{4} {c_2} {c_3} {c_4}}{{{\pi }^6}}+\frac{90704 c_{1}^{4} {c_2}
{c_3} {c_4}}{{{\pi }^5}}+\frac{1688 c_{1}^{4} {c_2} {c_3} {c_4}}{3 {{\pi }^4}}+  {}
{}
 \frac{1056 c_{1}^{4} {c_2} {c_3} {c_4}}{{{\pi }^3}}-\frac{9 c_{1}^{4} {c_2} {c_3}
{c_4}}{{{\pi }^2}}-\frac{1641920 c_{1}^{2} c_{2}^{2} {c_3} {c_4}}{{{\pi }^6}}+  {}
{}
 \frac{527008 c_{1}^{2} c_{2}^{2} {c_3} {c_4}}{{{\pi }^5}}-\frac{15996 c_{1}^{2} c_{2}^{2}
{c_3} {c_4}}{{{\pi }^4}}-\frac{720 c_{1}^{2} c_{2}^{2} {c_3} {c_4}}{{{\pi }^3}}+  {}
{}
 \frac{11 c_{1}^{2} c_{2}^{2} {c_3} {c_4}}{{{\pi }^2}}+\frac{16400 c_{2}^{3} {c_3}
{c_4}}{{{\pi }^5}}-\frac{3552 c_{2}^{3} {c_3} {c_4}}{{{\pi }^4}}-\frac{242 c_{2}^{3} {c_3} {c_4}}{3
{{\pi }^3}}-  {}
{}
 \frac{12 c_{2}^{3} {c_3} {c_4}}{{{\pi }^2}}-\frac{447808 c_{1}^{3} c_{3}^{2} {c_4}}{{{\pi
}^6}}+\frac{126528 c_{1}^{3} c_{3}^{2} {c_4}}{{{\pi }^5}}-\frac{344 c_{1}^{3} c_{3}^{2} {c_4}}{3
{{\pi }^4}}-  {}
{}
 \frac{164 c_{1}^{3} c_{3}^{2} {c_4}}{{{\pi }^3}}+\frac{6 c_{1}^{3} c_{3}^{2} {c_4}}{{{\pi
}^2}}+\frac{67328 {c_1} {c_2} c_{3}^{2} {c_4}}{{{\pi }^5}}-\frac{18576 {c_1} {c_2} c_{3}^{2}
{c_4}}{{{\pi }^4}}+  {}
{}
 \frac{292 {c_1} {c_2} c_{3}^{2} {c_4}}{{{\pi }^3}}-\frac{8 {c_1} {c_2} c_{3}^{2}
{c_4}}{{{\pi }^2}}-\frac{336 c_{3}^{3} {c_4}}{{{\pi }^4}}+\frac{40 c_{3}^{3} {c_4}}{{{\pi }^3}}+\frac{3 c_{3}^{3}
{c_4}}{{{\pi }^2}}+  {}
{}
 \frac{827136 c_{1}^{5} c_{4}^{2}}{{{\pi }^7}}-\frac{198848 c_{1}^{5} c_{4}^{2}}{{{\pi }^6}}-\frac{10112
c_{1}^{5} c_{4}^{2}}{{{\pi }^5}}-\frac{40 c_{1}^{5} c_{4}^{2}}{{{\pi }^4}}-\frac{107 c_{1}^{5} c_{4}^{2}}{10
{{\pi }^3}}-  {}
{}
 \frac{3 c_{1}^{5} c_{4}^{2}}{2 {{\pi }^2}}-\frac{469280 c_{1}^{3} {c_2} c_{4}^{2}}{{{\pi
}^6}}+\frac{134944 c_{1}^{3} {c_2} c_{4}^{2}}{{{\pi }^5}}-\frac{1592 c_{1}^{3} {c_2} c_{4}^{2}}{3
{{\pi }^4}}-  {}
{}
 \frac{584 c_{1}^{3} {c_2} c_{4}^{2}}{3 {{\pi }^3}}+\frac{5 c_{1}^{3} {c_2} c_{4}^{2}}{{{\pi
}^2}}+\frac{30384 {c_1} c_{2}^{2} c_{4}^{2}}{{{\pi }^5}}-\frac{7472 {c_1} c_{2}^{2} c_{4}^{2}}{{{\pi }^4}}-\frac{90
{c_1} c_{2}^{2} c_{4}^{2}}{{{\pi }^3}}-  {}
{}
 \frac{2 {c_1} c_{2}^{2} c_{4}^{2}}{{{\pi }^2}}+\frac{28736 c_{1}^{2} {c_3} c_{4}^{2}}{{{\pi
}^5}}-\frac{7040 c_{1}^{2} {c_3} c_{4}^{2}}{{{\pi }^4}}-\frac{60 c_{1}^{2} {c_3} c_{4}^{2}}{{{\pi }^3}}-\frac{4
c_{1}^{2} {c_3} c_{4}^{2}}{{{\pi }^2}}-  {}
{}
 \frac{1576 {c_2} {c_3} c_{4}^{2}}{{{\pi }^4}}+\frac{304 {c_2} {c_3} c_{4}^{2}}{{{\pi }^3}}+\frac{8
{c_2} {c_3} c_{4}^{2}}{{{\pi }^2}}-\frac{336 {c_1} c_{4}^{3}}{{{\pi }^4}}+\frac{40 {c_1} c_{4}^{3}}{{{\pi
}^3}}+  {}
{}
 \frac{3 {c_1} c_{4}^{3}}{{{\pi }^2}}+\frac{9857536 c_{1}^{8} {c_5}}{{{\pi }^9}}-\frac{1630720 c_{1}^{8}
{c_5}}{{{\pi }^8}}-\frac{307872 c_{1}^{8} {c_5}}{{{\pi }^7}}+\frac{3520 c_{1}^{8} {c_5}}{3 {{\pi }^6}}-  {}
{}
 \frac{76324 c_{1}^{8} {c_5}}{15 {{\pi }^5}}+\frac{10114 c_{1}^{8} {c_5}}{45 {{\pi }^4}}-\frac{7381
c_{1}^{8} {c_5}}{140 {{\pi }^3}}-\frac{6 c_{1}^{8} {c_5}}{{{\pi }^2}}-\frac{14874496 c_{1}^{6} {c_2}
{c_5}}{{{\pi }^8}}+  {}
{}
 \frac{3849600 c_{1}^{6} {c_2} {c_5}}{{{\pi }^7}}+\frac{335312 c_{1}^{6} {c_2} {c_5}}{3
{{\pi }^6}}+\frac{10224 c_{1}^{6} {c_2} {c_5}}{{{\pi }^5}}-\frac{10493 c_{1}^{6} {c_2} {c_5}}{15
{{\pi }^4}}+  {}
{}
 \frac{613 c_{1}^{6} {c_2} {c_5}}{15 {{\pi }^3}}+\frac{25 c_{1}^{6} {c_2} {c_5}}{2
{{\pi }^2}}+\frac{5628800 c_{1}^{4} c_{2}^{2} {c_5}}{{{\pi }^7}}-\frac{1713344 c_{1}^{4} c_{2}^{2} {c_5}}{{{\pi
}^6}}+  {}
{}
 \frac{20960 c_{1}^{4} c_{2}^{2} {c_5}}{{{\pi }^5}}+\frac{2248 c_{1}^{4} c_{2}^{2} {c_5}}{3
{{\pi }^4}}+\frac{1265 c_{1}^{4} c_{2}^{2} {c_5}}{6 {{\pi }^3}}-\frac{6 c_{1}^{4} c_{2}^{2} {c_5}}{{{\pi
}^2}}-  {}
{}
 \frac{468576 c_{1}^{2} c_{2}^{3} {c_5}}{{{\pi }^6}}+\frac{134016 c_{1}^{2} c_{2}^{3} {c_5}}{{{\pi
}^5}}-\frac{276 c_{1}^{2} c_{2}^{3} {c_5}}{{{\pi }^4}}-\frac{208 c_{1}^{2} c_{2}^{3} {c_5}}{{{\pi }^3}}+
 {}
{}
 \frac{5 c_{1}^{2} c_{2}^{3} {c_5}}{{{\pi }^2}}+\frac{2608 c_{2}^{4} {c_5}}{{{\pi }^5}}-\frac{352
c_{2}^{4} {c_5}}{{{\pi }^4}}-\frac{106 c_{2}^{4} {c_5}}{3 {{\pi }^3}}-\frac{4 c_{2}^{4} {c_5}}{{{\pi }^2}}+
 {}
{}
 \frac{1828160 c_{1}^{5} {c_3} {c_5}}{{{\pi }^7}}-\frac{482176 c_{1}^{5} {c_3} {c_5}}{{{\pi
}^6}}-\frac{9392 c_{1}^{5} {c_3} {c_5}}{{{\pi }^5}}-\frac{2272 c_{1}^{5} {c_3} {c_5}}{3 {{\pi }^4}}+
 {}
{}
 \frac{1751 c_{1}^{5} {c_3} {c_5}}{30 {{\pi }^3}}-\frac{6 c_{1}^{5} {c_3} {c_5}}{{{\pi
}^2}}-\frac{1010720 c_{1}^{3} {c_2} {c_3} {c_5}}{{{\pi }^6}}+\frac{306560 c_{1}^{3} {c_2} {c_3}
{c_5}}{{{\pi }^5}}-  {}
{}
 \frac{17276 c_{1}^{3} {c_2} {c_3} {c_5}}{3 {{\pi }^4}}-\frac{664 c_{1}^{3} {c_2}
{c_3} {c_5}}{3 {{\pi }^3}}+\frac{67856 {c_1} c_{2}^{2} {c_3} {c_5}}{{{\pi }^5}}-  {}
{}
 \frac{18672 {c_1} c_{2}^{2} {c_3} {c_5}}{{{\pi }^4}}+\frac{280 {c_1} c_{2}^{2} {c_3}
{c_5}}{{{\pi }^3}}-\frac{8 {c_1} c_{2}^{2} {c_3} {c_5}}{{{\pi }^2}}+\frac{31680 c_{1}^{2} c_{3}^{2}
{c_5}}{{{\pi }^5}}-  {}
{}
 \frac{8128 c_{1}^{2} c_{3}^{2} {c_5}}{{{\pi }^4}}-\frac{36 c_{1}^{2} c_{3}^{2} {c_5}}{{{\pi
}^3}}-\frac{1432 {c_2} c_{3}^{2} {c_5}}{{{\pi }^4}}+\frac{272 {c_2} c_{3}^{2} {c_5}}{{{\pi }^3}}+  {}
{}
 \frac{6 {c_2} c_{3}^{2} {c_5}}{{{\pi }^2}}-\frac{176960 c_{1}^{4} {c_4} {c_5}}{{{\pi }^6}}+\frac{42336
c_{1}^{4} {c_4} {c_5}}{{{\pi }^5}}+\frac{3908 c_{1}^{4} {c_4} {c_5}}{3 {{\pi }^4}}+  {}
{}
 \frac{208 c_{1}^{4} {c_4} {c_5}}{3 {{\pi }^3}}+\frac{9 c_{1}^{4} {c_4} {c_5}}{2
{{\pi }^2}}+\frac{69392 c_{1}^{2} {c_2} {c_4} {c_5}}{{{\pi }^5}}-\frac{19216 c_{1}^{2} {c_2} {c_4}
{c_5}}{{{\pi }^4}}+  {}
{}
 \frac{280 c_{1}^{2} {c_2} {c_4} {c_5}}{{{\pi }^3}}-\frac{4 c_{1}^{2} {c_2} {c_4}
{c_5}}{{{\pi }^2}}-\frac{1488 c_{2}^{2} {c_4} {c_5}}{{{\pi }^4}}+\frac{280 c_{2}^{2} {c_4} {c_5}}{{{\pi
}^3}}+  {}
{}
 \frac{8 c_{2}^{2} {c_4} {c_5}}{{{\pi }^2}}-\frac{3544 {c_1} {c_3} {c_4} {c_5}}{{{\pi
}^4}}+\frac{800 {c_1} {c_3} {c_4} {c_5}}{{{\pi }^3}}+\frac{44 c_{4}^{2} {c_5}}{{{\pi }^3}}-\frac{4
c_{4}^{2} {c_5}}{{{\pi }^2}}+  {}
{}
 \frac{7472 c_{1}^{3} c_{5}^{2}}{{{\pi }^5}}-\frac{1440 c_{1}^{3} c_{5}^{2}}{{{\pi }^4}}-\frac{278
c_{1}^{3} c_{5}^{2}}{3 {{\pi }^3}}-\frac{2 c_{1}^{3} c_{5}^{2}}{{{\pi }^2}}-\frac{1576 {c_1} {c_2}
c_{5}^{2}}{{{\pi }^4}}+  {}
{}
 \frac{304 {c_1} {c_2} c_{5}^{2}}{{{\pi }^3}}+\frac{8 {c_1} {c_2} c_{5}^{2}}{{{\pi }^2}}+\frac{44
{c_3} c_{5}^{2}}{{{\pi }^3}}-\frac{4 {c_3} c_{5}^{2}}{{{\pi }^2}}-\frac{1266304 c_{1}^{7} {c_6}}{{{\pi }^8}}+
 {}
{}
 \frac{202880 c_{1}^{7} {c_6}}{{{\pi }^7}}+\frac{103424 c_{1}^{7} {c_6}}{3 {{\pi }^6}}+\frac{592
c_{1}^{7} {c_6}}{{{\pi }^5}}+\frac{2592 c_{1}^{7} {c_6}}{5 {{\pi }^4}}+\frac{37 c_{1}^{7} {c_6}}{3
{{\pi }^3}}+  {}
{}
 \frac{349 c_{1}^{7} {c_6}}{56 {{\pi }^2}}+\frac{1618368 c_{1}^{5} {c_2} {c_6}}{{{\pi }^7}}-\frac{403328
c_{1}^{5} {c_2} {c_6}}{{{\pi }^6}}-\frac{9832 c_{1}^{5} {c_2} {c_6}}{{{\pi }^5}}-  {}
{}
 \frac{1408 c_{1}^{5} {c_2} {c_6}}{{{\pi }^4}}+\frac{211 c_{1}^{5} {c_2} {c_6}}{30
{{\pi }^3}}-\frac{15 c_{1}^{5} {c_2} {c_6}}{{{\pi }^2}}-\frac{449216 c_{1}^{3} c_{2}^{2} {c_6}}{{{\pi }^6}}+
 {}
{}
 \frac{126496 c_{1}^{3} c_{2}^{2} {c_6}}{{{\pi }^5}}-\frac{1304 c_{1}^{3} c_{2}^{2} {c_6}}{3
{{\pi }^4}}-\frac{224 c_{1}^{3} c_{2}^{2} {c_6}}{3 {{\pi }^3}}+\frac{10 c_{1}^{3} c_{2}^{2} {c_6}}{{{\pi
}^2}}+  {}
{}
 \frac{18224 {c_1} c_{2}^{3} {c_6}}{{{\pi }^5}}-\frac{4128 {c_1} c_{2}^{3} {c_6}}{{{\pi }^4}}-\frac{278
{c_1} c_{2}^{3} {c_6}}{3 {{\pi }^3}}-\frac{8 {c_1} c_{2}^{3} {c_6}}{{{\pi }^2}}-  {}
{}
 \frac{191040 c_{1}^{4} {c_3} {c_6}}{{{\pi }^6}}+\frac{47840 c_{1}^{4} {c_3} {c_6}}{{{\pi
}^5}}+\frac{2648 c_{1}^{4} {c_3} {c_6}}{3 {{\pi }^4}}+\frac{72 c_{1}^{4} {c_3} {c_6}}{{{\pi }^3}}+
 {}
{}
 \frac{6 c_{1}^{4} {c_3} {c_6}}{{{\pi }^2}}+\frac{69856 c_{1}^{2} {c_2} {c_3} {c_6}}{{{\pi
}^5}}-\frac{19280 c_{1}^{2} {c_2} {c_3} {c_6}}{{{\pi }^4}}+\frac{264 c_{1}^{2} {c_2} {c_3}
{c_6}}{{{\pi }^3}}-  {}
{}
 \frac{4 c_{1}^{2} {c_2} {c_3} {c_6}}{{{\pi }^2}}-\frac{1576 c_{2}^{2} {c_3} {c_6}}{{{\pi
}^4}}+\frac{304 c_{2}^{2} {c_3} {c_6}}{{{\pi }^3}}+\frac{8 c_{2}^{2} {c_3} {c_6}}{{{\pi }^2}}-  {}
{}
 \frac{1488 {c_1} c_{3}^{2} {c_6}}{{{\pi }^4}}+\frac{280 {c_1} c_{3}^{2} {c_6}}{{{\pi }^3}}+\frac{8
{c_1} c_{3}^{2} {c_6}}{{{\pi }^2}}+\frac{17936 c_{1}^{3} {c_4} {c_6}}{{{\pi }^5}}-  {}
{}
 \frac{4096 c_{1}^{3} {c_4} {c_6}}{{{\pi }^4}}-\frac{242 c_{1}^{3} {c_4} {c_6}}{3 {{\pi
}^3}}-\frac{8 c_{1}^{3} {c_4} {c_6}}{{{\pi }^2}}-\frac{3720 {c_1} {c_2} {c_4} {c_6}}{{{\pi }^4}}+
 {}
{}
 \frac{848 {c_1} {c_2} {c_4} {c_6}}{{{\pi }^3}}+\frac{116 {c_3} {c_4} {c_6}}{{{\pi
}^3}}-\frac{16 {c_3} {c_4} {c_6}}{{{\pi }^2}}-\frac{1432 c_{1}^{2} {c_5} {c_6}}{{{\pi }^4}}+  {}
{}
 \frac{272 c_{1}^{2} {c_5} {c_6}}{{{\pi }^3}}+\frac{6 c_{1}^{2} {c_5} {c_6}}{{{\pi }^2}}+\frac{116
{c_2} {c_5} {c_6}}{{{\pi }^3}}-\frac{16 {c_2} {c_5} {c_6}}{{{\pi }^2}}+\frac{44 {c_1} c_{6}^{2}}{{{\pi
}^3}}-  {}
{}
 \frac{4 {c_1} c_{6}^{2}}{{{\pi }^2}}+\frac{160640 c_{1}^{6} {c_7}}{{{\pi }^7}}-\frac{24704 c_{1}^{6}
{c_7}}{{{\pi }^6}}-\frac{10888 c_{1}^{6} {c_7}}{3 {{\pi }^5}}-\frac{520 c_{1}^{6} {c_7}}{3 {{\pi }^4}}-
 {}
{}
 \frac{4621 c_{1}^{6} {c_7}}{90 {{\pi }^3}}-\frac{14 c_{1}^{6} {c_7}}{3 {{\pi }^2}}-\frac{167104
c_{1}^{4} {c_2} {c_7}}{{{\pi }^6}}+\frac{39328 c_{1}^{4} {c_2} {c_7}}{{{\pi }^5}}+  {}
{}
 \frac{844 c_{1}^{4} {c_2} {c_7}}{{{\pi }^4}}+\frac{484 c_{1}^{4} {c_2} {c_7}}{3 {{\pi
}^3}}+\frac{9 c_{1}^{4} {c_2} {c_7}}{{{\pi }^2}}+\frac{30384 c_{1}^{2} c_{2}^{2} {c_7}}{{{\pi }^5}}-  {}
{}
 \frac{7472 c_{1}^{2} c_{2}^{2} {c_7}}{{{\pi }^4}}-\frac{90 c_{1}^{2} c_{2}^{2} {c_7}}{{{\pi
}^3}}-\frac{2 c_{1}^{2} c_{2}^{2} {c_7}}{{{\pi }^2}}-\frac{336 c_{2}^{3} {c_7}}{{{\pi }^4}}+\frac{40 c_{2}^{3}
{c_7}}{{{\pi }^3}}+  {}
{}
 \frac{3 c_{2}^{3} {c_7}}{{{\pi }^2}}+\frac{18224 c_{1}^{3} {c_3} {c_7}}{{{\pi }^5}}-\frac{4128
c_{1}^{3} {c_3} {c_7}}{{{\pi }^4}}-\frac{278 c_{1}^{3} {c_3} {c_7}}{3 {{\pi }^3}}-  {}
{}
 \frac{8 c_{1}^{3} {c_3} {c_7}}{{{\pi }^2}}-\frac{3720 {c_1} {c_2} {c_3} {c_7}}{{{\pi
}^4}}+\frac{848 {c_1} {c_2} {c_3} {c_7}}{{{\pi }^3}}+\frac{44 c_{3}^{2} {c_7}}{{{\pi }^3}}-\frac{4
c_{3}^{2} {c_7}}{{{\pi }^2}}-  {}
{}
 \frac{1576 c_{1}^{2} {c_4} {c_7}}{{{\pi }^4}}+\frac{304 c_{1}^{2} {c_4} {c_7}}{{{\pi }^3}}+\frac{8
c_{1}^{2} {c_4} {c_7}}{{{\pi }^2}}+\frac{116 {c_2} {c_4} {c_7}}{{{\pi }^3}}-\frac{16 {c_2} {c_4}
{c_7}}{{{\pi }^2}}+  {}
{}
 \frac{116 {c_1} {c_5} {c_7}}{{{\pi }^3}}-\frac{16 {c_1} {c_5} {c_7}}{{{\pi }^2}}-\frac{6
{c_6} {c_7}}{{{\pi }^2}}-\frac{20416 c_{1}^{5} {c_8}}{{{\pi }^6}}+\frac{2976 c_{1}^{5} {c_8}}{{{\pi }^5}}+  {}
{}
 \frac{368 c_{1}^{5} {c_8}}{{{\pi }^4}}+\frac{92 c_{1}^{5} {c_8}}{3 {{\pi }^3}}+\frac{91 c_{1}^{5}
{c_8}}{20 {{\pi }^2}}+\frac{16400 c_{1}^{3} {c_2} {c_8}}{{{\pi }^5}}-\frac{3552 c_{1}^{3} {c_2}
{c_8}}{{{\pi }^4}}-  {}
{}
 \frac{242 c_{1}^{3} {c_2} {c_8}}{3 {{\pi }^3}}-\frac{12 c_{1}^{3} {c_2} {c_8}}{{{\pi
}^2}}-\frac{1576 {c_1} c_{2}^{2} {c_8}}{{{\pi }^4}}+\frac{304 {c_1} c_{2}^{2} {c_8}}{{{\pi }^3}}+  {}
{}
 \frac{8 {c_1} c_{2}^{2} {c_8}}{{{\pi }^2}}-\frac{1576 c_{1}^{2} {c_3} {c_8}}{{{\pi }^4}}+\frac{304
c_{1}^{2} {c_3} {c_8}}{{{\pi }^3}}+\frac{8 c_{1}^{2} {c_3} {c_8}}{{{\pi }^2}}+\frac{116 {c_2} {c_3}
{c_8}}{{{\pi }^3}}-  {}
{}
 \frac{16 {c_2} {c_3} {c_8}}{{{\pi }^2}}+\frac{116 {c_1} {c_4} {c_8}}{{{\pi }^3}}-\frac{16
{c_1} {c_4} {c_8}}{{{\pi }^2}}-\frac{6 {c_5} {c_8}}{{{\pi }^2}}+\frac{2608 c_{1}^{4} {c_9}}{{{\pi }^5}}-
 {}
{}
 \frac{352 c_{1}^{4} {c_9}}{{{\pi }^4}}-\frac{106 c_{1}^{4} {c_9}}{3 {{\pi }^3}}-\frac{4 c_{1}^{4}
{c_9}}{{{\pi }^2}}-\frac{1488 c_{1}^{2} {c_2} {c_9}}{{{\pi }^4}}+\frac{280 c_{1}^{2} {c_2} {c_9}}{{{\pi
}^3}}+  {}
{}
 \frac{8 c_{1}^{2} {c_2} {c_9}}{{{\pi }^2}}+\frac{44 c_{2}^{2} {c_9}}{{{\pi }^3}}-\frac{4
c_{2}^{2} {c_9}}{{{\pi }^2}}+\frac{116 {c_1} {c_3} {c_9}}{{{\pi }^3}}-\frac{16 {c_1} {c_3} {c_9}}{{{\pi
}^2}}-\frac{6 {c_4} {c_9}}{{{\pi }^2}}-  {}
{}
 \frac{336 c_{1}^{3} {c_{10}}}{{{\pi }^4}}+\frac{40 c_{1}^{3} {c_{10}}}{{{\pi }^3}}+\frac{3 c_{1}^{3}
{c_{10}}}{{{\pi }^2}}+\frac{116 {c_1} {c_2} {c_{10}}}{{{\pi }^3}}-\frac{16 {c_1} {c_2} {c_{10}}}{{{\pi
}^2}}-  {}
{}
 \frac{6 {c_3} {c_{10}}}{{{\pi }^2}}+\frac{44 c_{1}^{2} {c_{11}}}{{{\pi }^3}}-\frac{4 c_{1}^{2}
{c_{11}}}{{{\pi }^2}}-\frac{6 {c_2} {c_{11}}}{{{\pi }^2}}-\frac{6 {c_1} {c_{12}}}{{{\pi }^2}}+\frac{{c_{13}}}{\pi }
-\frac{6085836800 c_{1}^{14}}{{{\pi }^{14}}}+\frac{421378048 c_{1}^{14}}{{{\pi }^{12}}}+\frac{11230336
c_{1}^{14}}{{{\pi }^{10}}}+\frac{7453160 c_{1}^{14}}{63 {{\pi }^8}}+  {}
{}
 \frac{31313 c_{1}^{14}}{30 {{\pi }^6}}+\frac{14813053 c_{1}^{14}}{110880 {{\pi }^4}}+\frac{163819127
c_{1}^{14}}{10762752 {{\pi }^2}}+\frac{21300428800 c_{1}^{12} {c_2}}{{{\pi }^{13}}}-  {}
{}
 \frac{2648662016 c_{1}^{12} {c_2}}{{{\pi }^{12}}}-\frac{1053445120 c_{1}^{12} {c_2}}{{{\pi }^{11}}}+\frac{19688448
c_{1}^{12} {c_2}}{{{\pi }^{10}}}-  {}
{}
 \frac{205766912 c_{1}^{12} {c_2}}{9 {{\pi }^9}}+\frac{11243232 c_{1}^{12} {c_2}}{5 {{\pi
}^8}}-\frac{17650204 c_{1}^{12} {c_2}}{63 {{\pi }^7}}+\frac{622888 c_{1}^{12} {c_2}}{9 {{\pi }^6}}-  {}
{}
 \frac{717229 c_{1}^{12} {c_2}}{150 {{\pi }^5}}+\frac{2702033 c_{1}^{12} {c_2}}{2520 {{\pi
}^4}}-\frac{13584955 c_{1}^{12} {c_2}}{133056 {{\pi }^3}}-\frac{4773 c_{1}^{12} {c_2}}{64 {{\pi }^2}}-
 {}
{}
 \frac{26792235008 c_{1}^{10} c_{2}^{2}}{{{\pi }^{12}}}+\frac{6019686400 c_{1}^{10} c_{2}^{2}}{{{\pi }^{11}}}+\frac{690114048
c_{1}^{10} c_{2}^{2}}{{{\pi }^{10}}}-  {}
{}
 \frac{20819968 c_{1}^{10} c_{2}^{2}}{3 {{\pi }^9}}+\frac{26628784 c_{1}^{10} c_{2}^{2}}{5
{{\pi }^8}}-\frac{4086368 c_{1}^{10} c_{2}^{2}}{3 {{\pi }^7}}-\frac{8369624 c_{1}^{10} c_{2}^{2}}{63 {{\pi
}^6}}-  {}
{}
 \frac{622744 c_{1}^{10} c_{2}^{2}}{45 {{\pi }^5}}-\frac{6366541 c_{1}^{10} c_{2}^{2}}{1200
{{\pi }^4}}+\frac{119963 c_{1}^{10} c_{2}^{2}}{504 {{\pi }^3}}+\frac{9453 c_{1}^{10} c_{2}^{2}}{64 {{\pi
}^2}}+  {}
{}
 \frac{14863994880 c_{1}^{8} c_{2}^{3}}{{{\pi }^{11}}}-\frac{4383055872 c_{1}^{8} c_{2}^{3}}{{{\pi }^{10}}}-\frac{307151104
c_{1}^{8} c_{2}^{3}}{3 {{\pi }^9}}+  {}
{}
 \frac{2737280 c_{1}^{8} c_{2}^{3}}{3 {{\pi }^8}}+\frac{3139760 c_{1}^{8} c_{2}^{3}}{{{\pi }^7}}+\frac{355040
c_{1}^{8} c_{2}^{3}}{{{\pi }^6}}+\frac{2366408 c_{1}^{8} c_{2}^{3}}{35 {{\pi }^5}}+  {}
{}
 \frac{332153 c_{1}^{8} c_{2}^{3}}{35 {{\pi }^4}}-\frac{42647 c_{1}^{8} c_{2}^{3}}{280 {{\pi
}^3}}-\frac{609 c_{1}^{8} c_{2}^{3}}{4 {{\pi }^2}}-\frac{3599382528 c_{1}^{6} c_{2}^{4}}{{{\pi }^{10}}}+  {}
{}
 \frac{1178591232 c_{1}^{6} c_{2}^{4}}{{{\pi }^9}}-\frac{3820544 c_{1}^{6} c_{2}^{4}}{3 {{\pi }^8}}-\frac{4754048
c_{1}^{6} c_{2}^{4}}{{{\pi }^7}}-  {}
{}
 \frac{4270124 c_{1}^{6} c_{2}^{4}}{9 {{\pi }^6}}-\frac{527888 c_{1}^{6} c_{2}^{4}}{5 {{\pi
}^5}}-\frac{33486 c_{1}^{6} c_{2}^{4}}{5 {{\pi }^4}}-\frac{2029 c_{1}^{6} c_{2}^{4}}{18 {{\pi }^3}}+  {}
{}
 \frac{1117 c_{1}^{6} c_{2}^{4}}{12 {{\pi }^2}}+\frac{325355520 c_{1}^{4} c_{2}^{5}}{{{\pi }^9}}-\frac{101803520
c_{1}^{4} c_{2}^{5}}{{{\pi }^8}}-\frac{1848704 c_{1}^{4} c_{2}^{5}}{{{\pi }^7}}+  {}
{}
 \frac{2307392 c_{1}^{4} c_{2}^{5}}{3 {{\pi }^6}}+\frac{175836 c_{1}^{4} c_{2}^{5}}{5 {{\pi
}^5}}+\frac{31126 c_{1}^{4} c_{2}^{5}}{9 {{\pi }^4}}+\frac{1169 c_{1}^{4} c_{2}^{5}}{12 {{\pi }^3}}-\frac{28
c_{1}^{4} c_{2}^{5}}{{{\pi }^2}}-  {}
{}
 \frac{7216384 c_{1}^{2} c_{2}^{6}}{{{\pi }^8}}+\frac{1697280 c_{1}^{2} c_{2}^{6}}{{{\pi }^7}}+\frac{404432
c_{1}^{2} c_{2}^{6}}{3 {{\pi }^6}}-\frac{3456 c_{1}^{2} c_{2}^{6}}{{{\pi }^5}}+  {}
{}
 \frac{1613 c_{1}^{2} c_{2}^{6}}{15 {{\pi }^4}}-\frac{170 c_{1}^{2} c_{2}^{6}}{3 {{\pi }^3}}+\frac{53
c_{1}^{2} c_{2}^{6}}{8 {{\pi }^2}}+\frac{8448 c_{2}^{7}}{{{\pi }^7}}-\frac{784 c_{2}^{7}}{3 {{\pi }^5}}-\frac{178
c_{2}^{7}}{15 {{\pi }^3}}+  {}
{}
 \frac{5 c_{2}^{7}}{112 \pi }-\frac{3787771904 c_{1}^{11} {c_3}}{{{\pi }^{12}}}+\frac{601968640 c_{1}^{11}
{c_3}}{{{\pi }^{11}}}+\frac{159883776 c_{1}^{11} {c_3}}{{{\pi }^{10}}}-  {}
{}
 \frac{11315200 c_{1}^{11} {c_3}}{3 {{\pi }^9}}+\frac{14593488 c_{1}^{11} {c_3}}{5 {{\pi }^8}}-\frac{999856
c_{1}^{11} {c_3}}{3 {{\pi }^7}}+\frac{1081624 c_{1}^{11} {c_3}}{63 {{\pi }^6}}-  {}
{}
 \frac{298096 c_{1}^{11} {c_3}}{45 {{\pi }^5}}-\frac{6115157 c_{1}^{11} {c_3}}{25200 {{\pi
}^4}}-\frac{8861 c_{1}^{11} {c_3}}{1008 {{\pi }^3}}+\frac{30567 c_{1}^{11} {c_3}}{1408 {{\pi }^2}}+  {}
{}
 \frac{8404408320 c_{1}^{9} {c_2} {c_3}}{{{\pi }^{11}}}-\frac{2150793216 c_{1}^{9} {c_2} {c_3}}{{{\pi
}^{10}}}-\frac{146418688 c_{1}^{9} {c_2} {c_3}}{{{\pi }^9}}+  {}
{}
 \frac{3284736 c_{1}^{9} {c_2} {c_3}}{{{\pi }^8}}-\frac{601224 c_{1}^{9} {c_2} {c_3}}{{{\pi
}^7}}+\frac{405056 c_{1}^{9} {c_2} {c_3}}{{{\pi }^6}}+  {}
{}
 \frac{2893636 c_{1}^{9} {c_2} {c_3}}{105 {{\pi }^5}}+\frac{2091538 c_{1}^{9} {c_2}
{c_3}}{315 {{\pi }^4}}+\frac{286243 c_{1}^{9} {c_2} {c_3}}{1120 {{\pi }^3}}-\frac{1263 c_{1}^{9}
{c_2} {c_3}}{16 {{\pi }^2}}-  {}
{}
 \frac{5944971264 c_{1}^{7} c_{2}^{2} {c_3}}{{{\pi }^{10}}}+\frac{1923584000 c_{1}^{7} c_{2}^{2}
{c_3}}{{{\pi }^9}}-\frac{14134848 c_{1}^{7} c_{2}^{2} {c_3}}{{{\pi }^8}}-  {}
{}
 \frac{1581632 c_{1}^{7} c_{2}^{2} {c_3}}{3 {{\pi }^7}}-\frac{4031728 c_{1}^{7} c_{2}^{2}
{c_3}}{3 {{\pi }^6}}-\frac{553784 c_{1}^{7} c_{2}^{2} {c_3}}{5 {{\pi }^5}}-  {}
{}
 \frac{4491401 c_{1}^{7} c_{2}^{2} {c_3}}{210 {{\pi }^4}}-\frac{81101 c_{1}^{7} c_{2}^{2}
{c_3}}{105 {{\pi }^3}}+\frac{421 c_{1}^{7} c_{2}^{2} {c_3}}{4 {{\pi }^2}}+\frac{1534288896 c_{1}^{5}
c_{2}^{3} {c_3}}{{{\pi }^9}}-  {}
{}
 \frac{544567296 c_{1}^{5} c_{2}^{3} {c_3}}{{{\pi }^8}}+\frac{17225312 c_{1}^{5} c_{2}^{3}
{c_3}}{{{\pi }^7}}+\frac{3734720 c_{1}^{5} c_{2}^{3} {c_3}}{3 {{\pi }^6}}+  {}
{}
 \frac{3551834 c_{1}^{5} c_{2}^{3} {c_3}}{15 {{\pi }^5}}+\frac{995264 c_{1}^{5} c_{2}^{3}
{c_3}}{45 {{\pi }^4}}+\frac{3921 c_{1}^{5} c_{2}^{3} {c_3}}{4 {{\pi }^3}}-\frac{90 c_{1}^{5} c_{2}^{3}
{c_3}}{{{\pi }^2}}-  {}
{}
 \frac{123065600 c_{1}^{3} c_{2}^{4} {c_3}}{{{\pi }^8}}+\frac{41809664 c_{1}^{3} c_{2}^{4}
{c_3}}{{{\pi }^7}}-\frac{2544128 c_{1}^{3} c_{2}^{4} {c_3}}{3 {{\pi }^6}}-  {}
{}
 \frac{173888 c_{1}^{3} c_{2}^{4} {c_3}}{{{\pi }^5}}-\frac{89675 c_{1}^{3} c_{2}^{4} {c_3}}{9
{{\pi }^4}}-\frac{1177 c_{1}^{3} c_{2}^{4} {c_3}}{3 {{\pi }^3}}+\frac{105 c_{1}^{3} c_{2}^{4} {c_3}}{4
{{\pi }^2}}+  {}
{}
 \frac{1710976 {c_1} c_{2}^{5} {c_3}}{{{\pi }^7}}-\frac{431872 {c_1} c_{2}^{5} {c_3}}{{{\pi
}^6}}-\frac{15208 {c_1} c_{2}^{5} {c_3}}{{{\pi }^5}}-\frac{848 {c_1} c_{2}^{5} {c_3}}{3 {{\pi }^4}}+
 {}
{}
 \frac{263 {c_1} c_{2}^{5} {c_3}}{30 {{\pi }^3}}-\frac{11 {c_1} c_{2}^{5} {c_3}}{{{\pi
}^2}}-\frac{583916544 c_{1}^{8} c_{3}^{2}}{{{\pi }^{10}}}+\frac{155697152 c_{1}^{8} c_{3}^{2}}{{{\pi }^9}}+  {}
{}
 \frac{8287552 c_{1}^{8} c_{3}^{2}}{{{\pi }^8}}-\frac{667264 c_{1}^{8} c_{3}^{2}}{3 {{\pi }^7}}-\frac{101956
c_{1}^{8} c_{3}^{2}}{3 {{\pi }^6}}-\frac{80768 c_{1}^{8} c_{3}^{2}}{5 {{\pi }^5}}-  {}
{}
 \frac{1691167 c_{1}^{8} c_{3}^{2}}{630 {{\pi }^4}}-\frac{13586 c_{1}^{8} c_{3}^{2}}{105 {{\pi
}^3}}+\frac{393 c_{1}^{8} c_{3}^{2}}{32 {{\pi }^2}}+\frac{680809472 c_{1}^{6} {c_2} c_{3}^{2}}{{{\pi }^9}}-
 {}
{}
 \frac{224014336 c_{1}^{6} {c_2} c_{3}^{2}}{{{\pi }^8}}+\frac{8523424 c_{1}^{6} {c_2} c_{3}^{2}}{3
{{\pi }^7}}+\frac{1070272 c_{1}^{6} {c_2} c_{3}^{2}}{3 {{\pi }^6}}+  {}
{}
 \frac{1643092 c_{1}^{6} {c_2} c_{3}^{2}}{15 {{\pi }^5}}+\frac{620732 c_{1}^{6} {c_2}
c_{3}^{2}}{45 {{\pi }^4}}+\frac{13823 c_{1}^{6} {c_2} c_{3}^{2}}{20 {{\pi }^3}}-\frac{21 c_{1}^{6}
{c_2} c_{3}^{2}}{{{\pi }^2}}-  {}  
{}
 \frac{199962112 c_{1}^{4} c_{2}^{2} c_{3}^{2}}{{{\pi }^8}}+\frac{71279104 c_{1}^{4} c_{2}^{2}
c_{3}^{2}}{{{\pi }^7}}   $

$-\frac{2661808 c_{1}^{4} c_{2}^{2} c_{3}^{2}}{{{\pi }^6}}-  {}
{}
 \frac{164992 c_{1}^{4} c_{2}^{2} c_{3}^{2}}{{{\pi }^5}}-\frac{21995 c_{1}^{4} c_{2}^{2} c_{3}^{2}}{{{\pi
}^4}}-\frac{3014 c_{1}^{4} c_{2}^{2} c_{3}^{2}}{3 {{\pi }^3}}+\frac{343 c_{1}^{4} c_{2}^{2} c_{3}^{2}}{8
{{\pi }^2}}+  {}
{}
 \frac{13590144 c_{1}^{2} c_{2}^{3} c_{3}^{2}}{{{\pi }^7}}-\frac{4568128 c_{1}^{2} c_{2}^{3}
c_{3}^{2}}{{{\pi }^6}}+\frac{134632 c_{1}^{2} c_{2}^{3} c_{3}^{2}}{{{\pi }^5}}+  {}
{}
 \frac{11120 c_{1}^{2} c_{2}^{3} c_{3}^{2}}{{{\pi }^4}}+\frac{417 c_{1}^{2} c_{2}^{3} c_{3}^{2}}{{{\pi
}^3}}-\frac{9 c_{1}^{2} c_{2}^{3} c_{3}^{2}}{{{\pi }^2}}-\frac{80864 c_{2}^{4} c_{3}^{2}}{{{\pi }^6}}+\frac{17856
c_{2}^{4} c_{3}^{2}}{{{\pi }^5}}+  {}
{}
 \frac{964 c_{2}^{4} c_{3}^{2}}{{{\pi }^4}}+\frac{24 c_{2}^{4} c_{3}^{2}}{{{\pi }^3}}+\frac{15 c_{2}^{4}
c_{3}^{2}}{2 {{\pi }^2}}-\frac{20920704 c_{1}^{5} c_{3}^{3}}{{{\pi }^8}}+\frac{6309888 c_{1}^{5} c_{3}^{3}}{{{\pi
}^7}}+  {}
{}
 \frac{83168 c_{1}^{5} c_{3}^{3}}{{{\pi }^6}}-\frac{25776 c_{1}^{5} c_{3}^{3}}{{{\pi }^5}}-\frac{10686
c_{1}^{5} c_{3}^{3}}{5 {{\pi }^4}}-\frac{619 c_{1}^{5} c_{3}^{3}}{3 {{\pi }^3}}+\frac{227 c_{1}^{5}
c_{3}^{3}}{40 {{\pi }^2}}+  {}
{}
 \frac{8545920 c_{1}^{3} {c_2} c_{3}^{3}}{{{\pi }^7}}-\frac{2728192 c_{1}^{3} {c_2} c_{3}^{3}}{{{\pi
}^6}}+\frac{97376 c_{1}^{3} {c_2} c_{3}^{3}}{3 {{\pi }^5}}+\frac{10576 c_{1}^{3} {c_2} c_{3}^{3}}{{{\pi
}^4}}+  {}
{}
 \frac{709 c_{1}^{3} {c_2} c_{3}^{3}}{3 {{\pi }^3}}-\frac{71 c_{1}^{3} {c_2} c_{3}^{3}}{3
{{\pi }^2}}-\frac{448000 {c_1} c_{2}^{2} c_{3}^{3}}{{{\pi }^6}}+\frac{126528 {c_1} c_{2}^{2} c_{3}^{3}}{{{\pi
}^5}}-  {}
{}
 \frac{224 {c_1} c_{2}^{2} c_{3}^{3}}{3 {{\pi }^4}}-\frac{148 {c_1} c_{2}^{2} c_{3}^{3}}{{{\pi
}^3}}+\frac{6 {c_1} c_{2}^{2} c_{3}^{3}}{{{\pi }^2}}-\frac{81696 c_{1}^{2} c_{3}^{4}}{{{\pi }^6}}+\frac{17984
c_{1}^{2} c_{3}^{4}}{{{\pi }^5}}+  {}
{}
 \frac{988 c_{1}^{2} c_{3}^{4}}{{{\pi }^4}}+\frac{24 c_{1}^{2} c_{3}^{4}}{{{\pi }^3}}+\frac{15 c_{1}^{2}
c_{3}^{4}}{2 {{\pi }^2}}+\frac{2608 {c_2} c_{3}^{4}}{{{\pi }^5}}-\frac{352 {c_2} c_{3}^{4}}{{{\pi }^4}}-\frac{106
{c_2} c_{3}^{4}}{3 {{\pi }^3}}-  {}
{}
 \frac{4 {c_2} c_{3}^{4}}{{{\pi }^2}}+\frac{569554944 c_{1}^{10} {c_4}}{{{\pi }^{11}}}-\frac{96405504
c_{1}^{10} {c_4}}{{{\pi }^{10}}}-\frac{65967616 c_{1}^{10} {c_4}}{3 {{\pi }^9}}+  {}
{}
 \frac{582400 c_{1}^{10} {c_4}}{{{\pi }^8}}-\frac{5769608 c_{1}^{10} {c_4}}{15 {{\pi }^7}}+\frac{47520
c_{1}^{10} {c_4}}{{{\pi }^6}}-\frac{123877 c_{1}^{10} {c_4}}{35 {{\pi }^5}}+  {}
{}
 \frac{19976 c_{1}^{10} {c_4}}{21 {{\pi }^4}}-\frac{2963701 c_{1}^{10} {c_4}}{50400 {{\pi
}^3}}-\frac{83 c_{1}^{10} {c_4}}{10 {{\pi }^2}}-\frac{1125061632 c_{1}^{8} {c_2} {c_4}}{{{\pi }^{10}}}+
 {}
{}
 \frac{299626496 c_{1}^{8} {c_2} {c_4}}{{{\pi }^9}}+\frac{14402752 c_{1}^{8} {c_2} {c_4}}{{{\pi
}^8}}-\frac{483392 c_{1}^{8} {c_2} {c_4}}{3 {{\pi }^7}}-  {}
{}
 \frac{91340 c_{1}^{8} {c_2} {c_4}}{3 {{\pi }^6}}-\frac{146232 c_{1}^{8} {c_2} {c_4}}{5
{{\pi }^5}}-\frac{607697 c_{1}^{8} {c_2} {c_4}}{126 {{\pi }^4}}-\frac{5399 c_{1}^{8} {c_2} {c_4}}{210
{{\pi }^3}}+  {}
{}
 \frac{659 c_{1}^{8} {c_2} {c_4}}{32 {{\pi }^2}}+\frac{668136448 c_{1}^{6} c_{2}^{2}
{c_4}}{{{\pi }^9}}-\frac{220100608 c_{1}^{6} c_{2}^{2} {c_4}}{{{\pi }^8}}+  {}
{}
 \frac{3813472 c_{1}^{6} c_{2}^{2} {c_4}}{{{\pi }^7}}+\frac{90976 c_{1}^{6} c_{2}^{2} {c_4}}{3
{{\pi }^6}}+\frac{2042404 c_{1}^{6} c_{2}^{2} {c_4}}{15 {{\pi }^5}}+  {}
{}
 \frac{144818 c_{1}^{6} c_{2}^{2} {c_4}}{15 {{\pi }^4}}+\frac{8248 c_{1}^{6} c_{2}^{2}
{c_4}}{15 {{\pi }^3}}-\frac{15 c_{1}^{6} c_{2}^{2} {c_4}}{{{\pi }^2}}-\frac{128738176 c_{1}^{4} c_{2}^{3}
{c_4}}{{{\pi }^8}}+  {}
{}
 \frac{44638464 c_{1}^{4} c_{2}^{3} {c_4}}{{{\pi }^7}}-\frac{1231760 c_{1}^{4} c_{2}^{3} {c_4}}{{{\pi
}^6}}-\frac{474112 c_{1}^{4} c_{2}^{3} {c_4}}{3 {{\pi }^5}}-  {}
{}
 \frac{8424 c_{1}^{4} c_{2}^{3} {c_4}}{{{\pi }^4}}-\frac{738 c_{1}^{4} c_{2}^{3} {c_4}}{{{\pi
}^3}}+\frac{41 c_{1}^{4} c_{2}^{3} {c_4}}{4 {{\pi }^2}}+\frac{5814144 c_{1}^{2} c_{2}^{4} {c_4}}{{{\pi
}^7}}-  {}
{}
 \frac{1750720 c_{1}^{2} c_{2}^{4} {c_4}}{{{\pi }^6}}+\frac{2304 c_{1}^{2} c_{2}^{4} {c_4}}{{{\pi
}^5}}+\frac{10456 c_{1}^{2} c_{2}^{4} {c_4}}{3 {{\pi }^4}}+\frac{326 c_{1}^{2} c_{2}^{4} {c_4}}{{{\pi
}^3}}-  {}
{}
 \frac{3 c_{1}^{2} c_{2}^{4} {c_4}}{2 {{\pi }^2}}-\frac{14784 c_{2}^{5} {c_4}}{{{\pi }^6}}+\frac{1792
c_{2}^{5} {c_4}}{{{\pi }^5}}+\frac{196 c_{2}^{5} {c_4}}{{{\pi }^4}}+\frac{136 c_{2}^{5} {c_4}}{3 {{\pi
}^3}}+  {}
{}
 \frac{41 c_{2}^{5} {c_4}}{20 {{\pi }^2}}+\frac{148124672 c_{1}^{7} {c_3} {c_4}}{{{\pi }^9}}-\frac{40535040
c_{1}^{7} {c_3} {c_4}}{{{\pi }^8}}-\frac{4717024 c_{1}^{7} {c_3} {c_4}}{3 {{\pi }^7}}+  {}
{}
 \frac{77120 c_{1}^{7} {c_3} {c_4}}{{{\pi }^6}}+\frac{29494 c_{1}^{7} {c_3} {c_4}}{15
{{\pi }^5}}+\frac{64228 c_{1}^{7} {c_3} {c_4}}{15 {{\pi }^4}}+\frac{51557 c_{1}^{7} {c_3} {c_4}}{420
{{\pi }^3}}-  {}
{}
 \frac{21 c_{1}^{7} {c_3} {c_4}}{2 {{\pi }^2}}-\frac{143310720 c_{1}^{5} {c_2} {c_3}
{c_4}}{{{\pi }^8}}+\frac{48166144 c_{1}^{5} {c_2} {c_3} {c_4}}{{{\pi }^7}}-  {}
{}
 \frac{1271536 c_{1}^{5} {c_2} {c_3} {c_4}}{{{\pi }^6}}-\frac{102176 c_{1}^{5} {c_2}
{c_3} {c_4}}{3 {{\pi }^5}}-\frac{110211 c_{1}^{5} {c_2} {c_3} {c_4}}{5 {{\pi }^4}}-  {}
{}
 \frac{11084 c_{1}^{5} {c_2} {c_3} {c_4}}{15 {{\pi }^3}}+\frac{30300288 c_{1}^{3} c_{2}^{2}
{c_3} {c_4}}{{{\pi }^7}}-\frac{10741824 c_{1}^{3} c_{2}^{2} {c_3} {c_4}}{{{\pi }^6}}+  {}
{}
 \frac{470568 c_{1}^{3} c_{2}^{2} {c_3} {c_4}}{{{\pi }^5}}+\frac{18792 c_{1}^{3} c_{2}^{2}
{c_3} {c_4}}{{{\pi }^4}}+\frac{3652 c_{1}^{3} c_{2}^{2} {c_3} {c_4}}{3 {{\pi }^3}}-  {}
{}
 \frac{9 c_{1}^{3} c_{2}^{2} {c_3} {c_4}}{{{\pi }^2}}-\frac{1005120 {c_1} c_{2}^{3}
{c_3} {c_4}}{{{\pi }^6}}+\frac{305472 {c_1} c_{2}^{3} {c_3} {c_4}}{{{\pi }^5}}-  {}
{}
 \frac{17396 {c_1} c_{2}^{3} {c_3} {c_4}}{3 {{\pi }^4}}-\frac{520 {c_1} c_{2}^{3}
{c_3} {c_4}}{3 {{\pi }^3}}+\frac{6039616 c_{1}^{4} c_{3}^{2} {c_4}}{{{\pi }^7}}-  {}
{}
 \frac{1871040 c_{1}^{4} c_{3}^{2} {c_4}}{{{\pi }^6}}+\frac{48584 c_{1}^{4} c_{3}^{2} {c_4}}{3
{{\pi }^5}}+\frac{13112 c_{1}^{4} c_{3}^{2} {c_4}}{3 {{\pi }^4}}+\frac{968 c_{1}^{4} c_{3}^{2} {c_4}}{3
{{\pi }^3}}-  {}
{}
 \frac{11 c_{1}^{4} c_{3}^{2} {c_4}}{2 {{\pi }^2}}-\frac{1677856 c_{1}^{2} {c_2} c_{3}^{2}
{c_4}}{{{\pi }^6}}+\frac{543840 c_{1}^{2} {c_2} c_{3}^{2} {c_4}}{{{\pi }^5}}-  {}
{}
 \frac{18152 c_{1}^{2} {c_2} c_{3}^{2} {c_4}}{{{\pi }^4}}-\frac{528 c_{1}^{2} {c_2}
c_{3}^{2} {c_4}}{{{\pi }^3}}+\frac{15 c_{1}^{2} {c_2} c_{3}^{2} {c_4}}{{{\pi }^2}}+\frac{30880 c_{2}^{2}
c_{3}^{2} {c_4}}{{{\pi }^5}}-  {}
{}
 \frac{7744 c_{2}^{2} c_{3}^{2} {c_4}}{{{\pi }^4}}-\frac{90 c_{2}^{2} c_{3}^{2} {c_4}}{{{\pi
}^3}}+\frac{16576 {c_1} c_{3}^{3} {c_4}}{{{\pi }^5}}-\frac{3616 {c_1} c_{3}^{3} {c_4}}{{{\pi }^4}}-  {}
{}
 \frac{102 {c_1} c_{3}^{3} {c_4}}{{{\pi }^3}}-\frac{12 {c_1} c_{3}^{3} {c_4}}{{{\pi }^2}}-\frac{8066688
c_{1}^{6} c_{4}^{2}}{{{\pi }^8}}+\frac{2066688 c_{1}^{6} c_{4}^{2}}{{{\pi }^7}}+  {}
{}
 \frac{309776 c_{1}^{6} c_{4}^{2}}{3 {{\pi }^6}}-\frac{8864 c_{1}^{6} c_{4}^{2}}{3 {{\pi }^5}}-\frac{3799
c_{1}^{6} c_{4}^{2}}{15 {{\pi }^4}}-\frac{252 c_{1}^{6} c_{4}^{2}}{5 {{\pi }^3}}+\frac{29 c_{1}^{6}
c_{4}^{2}}{4 {{\pi }^2}}+  {}
{}
 \frac{6179392 c_{1}^{4} {c_2} c_{4}^{2}}{{{\pi }^7}}-\frac{1938880 c_{1}^{4} {c_2} c_{4}^{2}}{{{\pi
}^6}}+\frac{81928 c_{1}^{4} {c_2} c_{4}^{2}}{3 {{\pi }^5}}+\frac{9424 c_{1}^{4} {c_2} c_{4}^{2}}{3
{{\pi }^4}}+  {}
{}
 \frac{824 c_{1}^{4} {c_2} c_{4}^{2}}{3 {{\pi }^3}}-\frac{805088 c_{1}^{2} c_{2}^{2}
c_{4}^{2}}{{{\pi }^6}}+\frac{246528 c_{1}^{2} c_{2}^{2} c_{4}^{2}}{{{\pi }^5}}-\frac{3368 c_{1}^{2} c_{2}^{2}
c_{4}^{2}}{{{\pi }^4}}-  {}
{}
 \frac{544 c_{1}^{2} c_{2}^{2} c_{4}^{2}}{{{\pi }^3}}+\frac{c_{1}^{2} c_{2}^{2} c_{4}^{2}}{2
{{\pi }^2}}+\frac{5824 c_{2}^{3} c_{4}^{2}}{{{\pi }^5}}-\frac{1008 c_{2}^{3} c_{4}^{2}}{{{\pi }^4}}-\frac{188
c_{2}^{3} c_{4}^{2}}{3 {{\pi }^3}}-  {}
{}
 \frac{4 c_{2}^{3} c_{4}^{2}}{{{\pi }^2}}-\frac{453888 c_{1}^{3} {c_3} c_{4}^{2}}{{{\pi }^6}}+\frac{128352
c_{1}^{3} {c_3} c_{4}^{2}}{{{\pi }^5}}-\frac{164 c_{1}^{3} {c_3} c_{4}^{2}}{3 {{\pi }^4}}-  {}
{}
 \frac{172 c_{1}^{3} {c_3} c_{4}^{2}}{{{\pi }^3}}+\frac{6 c_{1}^{3} {c_3} c_{4}^{2}}{{{\pi
}^2}}+\frac{70032 {c_1} {c_2} {c_3} c_{4}^{2}}{{{\pi }^5}}-\frac{19472 {c_1} {c_2} {c_3}
c_{4}^{2}}{{{\pi }^4}}+  {}
{}
 \frac{264 {c_1} {c_2} {c_3} c_{4}^{2}}{{{\pi }^3}}-\frac{4 {c_1} {c_2} {c_3}
c_{4}^{2}}{{{\pi }^2}}-\frac{624 c_{3}^{2} c_{4}^{2}}{{{\pi }^4}}+\frac{96 c_{3}^{2} c_{4}^{2}}{{{\pi }^3}}+\frac{8
c_{3}^{2} c_{4}^{2}}{{{\pi }^2}}+  {}
{}
 \frac{7472 c_{1}^{2} c_{4}^{3}}{{{\pi }^5}}-\frac{1440 c_{1}^{2} c_{4}^{3}}{{{\pi }^4}}-\frac{278
c_{1}^{2} c_{4}^{3}}{3 {{\pi }^3}}-\frac{2 c_{1}^{2} c_{4}^{3}}{{{\pi }^2}}-\frac{336 {c_2} c_{4}^{3}}{{{\pi
}^4}}+\frac{40 {c_2} c_{4}^{3}}{{{\pi }^3}}+  {}
{}
 \frac{3 {c_2} c_{4}^{3}}{{{\pi }^2}}-\frac{77483008 c_{1}^{9} {c_5}}{{{\pi }^{10}}}+\frac{13125632
c_{1}^{9} {c_5}}{{{\pi }^9}}+\frac{8293376 c_{1}^{9} {c_5}}{3 {{\pi }^8}}-  {}
{}
 \frac{190528 c_{1}^{9} {c_5}}{3 {{\pi }^7}}+\frac{416308 c_{1}^{9} {c_5}}{9 {{\pi }^6}}-\frac{22812
c_{1}^{9} {c_5}}{5 {{\pi }^5}}+\frac{20128 c_{1}^{9} {c_5}}{45 {{\pi }^4}}-  {}
{}
 \frac{3373 c_{1}^{9} {c_5}}{126 {{\pi }^3}}+\frac{563 c_{1}^{9} {c_5}}{64 {{\pi }^2}}+\frac{135451648
c_{1}^{7} {c_2} {c_5}}{{{\pi }^9}}-\frac{36062208 c_{1}^{7} {c_2} {c_5}}{{{\pi }^8}}-  {}
{}
 \frac{1339936 c_{1}^{7} {c_2} {c_5}}{{{\pi }^7}}-\frac{22848 c_{1}^{7} {c_2} {c_5}}{{{\pi
}^6}}+\frac{117274 c_{1}^{7} {c_2} {c_5}}{15 {{\pi }^5}}+  {}
{}
 \frac{20528 c_{1}^{7} {c_2} {c_5}}{15 {{\pi }^4}}+\frac{29831 c_{1}^{7} {c_2} {c_5}}{140
{{\pi }^3}}-\frac{45 c_{1}^{7} {c_2} {c_5}}{2 {{\pi }^2}}-\frac{65767296 c_{1}^{5} c_{2}^{2} {c_5}}{{{\pi
}^8}}+  {}
{}
 \frac{21214336 c_{1}^{5} c_{2}^{2} {c_5}}{{{\pi }^7}}-\frac{383408 c_{1}^{5} c_{2}^{2} {c_5}}{{{\pi
}^6}}-\frac{27968 c_{1}^{5} c_{2}^{2} {c_5}}{3 {{\pi }^5}}-  {}
{}
 \frac{117916 c_{1}^{5} c_{2}^{2} {c_5}}{15 {{\pi }^4}}-\frac{7327 c_{1}^{5} c_{2}^{2}
{c_5}}{15 {{\pi }^3}}+\frac{39 c_{1}^{5} c_{2}^{2} {c_5}}{2 {{\pi }^2}}+\frac{8878720 c_{1}^{3}
c_{2}^{3} {c_5}}{{{\pi }^7}}-  {}
{}
 \frac{2893184 c_{1}^{3} c_{2}^{3} {c_5}}{{{\pi }^6}}+\frac{171712 c_{1}^{3} c_{2}^{3} {c_5}}{3
{{\pi }^5}}+\frac{8752 c_{1}^{3} c_{2}^{3} {c_5}}{{{\pi }^4}}+\frac{2651 c_{1}^{3} c_{2}^{3} {c_5}}{9
{{\pi }^3}}-  {}
{}
 \frac{38 c_{1}^{3} c_{2}^{3} {c_5}}{3 {{\pi }^2}}-\frac{189696 {c_1} c_{2}^{4} {c_5}}{{{\pi
}^6}}+\frac{47680 {c_1} c_{2}^{4} {c_5}}{{{\pi }^5}}+\frac{2588 {c_1} c_{2}^{4} {c_5}}{3 {{\pi }^4}}+
 {}
{}
 \frac{80 {c_1} c_{2}^{4} {c_5}}{{{\pi }^3}}+\frac{13 {c_1} c_{2}^{4} {c_5}}{2 {{\pi
}^2}}-\frac{17414656 c_{1}^{6} {c_3} {c_5}}{{{\pi }^8}}+\frac{4792320 c_{1}^{6} {c_3} {c_5}}{{{\pi }^7}}+
 {}
{}
 \frac{337696 c_{1}^{6} {c_3} {c_5}}{3 {{\pi }^6}}+\frac{2096 c_{1}^{6} {c_3} {c_5}}{3
{{\pi }^5}}-\frac{59246 c_{1}^{6} {c_3} {c_5}}{45 {{\pi }^4}}-\frac{1394 c_{1}^{6} {c_3} {c_5}}{15
{{\pi }^3}}+  {}
{}
 \frac{23 c_{1}^{6} {c_3} {c_5}}{4 {{\pi }^2}}+\frac{13118400 c_{1}^{4} {c_2} {c_3}
{c_5}}{{{\pi }^7}}-\frac{4285248 c_{1}^{4} {c_2} {c_3} {c_5}}{{{\pi }^6}}+  {}
{}
 \frac{336224 c_{1}^{4} {c_2} {c_3} {c_5}}{3 {{\pi }^5}}+\frac{12680 c_{1}^{4} {c_2}
{c_3} {c_5}}{3 {{\pi }^4}}+\frac{814 c_{1}^{4} {c_2} {c_3} {c_5}}{{{\pi }^3}}-\frac{3 c_{1}^{4}
{c_2} {c_3} {c_5}}{{{\pi }^2}}-  {}
{}
 \frac{1720128 c_{1}^{2} c_{2}^{2} {c_3} {c_5}}{{{\pi }^6}}+\frac{558528 c_{1}^{2} c_{2}^{2}
{c_3} {c_5}}{{{\pi }^5}}-\frac{18028 c_{1}^{2} c_{2}^{2} {c_3} {c_5}}{{{\pi }^4}}-  {}
{}
 \frac{704 c_{1}^{2} c_{2}^{2} {c_3} {c_5}}{{{\pi }^3}}+\frac{6 c_{1}^{2} c_{2}^{2}
{c_3} {c_5}}{{{\pi }^2}}+\frac{16576 c_{2}^{3} {c_3} {c_5}}{{{\pi }^5}}-\frac{3616 c_{2}^{3} {c_3}
{c_5}}{{{\pi }^4}}-  {}
{}
 \frac{102 c_{2}^{3} {c_3} {c_5}}{{{\pi }^3}}-\frac{12 c_{2}^{3} {c_3} {c_5}}{{{\pi }^2}}-\frac{498240
c_{1}^{3} c_{3}^{2} {c_5}}{{{\pi }^6}}+\frac{147456 c_{1}^{3} c_{3}^{2} {c_5}}{{{\pi }^5}}-  {}
{}
 \frac{4112 c_{1}^{3} c_{3}^{2} {c_5}}{3 {{\pi }^4}}-\frac{728 c_{1}^{3} c_{3}^{2}
{c_5}}{3 {{\pi }^3}}+\frac{5 c_{1}^{3} c_{3}^{2} {c_5}}{{{\pi }^2}}+\frac{68384 {c_1} {c_2} c_{3}^{2}
{c_5}}{{{\pi }^5}}-  {}
{}
 \frac{18864 {c_1} {c_2} c_{3}^{2} {c_5}}{{{\pi }^4}}+\frac{280 {c_1} {c_2} c_{3}^{2}
{c_5}}{{{\pi }^3}}-\frac{8 {c_1} {c_2} c_{3}^{2} {c_5}}{{{\pi }^2}}-\frac{336 c_{3}^{3} {c_5}}{{{\pi }^4}}+
 {}
{}
 \frac{40 c_{3}^{3} {c_5}}{{{\pi }^3}}+\frac{3 c_{3}^{3} {c_5}}{{{\pi }^2}}+\frac{1789184 c_{1}^{5}
{c_4} {c_5}}{{{\pi }^7}}-\frac{457600 c_{1}^{5} {c_4} {c_5}}{{{\pi }^6}}-  {}
{}
 \frac{15360 c_{1}^{5} {c_4} {c_5}}{{{\pi }^5}}+\frac{96 c_{1}^{5} {c_4} {c_5}}{{{\pi }^4}}-\frac{159
c_{1}^{5} {c_4} {c_5}}{10 {{\pi }^3}}-\frac{11 c_{1}^{5} {c_4} {c_5}}{{{\pi }^2}}-  {}
{}
 \frac{1038272 c_{1}^{3} {c_2} {c_4} {c_5}}{{{\pi }^6}}+\frac{319616 c_{1}^{3} {c_2}
{c_4} {c_5}}{{{\pi }^5}}-\frac{7428 c_{1}^{3} {c_2} {c_4} {c_5}}{{{\pi }^4}}-  {}
{}
 \frac{256 c_{1}^{3} {c_2} {c_4} {c_5}}{3 {{\pi }^3}}+\frac{71504 {c_1} c_{2}^{2}
{c_4} {c_5}}{{{\pi }^5}}-\frac{19984 {c_1} c_{2}^{2} {c_4} {c_5}}{{{\pi }^4}}+  {}
{}
 \frac{260 {c_1} c_{2}^{2} {c_4} {c_5}}{{{\pi }^3}}+\frac{68384 c_{1}^{2} {c_3} {c_4}
{c_5}}{{{\pi }^5}}-\frac{18864 c_{1}^{2} {c_3} {c_4} {c_5}}{{{\pi }^4}}+  {}
{}
 \frac{280 c_{1}^{2} {c_3} {c_4} {c_5}}{{{\pi }^3}}-\frac{8 c_{1}^{2} {c_3} {c_4}
{c_5}}{{{\pi }^2}}-\frac{3672 {c_2} {c_3} {c_4} {c_5}}{{{\pi }^4}}+\frac{848 {c_2} {c_3} {c_4}
{c_5}}{{{\pi }^3}}-  {}
{}
 \frac{1432 {c_1} c_{4}^{2} {c_5}}{{{\pi }^4}}+\frac{272 {c_1} c_{4}^{2} {c_5}}{{{\pi }^3}}+\frac{6
{c_1} c_{4}^{2} {c_5}}{{{\pi }^2}}-\frac{81696 c_{1}^{4} c_{5}^{2}}{{{\pi }^6}}+  {}
{}
 \frac{17984 c_{1}^{4} c_{5}^{2}}{{{\pi }^5}}+\frac{988 c_{1}^{4} c_{5}^{2}}{{{\pi }^4}}+\frac{24
c_{1}^{4} c_{5}^{2}}{{{\pi }^3}}+\frac{15 c_{1}^{4} c_{5}^{2}}{2 {{\pi }^2}}+\frac{31504 c_{1}^{2} {c_2}
c_{5}^{2}}{{{\pi }^5}}-  {}
{}
 \frac{7888 c_{1}^{2} {c_2} c_{5}^{2}}{{{\pi }^4}}-\frac{96 c_{1}^{2} {c_2} c_{5}^{2}}{{{\pi
}^3}}-\frac{624 c_{2}^{2} c_{5}^{2}}{{{\pi }^4}}+\frac{96 c_{2}^{2} c_{5}^{2}}{{{\pi }^3}}+\frac{8 c_{2}^{2}
c_{5}^{2}}{{{\pi }^2}}-  {}
{}
 \frac{1576 {c_1} {c_3} c_{5}^{2}}{{{\pi }^4}}+\frac{304 {c_1} {c_3} c_{5}^{2}}{{{\pi }^3}}+\frac{8
{c_1} {c_3} c_{5}^{2}}{{{\pi }^2}}+\frac{44 {c_4} c_{5}^{2}}{{{\pi }^3}}-\frac{4 {c_4} c_{5}^{2}}{{{\pi
}^2}}+  {}
{}
 \frac{10000896 c_{1}^{8} {c_6}}{{{\pi }^9}}-\frac{1654016 c_{1}^{8} {c_6}}{{{\pi }^8}}-\frac{318656
c_{1}^{8} {c_6}}{{{\pi }^7}}+\frac{7328 c_{1}^{8} {c_6}}{3 {{\pi }^6}}-  {}
{}
 \frac{74494 c_{1}^{8} {c_6}}{15 {{\pi }^5}}+\frac{9574 c_{1}^{8} {c_6}}{45 {{\pi }^4}}-\frac{7451
c_{1}^{8} {c_6}}{140 {{\pi }^3}}-\frac{6 c_{1}^{8} {c_6}}{{{\pi }^2}}-\frac{15178240 c_{1}^{6} {c_2}
{c_6}}{{{\pi }^8}}+  {}
{}
 \frac{3938688 c_{1}^{6} {c_2} {c_6}}{{{\pi }^7}}+\frac{119328 c_{1}^{6} {c_2} {c_6}}{{{\pi
}^6}}+\frac{7792 c_{1}^{6} {c_2} {c_6}}{{{\pi }^5}}-\frac{9353 c_{1}^{6} {c_2} {c_6}}{15 {{\pi }^4}}+
 {}
{}
 \frac{733 c_{1}^{6} {c_2} {c_6}}{15 {{\pi }^3}}+\frac{25 c_{1}^{6} {c_2} {c_6}}{2
{{\pi }^2}}+\frac{5775040 c_{1}^{4} c_{2}^{2} {c_6}}{{{\pi }^7}}-\frac{1769664 c_{1}^{4} c_{2}^{2} {c_6}}{{{\pi
}^6}}+  {}
{}
 \frac{24152 c_{1}^{4} c_{2}^{2} {c_6}}{{{\pi }^5}}+\frac{2752 c_{1}^{4} c_{2}^{2} {c_6}}{3
{{\pi }^4}}+\frac{568 c_{1}^{4} c_{2}^{2} {c_6}}{3 {{\pi }^3}}-\frac{6 c_{1}^{4} c_{2}^{2} {c_6}}{{{\pi
}^2}}-  {}
{}
 \frac{482528 c_{1}^{2} c_{2}^{3} {c_6}}{{{\pi }^6}}+\frac{138336 c_{1}^{2} c_{2}^{3} {c_6}}{{{\pi
}^5}}-\frac{80 c_{1}^{2} c_{2}^{3} {c_6}}{{{\pi }^4}}-\frac{248 c_{1}^{2} c_{2}^{3} {c_6}}{{{\pi }^3}}+
 {}
{}
 \frac{4 c_{1}^{2} c_{2}^{3} {c_6}}{{{\pi }^2}}+\frac{2464 c_{2}^{4} {c_6}}{{{\pi }^5}}-\frac{336
c_{2}^{4} {c_6}}{{{\pi }^4}}-\frac{88 c_{2}^{4} {c_6}}{3 {{\pi }^3}}-\frac{4 c_{2}^{4} {c_6}}{{{\pi }^2}}+
 {}
{}
 \frac{1899008 c_{1}^{5} {c_3} {c_6}}{{{\pi }^7}}-\frac{506880 c_{1}^{5} {c_3} {c_6}}{{{\pi
}^6}}-\frac{9208 c_{1}^{5} {c_3} {c_6}}{{{\pi }^5}}-\frac{1280 c_{1}^{5} {c_3} {c_6}}{3 {{\pi }^4}}+
 {}
{}
 \frac{277 c_{1}^{5} {c_3} {c_6}}{10 {{\pi }^3}}-\frac{6 c_{1}^{5} {c_3} {c_6}}{{{\pi
}^2}}-\frac{1046720 c_{1}^{3} {c_2} {c_3} {c_6}}{{{\pi }^6}}+\frac{321600 c_{1}^{3} {c_2} {c_3}
{c_6}}{{{\pi }^5}}-  {}
{}
 \frac{7196 c_{1}^{3} {c_2} {c_3} {c_6}}{{{\pi }^4}}-\frac{400 c_{1}^{3} {c_2} {c_3}
{c_6}}{3 {{\pi }^3}}+\frac{71968 {c_1} c_{2}^{2} {c_3} {c_6}}{{{\pi }^5}}-  {}
{}
 \frac{20336 {c_1} c_{2}^{2} {c_3} {c_6}}{{{\pi }^4}}+\frac{360 {c_1} c_{2}^{2} {c_3}
{c_6}}{{{\pi }^3}}-\frac{4 {c_1} c_{2}^{2} {c_3} {c_6}}{{{\pi }^2}}+\frac{31504 c_{1}^{2} c_{3}^{2}
{c_6}}{{{\pi }^5}}-  {}
{}
 \frac{7888 c_{1}^{2} c_{3}^{2} {c_6}}{{{\pi }^4}}-\frac{96 c_{1}^{2} c_{3}^{2} {c_6}}{{{\pi
}^3}}-\frac{1488 {c_2} c_{3}^{2} {c_6}}{{{\pi }^4}}+\frac{280 {c_2} c_{3}^{2} {c_6}}{{{\pi }^3}}+\frac{8
{c_2} c_{3}^{2} {c_6}}{{{\pi }^2}}-  {}
{}
 \frac{190752 c_{1}^{4} {c_4} {c_6}}{{{\pi }^6}}+\frac{47776 c_{1}^{4} {c_4} {c_6}}{{{\pi
}^5}}+\frac{2732 c_{1}^{4} {c_4} {c_6}}{3 {{\pi }^4}}+\frac{80 c_{1}^{4} {c_4} {c_6}}{{{\pi }^3}}+
 {}
{}
 \frac{13 c_{1}^{4} {c_4} {c_6}}{2 {{\pi }^2}}+\frac{72032 c_{1}^{2} {c_2} {c_4}
{c_6}}{{{\pi }^5}}-\frac{20080 c_{1}^{2} {c_2} {c_4} {c_6}}{{{\pi }^4}}+\frac{248 c_{1}^{2} {c_2}
{c_4} {c_6}}{{{\pi }^3}}-  {}
{}
 \frac{1344 c_{2}^{2} {c_4} {c_6}}{{{\pi }^4}}+\frac{248 c_{2}^{2} {c_4} {c_6}}{{{\pi }^3}}+\frac{6
c_{2}^{2} {c_4} {c_6}}{{{\pi }^2}}-\frac{3672 {c_1} {c_3} {c_4} {c_6}}{{{\pi }^4}}+  {}
{}
 \frac{848 {c_1} {c_3} {c_4} {c_6}}{{{\pi }^3}}+\frac{44 c_{4}^{2} {c_6}}{{{\pi }^3}}-\frac{4
c_{4}^{2} {c_6}}{{{\pi }^2}}+\frac{16576 c_{1}^{3} {c_5} {c_6}}{{{\pi }^5}}-\frac{3616 c_{1}^{3} {c_5}
{c_6}}{{{\pi }^4}}-  {}
{}
 \frac{102 c_{1}^{3} {c_5} {c_6}}{{{\pi }^3}}-\frac{12 c_{1}^{3} {c_5} {c_6}}{{{\pi }^2}}-\frac{3672
{c_1} {c_2} {c_5} {c_6}}{{{\pi }^4}}+\frac{848 {c_1} {c_2} {c_5} {c_6}}{{{\pi }^3}}+  {}
{}
 \frac{116 {c_3} {c_5} {c_6}}{{{\pi }^3}}-\frac{16 {c_3} {c_5} {c_6}}{{{\pi }^2}}-\frac{624
c_{1}^{2} c_{6}^{2}}{{{\pi }^4}}+\frac{96 c_{1}^{2} c_{6}^{2}}{{{\pi }^3}}+\frac{8 c_{1}^{2} c_{6}^{2}}{{{\pi
}^2}}+  {}
{}
 \frac{44 {c_2} c_{6}^{2}}{{{\pi }^3}}-\frac{4 {c_2} c_{6}^{2}}{{{\pi }^2}}-\frac{1268096 c_{1}^{7}
{c_7}}{{{\pi }^8}}+\frac{203008 c_{1}^{7} {c_7}}{{{\pi }^7}}+\frac{103712 c_{1}^{7} {c_7}}{3 {{\pi }^6}}+  {}
{}
 \frac{592 c_{1}^{7} {c_7}}{{{\pi }^5}}+\frac{2592 c_{1}^{7} {c_7}}{5 {{\pi }^4}}+\frac{37
c_{1}^{7} {c_7}}{3 {{\pi }^3}}+\frac{349 c_{1}^{7} {c_7}}{56 {{\pi }^2}}+\frac{1622784 c_{1}^{5}
{c_2} {c_7}}{{{\pi }^7}}-  {}
{}
 \frac{403968 c_{1}^{5} {c_2} {c_7}}{{{\pi }^6}}-\frac{9992 c_{1}^{5} {c_2} {c_7}}{{{\pi }^5}}-\frac{1408
c_{1}^{5} {c_2} {c_7}}{{{\pi }^4}}+\frac{211 c_{1}^{5} {c_2} {c_7}}{30 {{\pi }^3}}-  {}
{}
 \frac{15 c_{1}^{5} {c_2} {c_7}}{{{\pi }^2}}-\frac{451776 c_{1}^{3} c_{2}^{2} {c_7}}{{{\pi
}^6}}+\frac{127072 c_{1}^{3} c_{2}^{2} {c_7}}{{{\pi }^5}}-\frac{1160 c_{1}^{3} c_{2}^{2} {c_7}}{3
{{\pi }^4}}-  {}
{}
 \frac{224 c_{1}^{3} c_{2}^{2} {c_7}}{3 {{\pi }^3}}+\frac{10 c_{1}^{3} c_{2}^{2} {c_7}}{{{\pi
}^2}}+\frac{18400 {c_1} c_{2}^{3} {c_7}}{{{\pi }^5}}-\frac{4192 {c_1} c_{2}^{3} {c_7}}{{{\pi }^4}}-  {}
{}
 \frac{278 {c_1} c_{2}^{3} {c_7}}{3 {{\pi }^3}}-\frac{8 {c_1} c_{2}^{3} {c_7}}{{{\pi
}^2}}-\frac{192544 c_{1}^{4} {c_3} {c_7}}{{{\pi }^6}}+\frac{48096 c_{1}^{4} {c_3} {c_7}}{{{\pi }^5}}+ 
{}
{}
 \frac{2792 c_{1}^{4} {c_3} {c_7}}{3 {{\pi }^4}}+\frac{72 c_{1}^{4} {c_3} {c_7}}{{{\pi
}^3}}+\frac{6 c_{1}^{4} {c_3} {c_7}}{{{\pi }^2}}+\frac{70912 c_{1}^{2} {c_2} {c_3} {c_7}}{{{\pi
}^5}}-  {}
{}
 \frac{19568 c_{1}^{2} {c_2} {c_3} {c_7}}{{{\pi }^4}}+\frac{252 c_{1}^{2} {c_2} {c_3}
{c_7}}{{{\pi }^3}}-\frac{4 c_{1}^{2} {c_2} {c_3} {c_7}}{{{\pi }^2}}-\frac{1552 c_{2}^{2} {c_3} {c_7}}{{{\pi
}^4}}+  {}
{}
 \frac{304 c_{2}^{2} {c_3} {c_7}}{{{\pi }^3}}+\frac{8 c_{2}^{2} {c_3} {c_7}}{{{\pi }^2}}-\frac{1552
{c_1} c_{3}^{2} {c_7}}{{{\pi }^4}}+\frac{304 {c_1} c_{3}^{2} {c_7}}{{{\pi }^3}}+\frac{8 {c_1} c_{3}^{2}
{c_7}}{{{\pi }^2}}+  {}
{}
 \frac{18400 c_{1}^{3} {c_4} {c_7}}{{{\pi }^5}}-\frac{4192 c_{1}^{3} {c_4} {c_7}}{{{\pi }^4}}-\frac{278
c_{1}^{3} {c_4} {c_7}}{3 {{\pi }^3}}-\frac{8 c_{1}^{3} {c_4} {c_7}}{{{\pi }^2}}-  {}
{}
 \frac{3848 {c_1} {c_2} {c_4} {c_7}}{{{\pi }^4}}+\frac{896 {c_1} {c_2} {c_4}
{c_7}}{{{\pi }^3}}+\frac{108 {c_3} {c_4} {c_7}}{{{\pi }^3}}-\frac{16 {c_3} {c_4} {c_7}}{{{\pi }^2}}-  {}
{}
 \frac{1552 c_{1}^{2} {c_5} {c_7}}{{{\pi }^4}}+\frac{304 c_{1}^{2} {c_5} {c_7}}{{{\pi }^3}}+\frac{8
c_{1}^{2} {c_5} {c_7}}{{{\pi }^2}}+\frac{108 {c_2} {c_5} {c_7}}{{{\pi }^3}}-  {}
{}
 \frac{16 {c_2} {c_5} {c_7}}{{{\pi }^2}}+\frac{108 {c_1} {c_6} {c_7}}{{{\pi }^3}}-\frac{16
{c_1} {c_6} {c_7}}{{{\pi }^2}}-\frac{2 c_{7}^{2}}{{{\pi }^2}}+\frac{160640 c_{1}^{6} {c_8}}{{{\pi }^7}}-  {}
{}
 \frac{24704 c_{1}^{6} {c_8}}{{{\pi }^6}}-\frac{10888 c_{1}^{6} {c_8}}{3 {{\pi }^5}}-\frac{520
c_{1}^{6} {c_8}}{3 {{\pi }^4}}-\frac{4621 c_{1}^{6} {c_8}}{90 {{\pi }^3}}-\frac{14 c_{1}^{6} {c_8}}{3
{{\pi }^2}}-  {}
{}
 \frac{167104 c_{1}^{4} {c_2} {c_8}}{{{\pi }^6}}+\frac{39328 c_{1}^{4} {c_2} {c_8}}{{{\pi
}^5}}+\frac{844 c_{1}^{4} {c_2} {c_8}}{{{\pi }^4}}+\frac{484 c_{1}^{4} {c_2} {c_8}}{3 {{\pi }^3}}+
 {}
{}
 \frac{9 c_{1}^{4} {c_2} {c_8}}{{{\pi }^2}}+\frac{30384 c_{1}^{2} c_{2}^{2} {c_8}}{{{\pi }^5}}-\frac{7472
c_{1}^{2} c_{2}^{2} {c_8}}{{{\pi }^4}}-\frac{90 c_{1}^{2} c_{2}^{2} {c_8}}{{{\pi }^3}}-\frac{2 c_{1}^{2}
c_{2}^{2} {c_8}}{{{\pi }^2}}-  {}
{}
 \frac{336 c_{2}^{3} {c_8}}{{{\pi }^4}}+\frac{40 c_{2}^{3} {c_8}}{{{\pi }^3}}+\frac{3 c_{2}^{3}
{c_8}}{{{\pi }^2}}+\frac{18224 c_{1}^{3} {c_3} {c_8}}{{{\pi }^5}}-\frac{4128 c_{1}^{3} {c_3} {c_8}}{{{\pi
}^4}}-  {}
{}
 \frac{278 c_{1}^{3} {c_3} {c_8}}{3 {{\pi }^3}}-\frac{8 c_{1}^{3} {c_3} {c_8}}{{{\pi
}^2}}-\frac{3720 {c_1} {c_2} {c_3} {c_8}}{{{\pi }^4}}+\frac{848 {c_1} {c_2} {c_3} {c_8}}{{{\pi
}^3}}+  {}
{}
 \frac{44 c_{3}^{2} {c_8}}{{{\pi }^3}}-\frac{4 c_{3}^{2} {c_8}}{{{\pi }^2}}-\frac{1576 c_{1}^{2}
{c_4} {c_8}}{{{\pi }^4}}+\frac{304 c_{1}^{2} {c_4} {c_8}}{{{\pi }^3}}+\frac{8 c_{1}^{2} {c_4} {c_8}}{{{\pi
}^2}}+  {}
{}
 \frac{116 {c_2} {c_4} {c_8}}{{{\pi }^3}}-\frac{16 {c_2} {c_4} {c_8}}{{{\pi }^2}}+\frac{116
{c_1} {c_5} {c_8}}{{{\pi }^3}}-\frac{16 {c_1} {c_5} {c_8}}{{{\pi }^2}}-\frac{6 {c_6} {c_8}}{{{\pi
}^2}}-  {}
{}
 \frac{20416 c_{1}^{5} {c_9}}{{{\pi }^6}}+\frac{2976 c_{1}^{5} {c_9}}{{{\pi }^5}}+\frac{368 c_{1}^{5}
{c_9}}{{{\pi }^4}}+\frac{92 c_{1}^{5} {c_9}}{3 {{\pi }^3}}+\frac{91 c_{1}^{5} {c_9}}{20 {{\pi }^2}}+  {}
{}
 \frac{16400 c_{1}^{3} {c_2} {c_9}}{{{\pi }^5}}-\frac{3552 c_{1}^{3} {c_2} {c_9}}{{{\pi }^4}}-\frac{242
c_{1}^{3} {c_2} {c_9}}{3 {{\pi }^3}}-\frac{12 c_{1}^{3} {c_2} {c_9}}{{{\pi }^2}}-  {}
{}
 \frac{1576 {c_1} c_{2}^{2} {c_9}}{{{\pi }^4}}+\frac{304 {c_1} c_{2}^{2} {c_9}}{{{\pi }^3}}+\frac{8
{c_1} c_{2}^{2} {c_9}}{{{\pi }^2}}-\frac{1576 c_{1}^{2} {c_3} {c_9}}{{{\pi }^4}}+  {}
{}
 \frac{304 c_{1}^{2} {c_3} {c_9}}{{{\pi }^3}}+\frac{8 c_{1}^{2} {c_3} {c_9}}{{{\pi }^2}}+\frac{116
{c_2} {c_3} {c_9}}{{{\pi }^3}}-\frac{16 {c_2} {c_3} {c_9}}{{{\pi }^2}}+\frac{116 {c_1} {c_4}
{c_9}}{{{\pi }^3}}-  {}
{}
 \frac{16 {c_1} {c_4} {c_9}}{{{\pi }^2}}-\frac{6 {c_5} {c_9}}{{{\pi }^2}}+\frac{2608 c_{1}^{4}
{c_{10}}}{{{\pi }^5}}-\frac{352 c_{1}^{4} {c_{10}}}{{{\pi }^4}}-\frac{106 c_{1}^{4} {c_{10}}}{3 {{\pi }^3}}- 
{}
{}
 \frac{4 c_{1}^{4} {c_{10}}}{{{\pi }^2}}-\frac{1488 c_{1}^{2} {c_2} {c_{10}}}{{{\pi }^4}}+\frac{280
c_{1}^{2} {c_2} {c_{10}}}{{{\pi }^3}}+\frac{8 c_{1}^{2} {c_2} {c_{10}}}{{{\pi }^2}}+  {}
{}
 \frac{44 c_{2}^{2} {c_{10}}}{{{\pi }^3}}-\frac{4 c_{2}^{2} {c_{10}}}{{{\pi }^2}}+\frac{116 {c_1}
{c_3} {c_{10}}}{{{\pi }^3}}-\frac{16 {c_1} {c_3} {c_{10}}}{{{\pi }^2}}-\frac{6 {c_4} {c_{10}}}{{{\pi }^2}}-
 {}
{}
 \frac{336 c_{1}^{3} {c_{11}}}{{{\pi }^4}}+\frac{40 c_{1}^{3} {c_{11}}}{{{\pi }^3}}+\frac{3 c_{1}^{3}
{c_{11}}}{{{\pi }^2}}+\frac{116 {c_1} {c_2} {c_{11}}}{{{\pi }^3}}-\frac{16 {c_1} {c_2} {c_{11}}}{{{\pi
}^2}}-  {}
{}
 \frac{6 {c_3} {c_{11}}}{{{\pi }^2}}+\frac{44 c_{1}^{2} {c_{12}}}{{{\pi }^3}}-\frac{4 c_{1}^{2}
{c_{12}}}{{{\pi }^2}}-\frac{6 {c_2} {c_{12}}}{{{\pi }^2}}-\frac{6 {c_1} {c_{13}}}{{{\pi }^2}}+\frac{{c_{14}}}{\pi }
$


\begin{references}
\bibitem{expansion} The complete expansion to $14^{th}$ order is  
displayed in the Appendix, beginning on page 6.
\end{references}
\end{document}